\documentclass[%
 reprint,
superscriptaddress,
nofootinbib,
 amsmath,amssymb,
 aps,
 prd,
floatfix,
]{revtex4-2}
\usepackage{tabularx}
\usepackage[table]{xcolor}
\definecolor{tableShade}{gray}{0.9}
\usepackage{graphicx}% Include figure files
\usepackage{dcolumn}% Align table columns on decimal point
\usepackage{bm}% bold math
\usepackage[acronym]{glossaries}

\preto{\abstractkeywords}{\nolinenumbers}
\usepackage{enumitem}
\usepackage{amsmath}
\usepackage[normalem]{ulem}
\usepackage{dcolumn}% Align table columns on decimal point
\usepackage{tabularx}
\usepackage{multirow}
\usepackage{booktabs}
\usepackage{orcidlink}
\usepackage{import}
\usepackage{amsmath}
\usepackage{amssymb}
\usepackage{graphicx}% Include figure files
\usepackage[normalem]{ulem}
\usepackage{dcolumn}% Align table columns on decimal point
\usepackage{afterpage}
\usepackage{multirow}
\usepackage{booktabs}
\usepackage{hhline}
\usepackage{makecell}
\usepackage{array}
\usepackage{float}
\usepackage{soul}
\usepackage{comment}
\usepackage{bm}
\usepackage{hyperref}
\hypersetup{
    colorlinks=true,
    linkcolor=blue,
    filecolor=magenta,      
    urlcolor=blue,
    citecolor=violet
}
\usepackage{longtable}
\def\la{\lambda}

\newcommand{\beq}{\begin{equation}}
\newcommand{\eeq}{\end{equation}}
\newcommand{\bea}{\begin{eqnarray}}
\newcommand{\eea}{\end{eqnarray}}

\def\la{\ell^a_{\text{max}}}
\def\lb{\ell^b_{\text{max}}}
\def\lmax{\ell_{\text{max}}}

\newcommand{\blipp}{{\tt BLIP}}
\newcommand{\blip}{{\tt BLIP} }
\newcommand{\blips}{{\tt BLIP}'s\ }

\begin{document}

\title{Angular Resolution of a Bayesian Search for Anisotropic Stochastic Gravitational Wave Backgrounds with LISA}
% \title[Angular Resolution of ASGWB searches with LISA]{Angular Resolution of the Search for Anisotropic Stochastic Gravitational Wave Backgrounds with LISA
% }

% Searching for anisotropic symmetry breaking in mass-dimension 6 gravitational wave dispersion
% Force line breaks with \\

\author{Malachy Bloom\,\orcidlink{0000-0001-5489-4204}}
\email{bloomm@carleton.edu}
\affiliation{Carleton College, Northfield, MN 55057, USA}

\author{Alexander W. Criswell\,\orcidlink{0000-0002-9225-7756}}
\affiliation{Minnesota Institute for Astrophysics, University of Minnesota, Minneapolis, MN 55455, USA}
\affiliation{School of Physics and Astronomy, University of Minnesota, Minneapolis, MN 55455, USA}
\affiliation{Department of Physics and Astronomy, Vanderbilt University, Nashville, TN 37240}
\affiliation{Department of Life and Physical Sciences, Fisk University, Nashville, TN 37208}

\author{Vuk Mandic\,\orcidlink{0000-0001-6333-8621}}
\affiliation{Minnesota Institute for Astrophysics, University of Minnesota, Minneapolis, MN 55455, USA}
\affiliation{School of Physics and Astronomy, University of Minnesota, Minneapolis, MN 55455, USA}

\date{\today}% It is always \today, today,

\begin{abstract}
The Laser Interferometer Space Antenna (LISA), a spaceborne gravitational wave (GW) detector set to launch in 2035, will observe several stochastic GW backgrounds in the mHz frequency band. At least one of these signals --- arising from the tens of millions of unresolved white dwarf binaries in the Milky Way --- is expected to be highly anisotropic on the sky. We evaluate the angular resolution of LISA and its ability to characterize anisotropic stochastic GW backgrounds (ASGWBs) using the Bayesian Spherical Harmonic formalism in the Bayesian LISA Inference Package (\blipp). We use \blip to simulate and analyze ASGWB signals in LISA across a large grid in total observing time, ASGWB amplitude, and angular size. We consider the ability of  the \blip anisotropic search algorithm to both characterize single point sources and to separate two point sources on the sky, using a full-width half-max (FWHM) metric to measure the quality and spread of the recovered spatial distributions. We find that the number of spherical harmonic coefficients used in the anisotropic search model is the primary factor that limits the search's angular resolution. Notably, this trend continues until computational limitations become relevant around $\ell_{\mathrm{max}}=16$; this exceeds the maximum angular resolution achieved by other map-making techniques for LISA ASGWBs.

\end{abstract}

% \keywords{Lorentz invariance violation, CPT symmetry breaking, spacetime birefringence, gravitational waves, gravity}%Use showkeys class option if keyword
\maketitle

\section{Introduction}

% Outline
% \begin{itemize}
%     \item what is a LISA
%     \item what is a SGWB/LISA expected to observe SGWBs
%     \item intro to anisotropic SGWBs. MW, LMC.
%     \item previous (analytic) expectations of LISA sensitivity to angular modes
%     \item extant anisotropic searches for e.g. LIGO/PTA? (in brief if at all)
%     \item enter BLIP. Only extant anisotropic SGWB search for LISA (to our knowledge) 
%     \item Description of BLIP
%     \item Description of BLIP spherical harmonic search, detailing blm/alms, motivation behind this setup \textcolor{red}{[AC: maybe this belongs in methods, we'll see]}
%     \item Motivation for understanding angular resolution \textcolor{red}{[AC: would be good to ask Vuk here, there are probably some additional motivations I'm unaware of.]}
% \end{itemize}

\subsection{Anisotropic Stochastic Gravitational Wave Backgrounds in LISA}
The Laser Interferometer Space Antenna (LISA) is a future spaceborne mHz-band gravitational-wave (GW) observatory set to launch in 2035 \citep{amaro-seoane_laser_2017}. A wide variety of astrophysical sources are expected to emit GWs in the LISA band (see \citet{amaro-seoane_astrophysics_2023} for a review); while some of these signals are expected to be individually detectable, many others will remain unresolved, instead contributing to one of several stochastic gravitational wave backgrounds (SGWBs). SGWBs are confusion noise formed by the superposition of many unresolved astrophysical or cosmological GW sources. Recently, \citet{agazie_the_2023a} reported evidence for such a signal in the nHz band using pulsar timing arrays. LISA data is expected to feature a number of mHz SGWBs, including those from galactic (and extragalactic) white dwarf binary systems \citep{edlund_white_2005,ruiter_lisa_2010,rieck_lisa_2024}, extreme-mass-ratio inspirals \citep{pozzoli_computation_2023c}, and stellar-origin compact binaries (e.g., \citep{babak_stochastic_2023}), among others. Notably, the SGWB from unresolved galactic white dwarf binaries in the Milky Way is expected to be so loud as to sit above the LISA noise curve, and is therefore termed the ``galactic foreground" \citep{edlund_white_2005,ruiter_lisa_2010}. Both the galactic foreground and potential SGWBs from white dwarf binaries in satellite galaxies of the Milky Way \citep{rieck_lisa_2024} will be highly anisotropic due to the spatial distribution of their component sources. Additionally, it is possible LISA may contain SGWBs from a number of cosmological sources, such as inflation, phase transitions, cosmic (super)strings, and primordial black holes; see \citet{caprini_cosmological_2018} for a review. All SGWBs from non-local sources, astrophysical and cosmological alike, can be expected to contain some level of anisotropy due to the large scale structure of the universe, deriving from both the specific source distribution and propagation effects \citep{bethke_anisotropies_2013,geller_primordial_2018,bartolo_gravitational_2020,contaldi_anisotropies_2017,bartolo_anisotropies_2019,bartolo_characterizing_2020}. Characterizing the spatial distribution on the sky of anisotropic SGWBs (ASGWBs) is crucial; not only does proper subtraction of an ASGWB require accurate description of its spatial distribution, but accurate spatial characterization could also enable scientific gains from the ASGWBs in question\citep[e.g.,][]{breivik_constraining_2020}. Moreover, constraining the level of anisotropy of astrophysical SGWBs can inform our understanding of the large-scale structure of the Universe \citep{cusin_the_2018,cusin_first_2018}. Finally, constraining the anisotropy of potential cosmological SGWBs could provide a novel window into the physics of the early Universe \citep{bartolo_probing_2022}.

\subsection{Anisotropic SGWB Searches}
Several analyses exist to search for and characterize ASGWBs in ground-based GW detectors \citep{floden_angular_2022,abbott_search_2021} and Pulsar Timing Arrays \citep{ali-haimoud_fisher_2020,mingarelli_characterizing_2013,taylor_searching_2013,taylor_from_2020,cornish_mapping_2014,agazie_the_2023d}. For LISA, there are several studies that use Fisher-matrix or other frequentist approaches~\citep{Cornish:2001hg, Ungarelli:2001xu, Kudoh:2004he, Taruya:2005yf, Taruya:2006kqa, Renzini:2018vkx, bartolo_probing_2022}. Additionally, Bayesian methods have been developed for LISA analyses of the anisotropic Galactic foreground and other local-universe ASGWBs \citep{buscicchio_a_2024a, piarulli_a_2024b, pozzoli_cyclostationary_2024,criswell_templated_2024}. However, the only extant analysis infrastructure capable of performing a generic Bayesian anisotropic search for SGWBs in LISA is, to the authors' knowledge at time of writing, the Bayesian LISA Inference Package (\blipp). \blip is an open souce Python package capable of end-to-end simulation and Bayesian inference of isotropic and anisotropic SGWBs in LISA, first presented in \citet{banagiri_mapping_2021b}.\footnote{\href{https://github.com/sharanbngr/blip}{https://github.com/sharanbngr/blip}} A full description of \blip and its spherical harmonic ASGWB search can be found in \citet{banagiri_mapping_2021b}; details relevant to this study are presented in brief in \S\ref{sec:blip}.

Given the potential scientific gains that accompany accurate characterization of ASGWBs in LISA, it is important to understand the capabilities of LISA for characterizing anisotropic sources. LISA's angular resolution has been previously considered: first by \citet{peterseim_angular_1997,cutler_angular_1998,moore_angular_2002} and followed thereafter with studies of angular resolution to ASGWBs for frequentist approaches specifically \citep{kudoh_probing_2005,taruya_probing_2005,contaldi_maximum_2020,bartolo_probing_2022,mentasti_probing_2024a}. While earlier studies \citep{kudoh_probing_2005,taruya_probing_2005} found that LISA's angular resolution would be limited to $\lmax\leq4$, the improved formalism of \citet{contaldi_maximum_2020} may be able to achieve a limit of $\lmax\lesssim15$ for the optimal case ($\lmax\lesssim8$ for more realistic considerations \citep{bartolo_probing_2022,mentasti_probing_2024a}). However, the angular resolution of Bayesian ASGWB searches with LISA has not previously been considered. Doing so is especially important given that LISA ASGWB searches are likely to take place in a global fit setting \citep[e.g.,][]{littenberg_prototype_2023,katz_an_2024} and as such will necessarily be Bayesian in nature. 

We therefore investigate the angular resolution of LISA and its ability to characterize ASGWBs with a Bayesian approach using the Bayesian spherical harmonic search in \blip. We consider the ability of the spherical harmonic Bayesian search to characterize both single and double point-source ASGWBs across a broad range of signal amplitudes, spot sizes, and choice of spherical harmonic $\ell_{\mathrm{max}}$ used to parameterize the analysis. We discuss the \blip Bayesian spherical harmonic search in \S\ref{sec:blip}; the ASGWB simulation method is outlined in \S\ref{sec:sims}. Analyses of single-point and two-point source configurations are presented in \S\ref{sec:sp} and \S\ref{sec:tp}, respectively. Discussion of these results and their implications for LISA's angular resolution with Bayesian ASGWB searches is given in \S\ref{sec:conclusion}, alongside directions of future inquiry.

\section{Methodology}\label{sec:method}

\subsection{The \blip Anisotropic Analysis}\label{sec:blip}
\blips Bayesian spherical harmonic ASGWB search performs simultaneous inference of the LISA noise spectrum, the ASGWB spectrum, and the ASGWB's spatial distribution as represented in a spherical harmonic expansion. The underlying mechanics of the \blip ASGWB search algorithm are based on formalisms developed in \citet{cornish_space_2001} and \citet{cornish_detecting_2001}, with the central addition of the spherical harmonic treatment described below and discussed in detail in \citet{banagiri_mapping_2021b}. Importantly, the \blip anisotropic search explicitly models the motion of the LISA detector through space and, accordingly, the time-dependence of its directional response to ASGWBs. A full derivation of this response in the spherical harmonic basis can be found in \citet{banagiri_mapping_2021b}. We use \texttt{dynesty} \citep{speagle_dynesty:_2020} nested sampling to produce Bayesian posterior distributions on all model parameters. The LISA instrumental noise is modelled with the two-component position and acceleration noise model given in the LISA proposal \citep{amaro-seoane_laser_2017}:
\begin{equation}\label{eq:instr_noise}
	\begin{split}
		S_p (f) = & N_p \left[1 + \left(\frac{2 \, \text{mHz}}{f}\right)^4 \right] \text{Hz}^{-1}, \\
		S_a (f) = &  \left[1 + \left(\frac{0.4 \, \text{mHz}}{f} \right)^2\right]\\
        &\qquad\times \left[1 + \left(\frac{f}{8 \, \text{mHz}}  \right)^4 \right] \times \frac{N_a}{(2 \pi f)^4 } \text{Hz}^{-1},
	\end{split}
\end{equation}
where $N_a$ and $N_p$ are independent parameters to be inferred. We assume the spectral and spatial dependence of the ASGWB can be separated; that is we model the ASGWB as
\begin{equation}\label{eq:spectral_spatial}
    \Omega_{\mathrm{GW}}(f,\mathbf{n};\vec\theta) = \Omega(f; \vec{\theta}) \mathcal{P}(\mathbf{n}),
\end{equation}
where $\Omega(f; \vec{\theta})$ is the spectral shape of the SGWB parameterized by $\vec{\theta}$, given in terms of the dimensionless GW energy density\footnote{This is related to the strain power spectrum of the ASGWB, $S_{\mathrm{GW}}$, by $\Omega(f) = (2\pi^2f^3/3H_0^2)S_{\mathrm{GW}}(f)$. We assume a value of $H_0 = 67.88$ km/s/Mpc.}, and $\mathcal{P}(\mathbf{n})$ is the corresponding spatial distribution (whose integral over the sky is normalized to 1). For this study, we assume a power law spectral shape for the ASGWB:
\begin{equation}\label{eq:powerlaw}
    \Omega(f; \vec{\theta}) = \Omega_{\text{ref}} \left ( \frac{f}{f_\text{ref}} \right )^{\alpha}, 
\end{equation}
where $\vec{\theta} = \{\Omega_{\mathrm{ref}},\alpha\}$, $\Omega_{\mathrm{ref}}$ is the dimensionless GW energy density evaluated at the reference frequency $f_{\mathrm{ref}}=25$ Hz, and $\alpha$ is the power law spectral index. 

The ASGWB spatial distribution $\mathcal{P}(\mathbf{n})$ is modelled using a spherical harmonic expansion up to a user-defined value of $\ell_{\mathrm{max}}$. Specifically, \blip infers the spherical harmonic expansion coefficients $b_{\ell m}$ of the \textit{square root} of the power on the sky, $\mathcal{S}(\mathbf{n})$, such that
\begin{equation}\label{sph-harm}
\begin{split}
    \mathcal{S}(\mathbf{n}) = \sqrt{\mathcal{P}(\mathbf{n})} &= \left[\sum_{\ell=0}^{\ell_{\rm max}^a} \sum_{m = -\ell}^{\ell} a_{\ell m}Y_{\ell m}(\mathbf{n}) \right]^{1/2} \\
    &= \sum_{\ell=0}^{\ell_{\rm max}^b} \sum_{m = -\ell}^{\ell} b_{\ell m}Y_{\ell m}(\mathbf{n}).
\end{split}
\end{equation}
Fitting the square root of the distribution mathematically ensures that all possible inferred distributions of power on the sky are not only real but also non-negative (a non-trivial problem; see \citet{banagiri_mapping_2021b}). The $b_{\ell m}$s are related to the usual spherical harmonic $a_{\ell m}$s via a Clebsch-Gordan decomposition:
\begin{equation}\label{cg_decomp}
\begin{split}
    a_{\ell m}= \sum_{L,M} \sum_{L',M'}  b_{LM} b_{L'M'} & \sqrt{\frac{(2L+1)(2L'+1)}{4\pi(2L+1)}} \\ & \times C_{\ell m}^{LM,L'M'}C_{\ell0}^{L0,L'0},
\end{split}
\end{equation}
with associated selection rules \citep{banagiri_mapping_2021b}. An important consequence of this is that the truncation $\ell_{\mathrm{max}}$ for each expansion are related by $\la = 2 \lb$. 

Each analysis therefore infers the log power law amplitude $\log_{10}\Omega_{\mathrm{ref}}$, the power law index $\alpha$, and the spherical harmonic coefficients $b_{\ell m}$. These latter are parameterized in terms of their amplitude $|b_{\ell m}|$ (for all $\ell,m$) and their phase $\phi_{\ell m}$ (for complex $b_{\ell m}$ coefficients, i.e. those with $m\neq0$). The priors used in this work are as follows:
\begin{itemize}
    \item $\pi(\log_{10}\Omega_{\mathrm{ref}}) = \mathcal{U}(-14,8)$
    \item $\pi(\alpha) = \mathcal{U}(-5,5)$
    \item $\pi(b_{\ell 0}) = \mathcal{U}(-3,3)$ for real $b_{\ell 0}$
    \item $\pi(|b_{\ell m}|) = \mathcal{U}(0,3)$ for complex $b_{\ell m}$, $m\neq0$
    \item $\pi(\phi_{\ell m}) = \mathcal{U}(-\pi,\pi)$ for complex $b_{\ell m}$, $m\neq0$
\end{itemize}

For further details on the \blip anisotropic treatment, its Fourier-domain likelihood, and the underlying mechanics of the \blip analysis infrastructure, refer to \citet{banagiri_mapping_2021b}.

\subsection{ASGWB Simulations}\label{sec:sims}

We consider two types of simulated ASGWB: one consisting of a single, localized source (``single point" simulations), and one consisting of two such sources (``two point" simulations). Each simulated ASGWB has a power-law spectrum with spectral index $\alpha = 2/3$ and has its spatial distribution represented in the spherical harmonic basis for direct comparison to the \blip ASGWB analysis. For the single-point case, we perform simulations of LISA ASGWB data across a grid in three major parameters:  $\ell^a_{\text{max, inj}}$, the spherical harmonic cutoff used to simulate the ASGWB in the spherical harmonic basis; $\Omega_{\textrm{ref}}$, the amplitude of the simulated power-law ASGWB spectrum; and $ T_{\mathrm{obs}}$, the duration of the simulated data. For the two-point case, we additionally vary $\Delta \phi$, the angular separation (in radians) between the two component sources. All single-point sources are simulated at an ecliptic sky position of $(\theta,\phi) = (\frac{\pi}{2},\frac{\pi}{2})$. The sky positions for the two-point case vary due to changing their separation $\Delta\phi$; the sky positions $(\theta_1,\phi_1),(\theta_2,\phi_2)$ for all two-point simulations are given in Table~\ref{table:tp_sims}. We do not otherwise vary sky location as a parameter of interest as the directional sensitivity of LISA is known and can be accounted for analytically.

% We perform simulations of LISA ASGWB data across a grid in four major parameters:  $\ell^a_{\text{max, inj}}$, the spherical harmonic cutoff used to simulate ASGWBs in the spherical harmonic basis; $\Omega_{\textrm{ref}}$, the amplitude of the simulated power-law ASGWB spectrum; $ T_{\mathrm{obs}}$, the length of the simulated data; and for two-point simulations $\Delta \phi$, the angular separation (in radians) between the two point sources. We consider two types of simulation: an ASGWB consisting of a single point source, and one consisting of two point sources. 

% Each simulated point source is represented in the spherical harmonic basis with $\ell^a_{\text{max, inj}}$ for direct comparison to the \blip ASGWB analysis. 
% For the two-point configuration, we vary the angular separation of the two point sources, $\Delta \phi$. 
We analyze each simulation at several different values of $\la$, the spherical harmonic $\lmax$ in the recovery model.\footnote{Note that $\la$ and $\ell^a_{\text{max, inj}}$ need not be the same; we in fact consider the effects of mismatch between these parameters in \S\ref{sec:sp}. We restrict analysis-simulation pairs such that $\la \leq \ell^a_{\text{max, inj}}$.} All simulations and analyses in this work use skymaps with a Healpy \citep{gorski_healpix_2005} {\tt nside} (pixel map resolution) of 32 (corresponding angular scale of $\ell_a\approx98$), and simulate/analyze data in the $\mathrm{X}-\mathrm{Y}-\mathrm{Z}$ time-delay interferometry channels \citep{tinto_time-delay_2020}. A summary of the simulation grid can be found in Table~\ref{table:params}; a full list of all simulations and associated parameters can be found in Appendix~\ref{appendix:simdetails}. The maximum values of $\la$ and $\ell^a_{\text{max, inj}}$ are driven by computational limitations due to both the memory requirements and convolution cost of the $3\times 3 \times n_f \times n_t \times n_{\mathbf{n}} \times \lb(\lb+1)$ LISA spherical harmonic response, where $n_f$, $n_t$, and $n_{\mathbf{n}}$ are the number of frequency bins, time segments, and sky pixels,\footnote{Note that the final response used at sample-time is integrated over sky direction, so the convolutional cost at sample time is ``only" driven by matrix operations with a $3\times 3 \times n_f \times n_t \times \lb(\lb+1)$ array.} respectively.\footnote{``Computational limits" can be taken to mean either exceeding $\sim2$ TB of RAM or having such a high likelihood evaluation cost that the \texttt{dynesty} sampler used in this work would not converge on a timescale of $\lesssim1$ month (wall time). } The maximum of $\Omega_{\mathrm{ref}}$ is chosen to be comparable to the amplitude of the Galactic white dwarf binary foreground.  We simulated and analyzed  ASGWBs with values of $\Omega_{\mathrm{ref}}$ below the listed minimum, but were unable to obtain consistent recoveries with the observing durations considered due to the simulated signals' low signal-to-noise ratio (SNR); see \S\ref{sec:sp}. We therefore did not fill out the full grid at these subthreshold values of $\Omega_{\mathrm{ref}}$ to avoid undue waste of computational resources. It will be important in future to explore the realistic case of using a Bayesian spherical harmonic search to characterize low-SNR ASGWBs in the presence of the Galactic foreground over the nominal LISA mission duration. However, as this study is concerned with the general performance and limitations of Bayesian LISA ASGWB analyses, we leave a dedicated study of this more realistic case to future work (and note that a proof-of-concept of such a search can be found in \citet{rieck_lisa_2024}). The range of $ T_{\mathrm{obs}}$ is chosen so as to probe the effects of LISA's orbit on the angular resolution of our anisotropic search. While the nominal LISA mission duration is 4 years, the scaling of SGWB sensitivity with observing time is well-understood \citep{adams_discriminating_2010} and further complete orbits are expected to follow the standard $\sqrt{T_{\mathrm{obs}}}$. Finally, the range of $\Delta \phi$ spans $(\pi/5,\pi)$ radians at an angular step of $\pi/5$, chosen to maintain feasibility in total number of simulations performed while still exploring the full gamut of two-point separations on the sky.

\begin{table}
\scriptsize
\centering
\rowcolors{2}{}{lightgray}
\renewcommand{\arraystretch}{2}
\newcolumntype{S}{@{\centering\arraybackslash}m{0.45cm}}
\begin{tabularx}{\linewidth}{>{\centering\arraybackslash}m{3.08cm}S@{\centering\arraybackslash}m{2.3cm}S@{\centering\arraybackslash}m{2.3cm}}
\hline
Parameter & \hspace{1.25cm}& Minimum &\hspace{1.25cm} & Maximum   \\
\hline
$\la$ && 4 && 16  \\
$\Omega_{\text{ref}}$ && $1.6\times 10^{-9}$ && $4.0\times 10^{-7}$  \\
$ T_{\mathrm{obs}}$ && 3 months && 1 year \\
$\ell^a_{\text{max, inj}}$ && 4 && 16  \\
$\Delta \phi$ && $\pi/5$ && $\pi$  \\
\hline
\end{tabularx}
\caption{Summary of simulation parameters.}
\label{table:params}
\end{table}

\subsection{Angular Resolution Metrics}\label{sec:metrics}
\begin{figure}
    \includegraphics[width=\columnwidth]{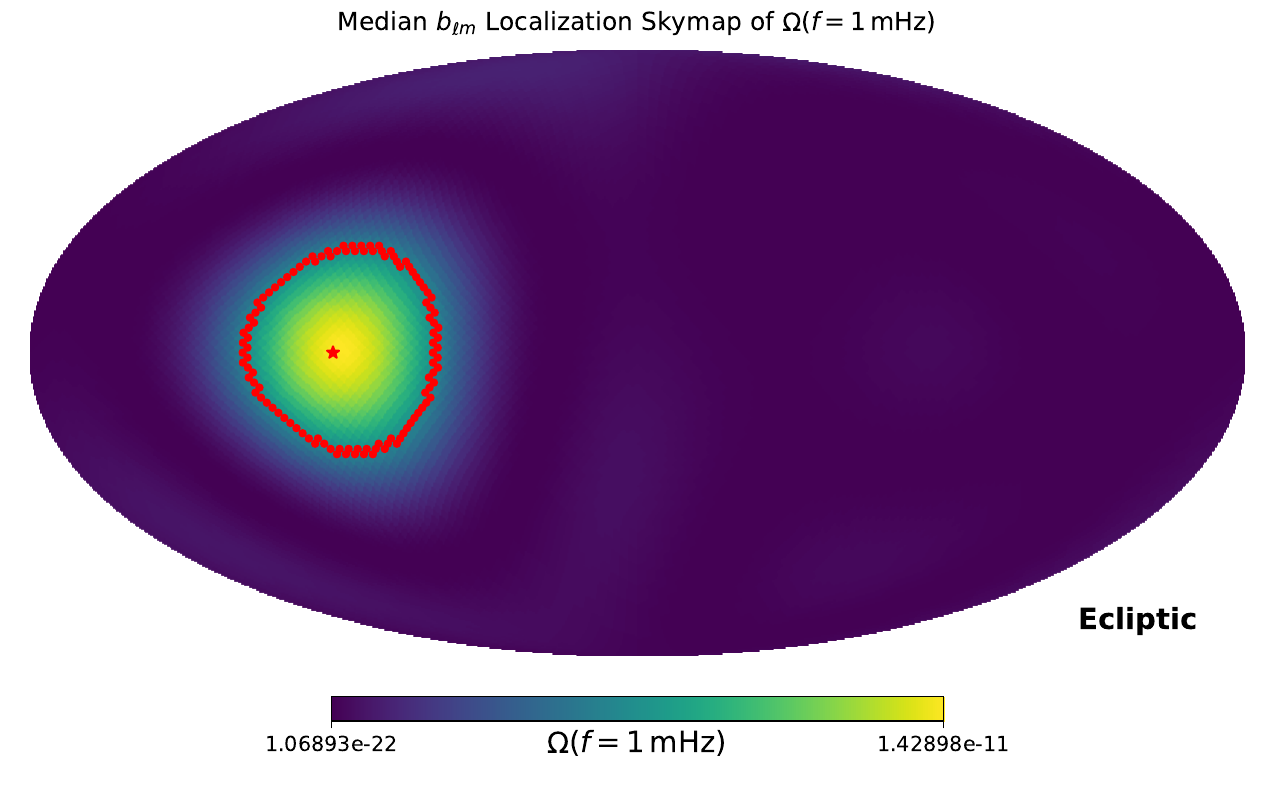}
    \includegraphics[width=\columnwidth]{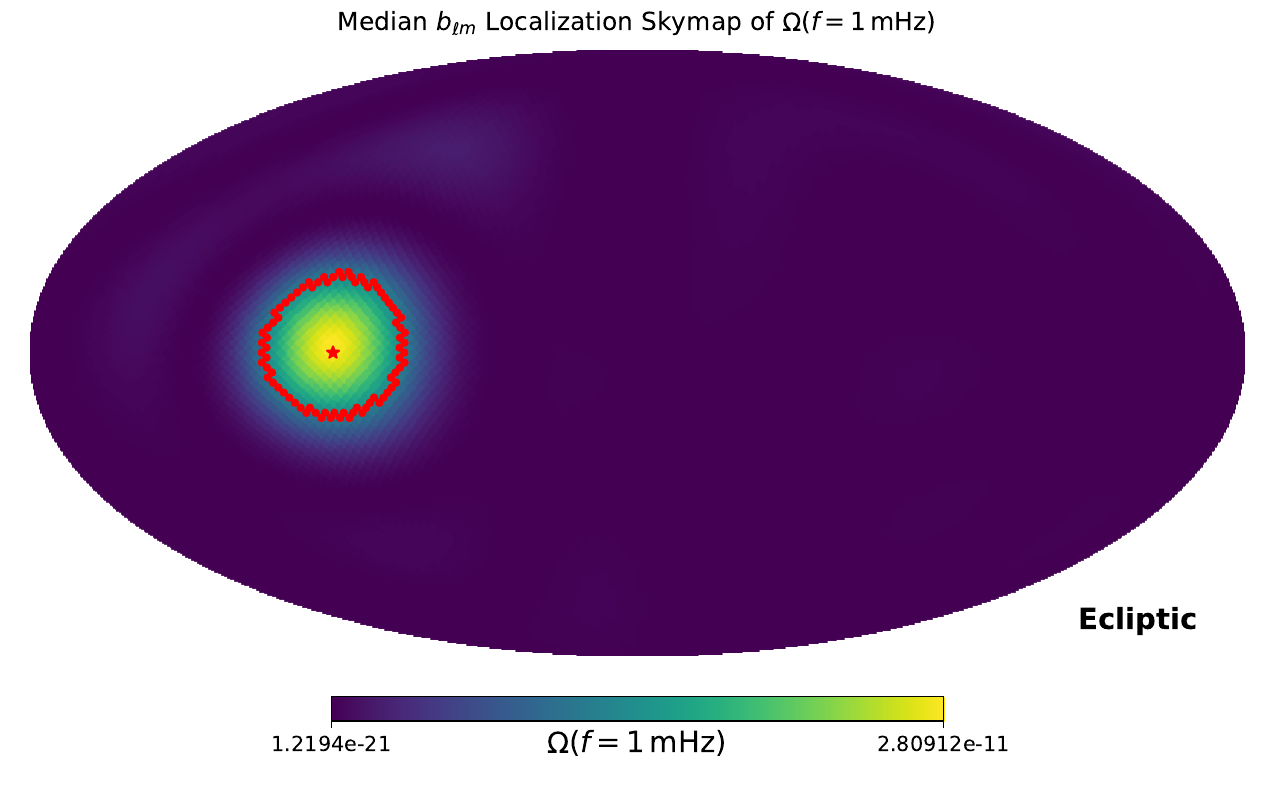}
    \includegraphics[width=\columnwidth]{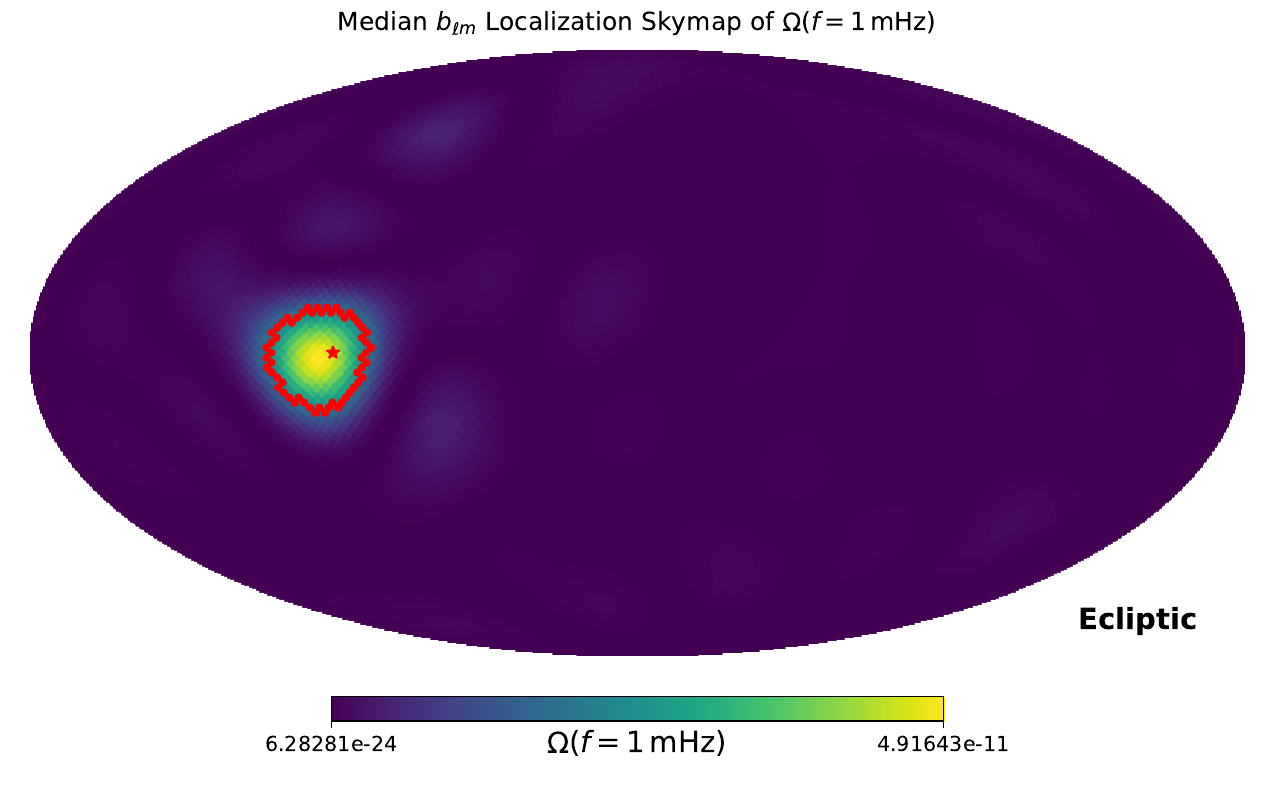}
    \caption{Posterior median ASGWB energy density skymaps with FWHM contours shown in red and the true (simulated) location of the point source indicated with a red star. These plots were produced via the analyses discussed in \S\ref{sec:sims}-\ref{sec:metrics} for simulated ASGWBs with  $\Omega_{\text{ref}} = 8\times 10^{-9}$, $T_{\mathrm{obs}}=3$ months, and (from top to bottom) $\la=6,10,16$. As $\la$ increases, the ability of the Bayesian ASGWB analysis to resolve the point source increases.}
    \label{fig:sp_skymaps}
\end{figure}

We employ two metrics based on the full-width-half-maximum (FWHM) measure to quantitatively evaluate \blips recovery of the simulated ASGWB spatial distributions for the single point and two point cases, respectively. To calculate the FWHM of a point source recovery described by a given posterior sample, we compute the Healpy skymap corresponding to its $b_{\ell m}$s, locating the maximum energy-density pixel within that skymap, and iteratively combining adjacent pixels with energy-density greater than half of the peak value until no such pixels remain. This process creates a contour around the peak; an example of these contours can be seen in Fig.~\ref{fig:sp_skymaps}. For the single point case, we then compute the fraction of the sky that lies within the contour, yielding a metric --- hereafter the ``SP metric" --- that corresponds to the spot size; the SP metric value is therefore inversely proportional to the angular resolution of the search. This metric is computed for every posterior sample of a given single-point \blip analysis; the data points and error bars shown in Figs.~\ref{fig:sp_svslmax} and~\ref{fig:sp_svsomega} are the means and 95\% credible intervals of the resulting distribution for each analysis.

\begin{figure}
    \includegraphics[width=\columnwidth]{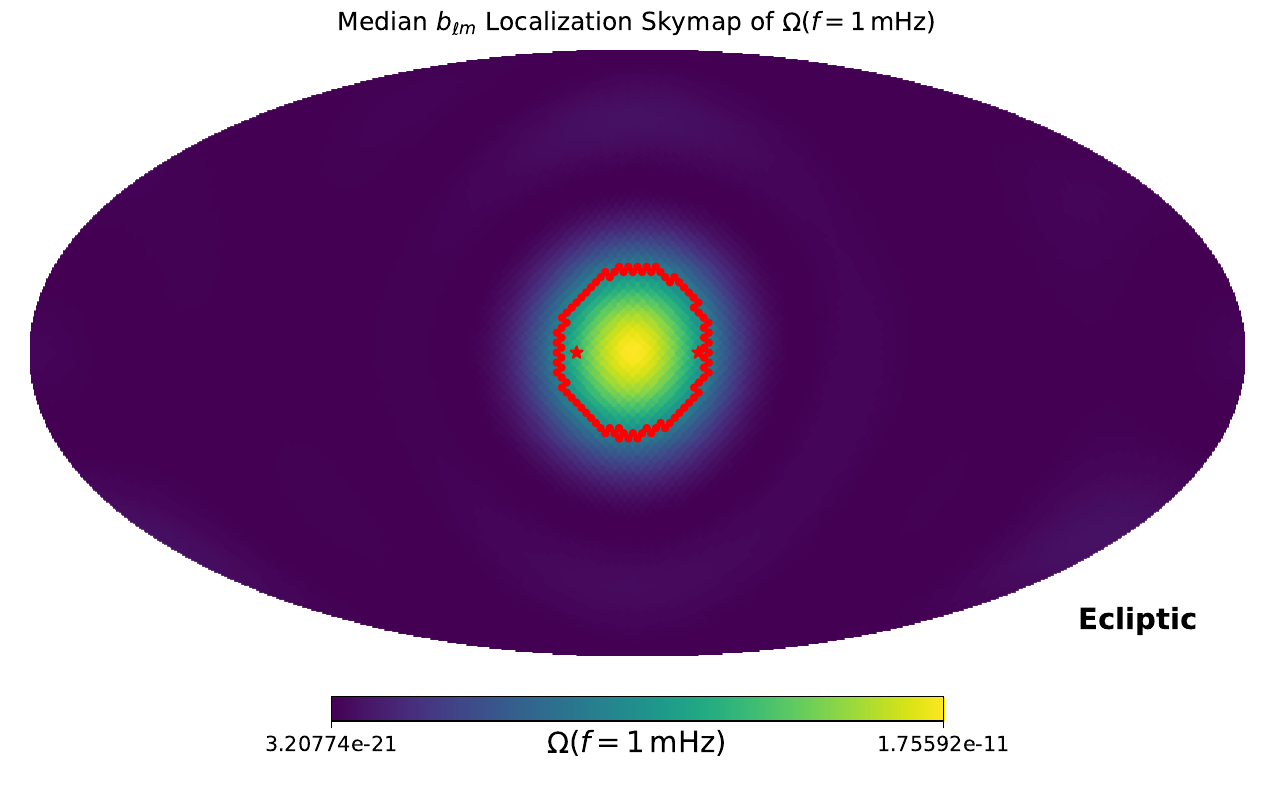}
    \includegraphics[width=\columnwidth]{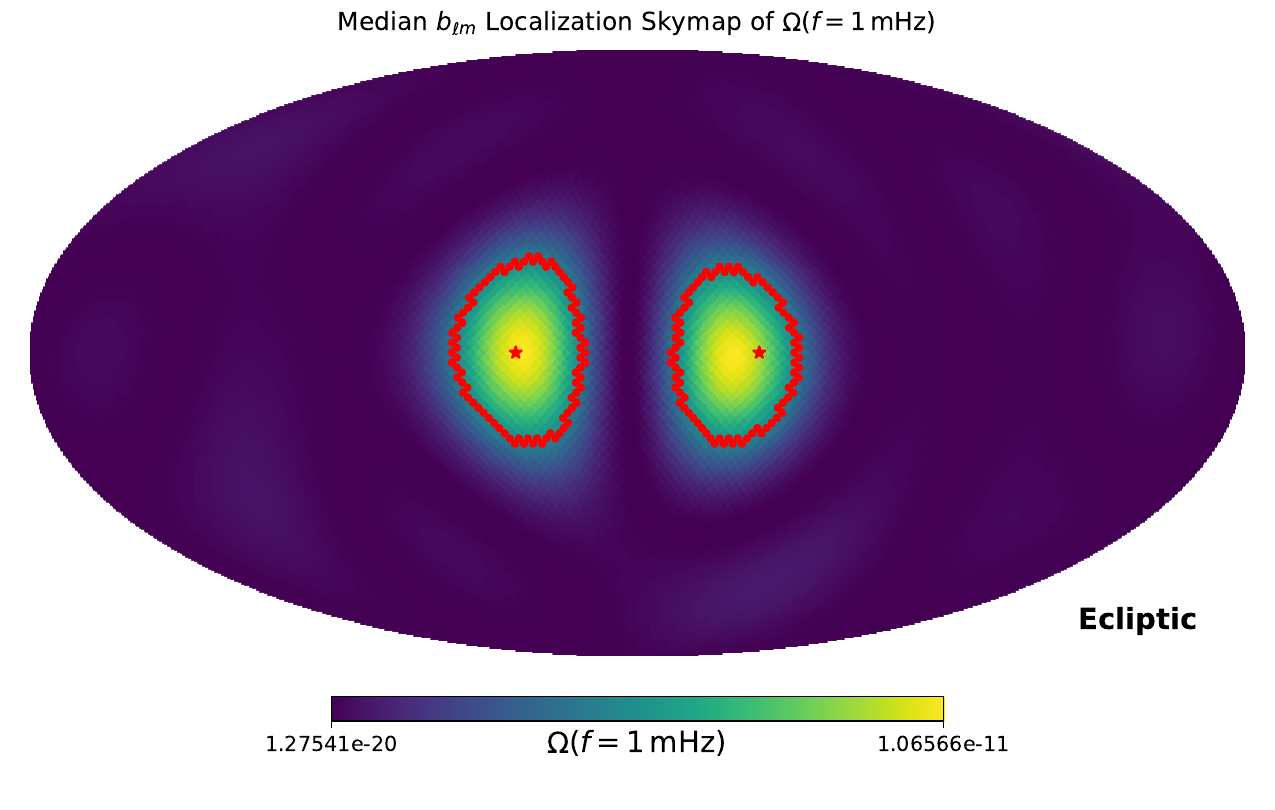}
    \includegraphics[width=\columnwidth]{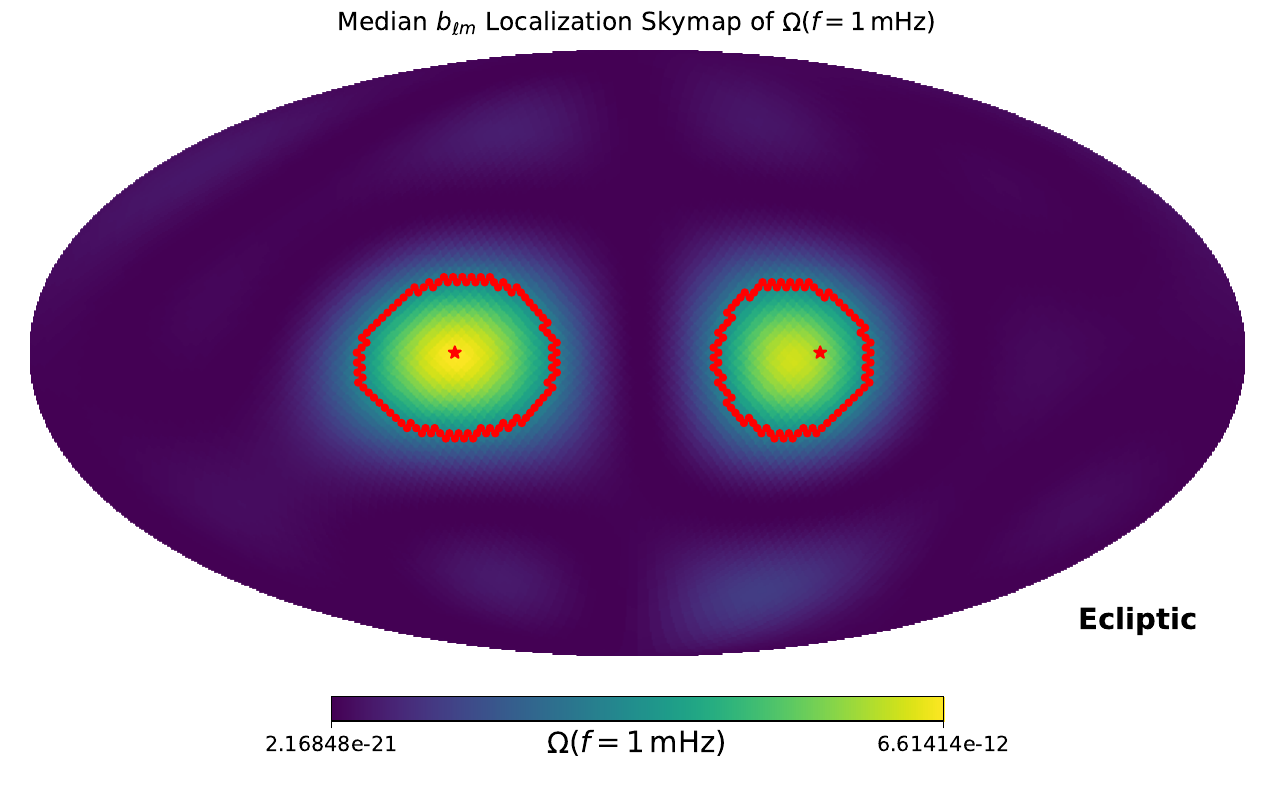}
    \caption{Posterior median ASGWB energy density skymaps with FWHM contours shown in red and the true (simulated) locations of the point sources indicated with a red star. These plots were produced via the analyses discussed in \S\ref{sec:sims}-\ref{sec:metrics} for simulated ASGWBs with $\Omega_{\text{ref}} = 8\times10^{-9}$, $\la=8$, $T_{\mathrm{obs}}=6$ months, and (from top to bottom) $\Delta \phi = \pi/5,\ 2\pi/5,\ 3\pi/5$. As $\Delta \phi$ decreases, the ability to resolve two point sources diminishes.}
    \label{fig:tp_skymaps}
\end{figure}

For the two point case, we compute the ratio of the (minimum) angular separation between the FWHM contours of each source separately and the combined FWHM fraction of the sky, following a similar approach to that of \citep{floden_angular_2022}. The two-point metric (TP metric) therefore quantifies the angular resolution by the ratio of separation to the total spot size; a larger value of the TP metric corresponds to better angular resolution. If two distinct FWHM contours cannot be found for a given skymap, the separation is taken to be zero (i.e., the sources are not resolved from one another) and the TP metric is therefore equal to zero. Examples of the TP metric contours for several different intrinsic angular separations are shown in Fig.~\ref{fig:tp_skymaps}. As for the SP metric, this calculation is performed for every posterior sample and the TP metric values and error bars shown in Figs.~\ref{fig:tp_metric_vs_sepangl} and~\ref{fig:tp_metric_vs_sepang} are the means and 95\% credible intervals of the resulting distributions. 

For each metric, we consider an approximate heuristic for comparison's sake. The approximate angular scale that can be probed by a spherical harmonic parameterization with a given $\la$ is
\begin{equation}
    \theta \sim \frac{\pi}{\la}.
\end{equation}
The approximate limit of the SP metric, assuming a 2D Gaussian point source with angular standard deviation $\frac{\theta}{2}$, will then be
\begin{equation}\label{eq:sp_heuristic}
    \mathrm{SP}_{\mathrm{min}} = \sin^2\left(\frac{1.1809}{4} \frac{\pi}{\la} \right),
\end{equation}
i.e., the fractional solid angle subtended by the FWHM of the Gaussian at an angular radius of $\theta_{\mathrm{FWHM}} \simeq 1.1809\frac{\theta}{2}$.
The equivalent heuristic for the TP metric for a given simulated angular separation $\Delta\phi$ will then be the ratio of their angular separation, less $2\times\theta_{\mathrm{FWHM}}$, to the combined area of each point source recovery (i.e., the sum of the SP metric of each source):
\begin{equation}\label{eq:tp_heuristic}
    \mathrm{TP}_{\mathrm{max}} = \frac{\max\left(\Delta\phi-1.1809\frac{\pi}{\la}\,,\ 0\right)}{2\,\mathrm{SP}_{\mathrm{min}}}.
\end{equation}
The resulting heuristic curves for $\mathrm{SP}_{\mathrm{min}}(\la)$ and $\mathrm{TP}_{\mathrm{max}}(\Delta\phi)_{|\la=8}$ are shown for comparison to our results in Figs.~\ref{fig:sp_svslmax} and~\ref{fig:tp_metric_vs_sepang}, respectively. We stress that these are necessarily approximate measures, and serve primarily to contextualize the trends seen in the data.

\section{Results}\label{sec:results}
We present Bayesian spherical harmonic analyses and their associated SP/TP metrics for all single-point and two-point simulations as described in Tables~\ref{table:sp_sims} and~\ref{table:tp_sims}, respectively. Each simulation is analyzed with \blip as described in \S\ref{sec:blip}. For visualization purposes, all skymaps shown are evaluated for $\Omega(f=1\text{mHz})$. All data points and error bars are means and 95\% credible intervals of their respective metric.

\subsection{Single Point Source}\label{sec:sp}

% For single-point source injections, we consider various values of $ T_{\mathrm{obs}}$  and  $\Omega_{\text{ref}}$ for injection parameters, and various cut-off $\la$ for recovery parameters. For Figs. \ref{fig:sp_skymaps} and \ref{fig:sp_svslmax}, $\ell^a_{\text{max, inj}} = 16$. For Fig. \ref{fig:sp_svsomega}, $\la = l^a_{\text{max, inj}}$.

\begin{figure}
    \includegraphics[width=\columnwidth]{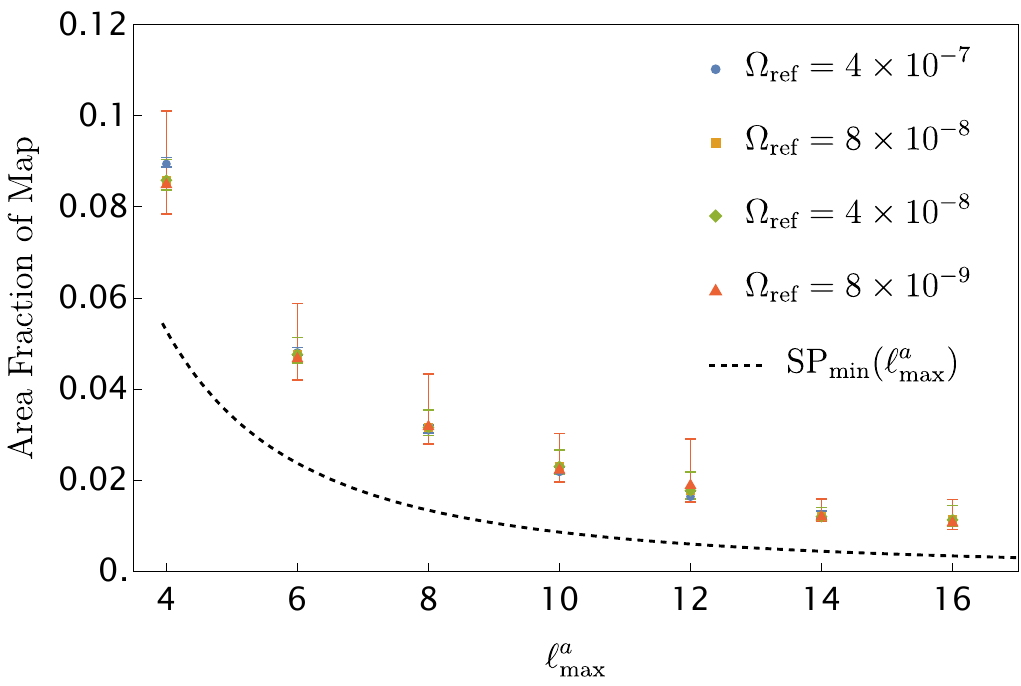}
    \caption{Spot size given by FWHM vs. $\la$ for various choices of $\Omega_{\text{ref}}$ during $ T_{\mathrm{obs}}=3$ months. Error bars mark 95\% confidence intervals. The dashed line shows the approximate minimum value of the SP metric for a spherical harmonic description with a given $\la$; see Eq.~\eqref{eq:sp_heuristic} and surrounding discussion.}
    \label{fig:sp_svslmax}
\end{figure}

\subsubsection{Dependence on $\la$}
To investigate angular resolution dependence on the choice of $\la$ used in the analysis, we analyze single-point sources with 4 different values of $\Omega_{\text{ref}}$ and vary $\la$ from 4 to 16. Figure \ref{fig:sp_svslmax} shows the associated SP metric for each analysis. As $\la$ increases the angular resolution improves drastically, whereas there is no obvious dependence on the ASGWB amplitude $\Omega_{\text{ref}}$ apart from improving the error bars. The overall trend follows that of the approximate limit's dependence on $\la$, albeit at a slight positive offset in the value of the SP metric.

% We find that there is no significant difference between setting $\ell^a_{\text{max, inj}}$ to $16$ or having it equal to the recovery. Thus the analysis works as long as $\ell^a_{\text{max, inj}} \geq \la$.  This confirms the expectation that the \blip spherical harmonic ASGWB search is only sensitive to spatial variations on scales it parameterizes with its spherical harmonic expansion. 

\subsubsection{Dependence on ASGWB Amplitude/Observing Time}
\begin{figure}
    \includegraphics[width=\columnwidth]{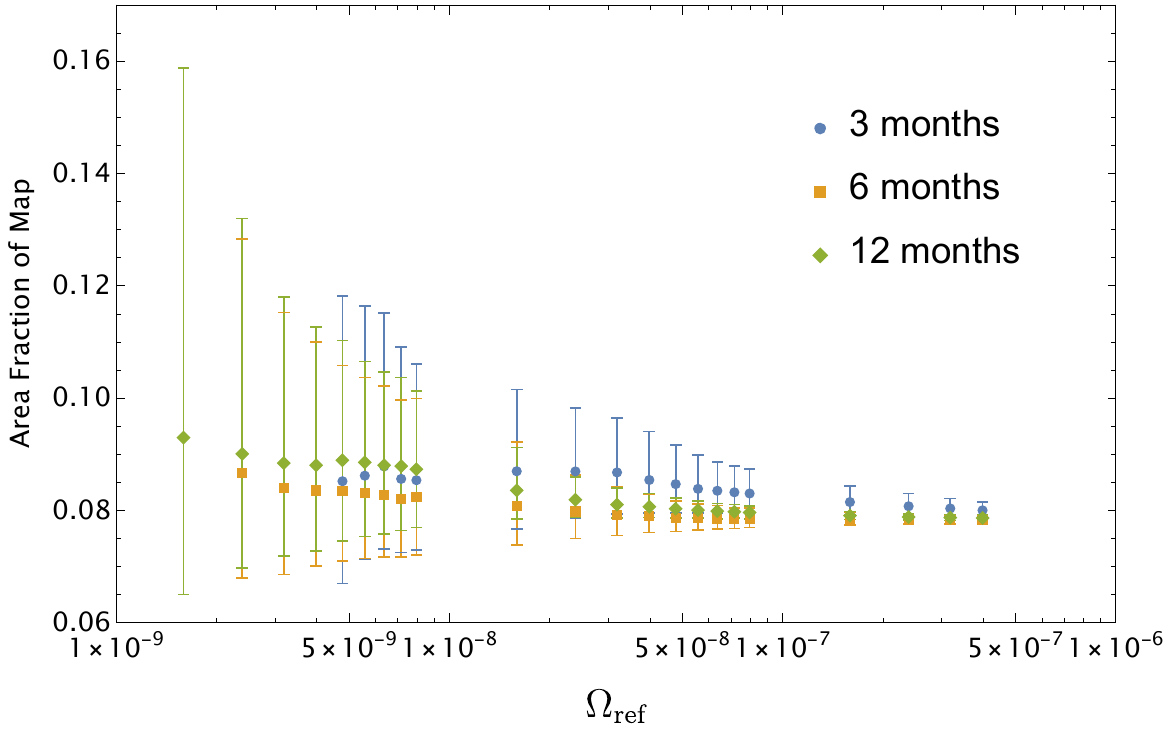}
    \caption{Spot size given by FWHM vs. $\Omega_{\text{ref}}$ for various choices of $T_{\mathrm{obs}}$ with $\la=4$.}
    \label{fig:sp_svsomega}
\end{figure}

The dependence on $\Omega_{\text{ref}}$ and $T_{\mathrm{obs}}$ is more closely inspected in Fig. \ref{fig:sp_svsomega}. We vary $\Omega_{\text{ref}}$ from $1.6 \times 10^{-9}$ to $4 \times 10^{-7}$ for total observing time, $ T_{\mathrm{obs}}$, ranging from 3 months to 1 year. No significant dependence of the SP metric on $\Omega_{\text{ref}}$ or $ T_{\mathrm{obs}}$ is apparent for recovered sources. However, we find that the recovery of a single point source signal has a binary dependence on the overall ASGWB signal amplitude for a given observing duration (equivalently, the signal-to-noise ratio of the ASGWB). If $\Omega_{\text{ref}} \gtrsim 10^{-8}$ for 1 year of data, the spatial distribution of the signal is recovered for $\la=4$, but increasing $\Omega_{\text{ref}}$ beyond that point does not improve angular resolution as measured by the SP metric (although it does reduce the parameter uncertainty and therefore the overall skymap/SP metric uncertainty). This behavior effectively defines a noise floor for each spherical harmonic mode. It will be important to characterize this noise floor; however, doing so will require meaningful consideration of low-amplitude ASGWBs and thus inclusion of the anisotropic Galactic foreground. As such, further exploration of this point is beyond the scope of this study and should be pursued in future work.

% As would be expected, increasing $ T_{\mathrm{obs}}$ allows for recoveries at slightly lower $\Omega_{\text{ref}}$, and reduces the overall uncertainty of the recovery, but doing so does not result in an improvement in the intrinsic angular resolution of the search. 

% Figures and tables should be placed at logical positions in the text. Don't
% worry about the exact layout, which will be handled by the publishers.

% Figures are referred to as e.g. Fig.~\ref{fig:example_figure}, and tables as
% e.g. Table~\ref{tab:example_table}.

% % Example figure
% \begin{figure}
% 	% To include a figure from a file named example.*
% 	% Allowable file formats are eps or ps if compiling using latex
% 	% or pdf, png, jpg if compiling using pdflatex
% 	\includegraphics[width=\columnwidth]{example}
%     \caption{This is an example figure. Captions appear below each figure.
% 	Give enough detail for the reader to understand what they're looking at,
% 	but leave detailed discussion to the main body of the text.}
%     \label{fig:example_figure}
% \end{figure}

% % Example table
% \begin{table}
% 	\centering
% 	\caption{This is an example table. Captions appear above each table.
% 	Remember to define the quantities, symbols and units used.}
% 	\label{tab:example_table}
% 	\begin{tabular}{lccr} % four columns, alignment for each
% 		\hline
% 		A & B & C & D\\
% 		\hline
% 		1 & 2 & 3 & 4\\
% 		2 & 4 & 6 & 8\\
% 		3 & 5 & 7 & 9\\
% 		\hline
% 	\end{tabular}
% \end{table}

\subsection{Two Point Source}\label{sec:tp}
For the two-point case, the orbital motion of the LISA constellation becomes highly relevant until completing at least half an orbit; for three months of data, the two point sources were largely unresolvable for all considered amplitudes of the simulated signal. We therefore report results for $T_{\mathrm{obs}}=6$ months. 

% Each two-point source is injected with $\ell^a_{\text{max, inj}} = 12$. The TP metric is computed for each posterior sample in the ASGWB recovery with 95\% confidence intervals reported in Figs. \ref{fig:tp_metric_vs_sepangl} and \ref{fig:tp_metric_vs_sepang}.

\subsubsection{Dependence on $\la$}
\begin{figure}%add point with astrix or dagger
    \includegraphics[width=\columnwidth]{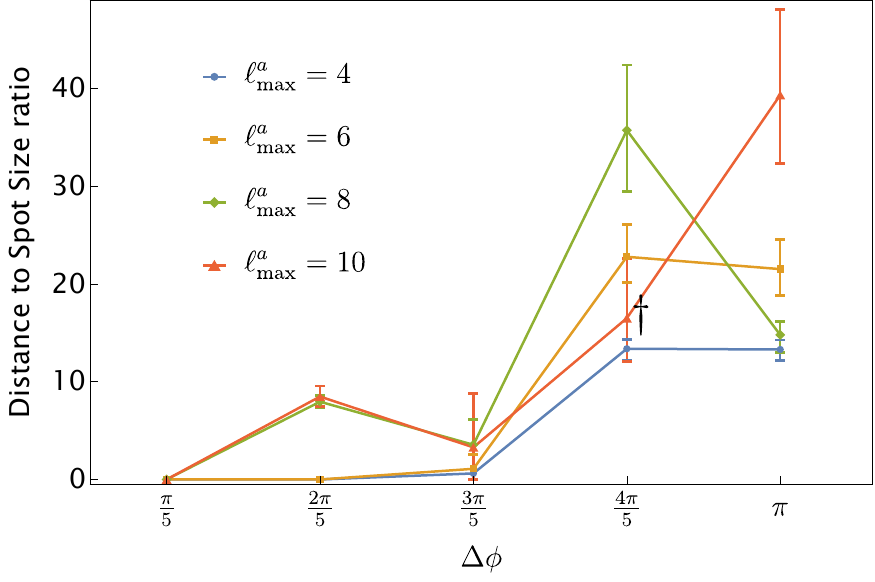}
    \caption{Distance to spot size ratio (TP metric) vs. injected angular separation for various choices of $\la$ during a 6 month observation. The point marked with $\dagger$ is simulated with different sky locations (see Table \ref{table:tp_sims}).}
    \label{fig:tp_metric_vs_sepangl}
\end{figure}

% The general trend in Fig. \ref{fig:tp_metric_vs_sepangl} follows what we would expect: angular resolution improves with increased angular separation. We see a relatively dramatic improvement at $\Delta\phi = 4\pi/5$. 
Consistent with the results in \S\ref{sec:sp}, the choice of a higher $\la$ increases LISA's ability to resolve two ASGWB sources as measured by the TP metric, with the exception of $\la=8$ at $\Delta\phi = \pi$. Note that the data point for $\la=10$ at $\Delta\phi = 4\pi/5$ is marked with a $\dagger$; analysis of the original simulation recovered a significant bimodality in the two-point metric. This resulted in disproportionately large error in our measurement of the two-point metric at this $\la$ and separation. Repeating the simulation with identical parameters, save that the injected points sources are translated on the sky (see Table \ref{table:tp_sims}), results in a non-bimodal recovery with uncertainty that is consistent with the overall trend. This indicates that the original bimodality was likely a result of interaction with the directional LISA response due to the specific locations of the point sources on the sky and the relatively short $T_{\mathrm{obs}}$ used in this simulation. We therefore use the measurement of the two-point metric obtained from the second, translated simulation for the data point marked with the $\dagger$ in Fig.~\ref{fig:tp_metric_vs_sepangl}. 

% resulting in extremely large error bars.} The presented result is for the same choice of parameters except that the injected $\phi_1$ and $\phi_2$ are shifted to the left (see Table \ref{table:tp_sims}). 

\subsubsection{Dependence on ASGWB Amplitude}
\begin{figure}
    \includegraphics[width=\columnwidth]{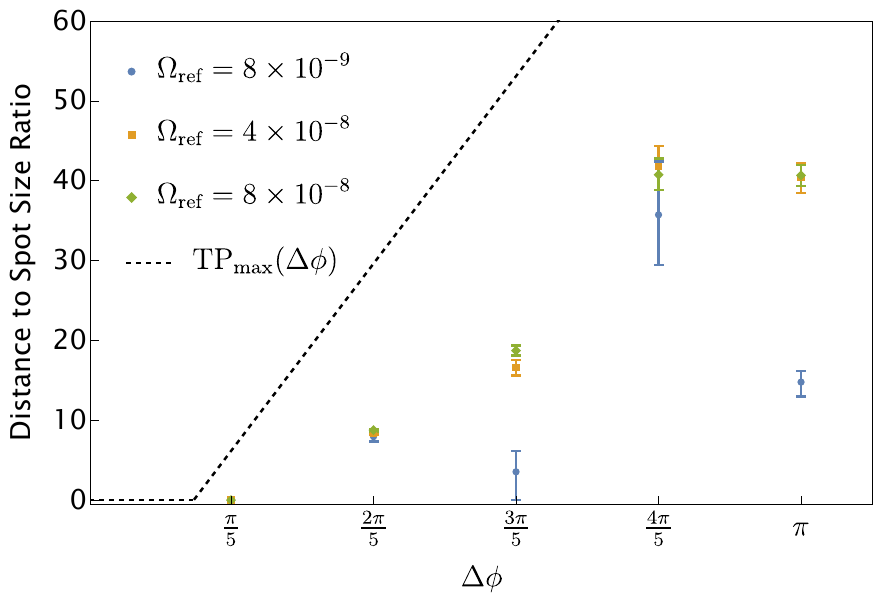}
    \caption{Distance to spot size ratio (TP metric) vs. injected angular separation for various choices of $\Omega_{\text{ref}}$ for $\la = 8$. The dashed line shows the approximate maximum value of the TP metric at $\la=8$ for a given $\Delta\phi$; see Eq.~\eqref{eq:tp_heuristic} and surrounding discussion.}
    \label{fig:tp_metric_vs_sepang}
\end{figure}

We additionally consider different values of the simulated ASGWB amplitude $\Omega_{\text{ref}}$ for $\la=8$ in Fig.~\ref{fig:tp_metric_vs_sepang}.  The drop in the TP metric from $\Delta\phi=2\pi/5$ to $\Delta\phi=3\pi/5$ present in Fig. \ref{fig:tp_metric_vs_sepangl} is no longer present with higher $\Omega_{\text{ref}}$, and the low TP metric of the $\la=8$ case from above is greatly improved.  In conjunction with the single-point results, this behavior again implies a statistical noise floor that depends both on the SNR of the ASGWB and the number of parameters used to model it; recall that for a given $\la = 2\lb$, the spherical harmonic spatial model has $\sum_{\ell=1}^{\lb}(2\ell+1)$ free parameters. 

The recovered TP metric is compared to the TP heuristic in Fig.~\ref{fig:sp_svslmax}. The observed trend is somewhat shallower in slope, but similarly linear (modulo some scatter at lower amplitudes). The difference in slope is likely in part explained by the offset of the SP metric heuristic from the data as seen in Fig.~\ref{fig:sp_svslmax}.

%invert order of key

\section{Discussion and Conclusions}\label{sec:conclusion}
We investigate the angular resolution of a Bayesian spherical harmonic search for ASGWBs in LISA. We use \blip to simulate and analyze single-point and two-point source ASGWBs across a broad parameter space for both signal simulation and recovery. These include ASGWB amplitude, total observing time, choice of spherical harmonic truncation $\la$, and --- for the two-point case --- angular separation of the component sources on the angular resolution of our Bayesian spherical harmonic ASGWB analysis. We quantitatively characterize these factors' associated trends in angular resolution using two metrics, both based off of the FWHM measure.

For middling-to-high-amplitude\footnote{(i.e., somewhat less than the amplitude of the Galactic foreground, although we only consider the case of a $\alpha=2/3$ power law; the impact of assumed spectral form on Bayesian ASGWB searches should be considered in future work.)} ASGWBs ($\Omega_{\mathrm{ref}}(f=25\mathrm{\, Hz}) > 10^{-9}$ for 1 year of observing time, given the LISA detector noise assumed in this work), we find that the choice of $\la$ is the primary determining factor for the search's angular resolution. This remains true up to the highest $\la$ considered in this work, $\la=16$. Higher choices of $\la$ would presumably continue this trend (although not indefinitely), but are not currently feasible due to the intensive computational cost of modelling the LISA detector response to each additional spherical harmonic mode and the difficulty of efficiently sampling such a high-dimensional parameter space.

The ASGWB amplitude, parameterized by  the dimensionless energy density $\Omega_{\text{ref}}$, has no discernible impact on the angular resolution of the search as long as the overall SNR for a given observing duration is sufficiently high to recover a signal. That being said, larger amplitudes/longer observing durations do improve the uncertainty of the recovery, as would be expected. It is worth noting that particularly short observing durations have severe negative impacts on the ASGWB recovery quality. As this effect is due to the fact that short observing times do not gain the full advantage of the LISA constellation's orbital motion, the negative impacts are mitigated by $T_{\mathrm{obs}}\ge6$ months and can therefore be expected to be negligible when considering the full LISA mission duration. However, early iterations of global fit analyses during the first $3-6$ months of the LISA mission could be impacted.  

At low SNR, however, we observe a threshold below which the combination of instrumental and statistical noise render the search entirely unable to characterize the ASGWB spatial distribution with a given $\la$. It is reasonable to infer that for low-SNR ASGWBs, the angular resolution of the search will be information-limited. That is, for a Bayesian search where the angular resolution is primarily determined by the choice of $\la$ --- and accordingly, the number of parameters in the ASGWB spatial model --- the amount of information contained in the signal will limit how many such parameters can be constrained, therefore limiting the angular resolution. A detailed study of this SNR threshold and its dependence on $\la$ is a promising direction for future work.

Furthermore, the behavior in Fig. \ref{fig:tp_metric_vs_sepangl} highlights some imperfections in the method used to discern between multiple overlapping ASGWBs. As the TP metric contours are computed at one half of the local maximum energy density, this method will not always be able to distinguish between two peaks even if they are present in the posterior results. One could instead try computing 90\% of the peak energy density, but this confronts separate concerns such as whether that metric fully characterizes the point-spread of the source, or more technical difficulties relating to the discreteness of the choice in {\tt nside}.  While the methods used in this work suffice for the spatial scales that are currently computationally accessible, more robust methods should be developed to distinguish between multiple ASGWB contributors at the finer scales that computational advances will grant access to. Such methods would then allow for ASGWB separation through (for instance) targeted directional searches with independent spectral models.

% For two-point source injections, we ultimately find that \blip can recover distinct signals with injected angular separations as low as $2\pi/5$ at $\Omega_{\text{ref}} = 10^{-7}$. Since increasing the choice of $\la$ seems to improve the recovery quality of two-point sources, we expect this is not the angular resolution limit of LISA. As computational improvements are made, probing higher $\la$ will become more feasible allowing the \blip anisotropic analysis to recover smaller angular separations. Furthermore, the curious behavior in Fig. \ref{fig:tp_metric_vs_sepangl} highlights the imperfections in the metric used to characterize angular resolution for ASGWBs with multiple peaks. Since contours are computed at one half of the local maximum energy density, this method will not always be able to distinguish between two peaks even if they are present in the posterior results. One could instead try computing 90\% of the peak energy density, but this confronts separate concerns such as whether that metric fully characterizes the point-spread of the source, or more technical difficulties relating to the discreteness of the choice in {\tt nside}.  While the methods used in this work suffice for the spatial scales \blip is currently able to handle computationally, more robust methods should be developed to characterize LISA's angular resolution for ASGWB searches at the finer scales that computational advances will grant access to. 

Our primary conclusion is that, while the limitations of LISA's angular response are expected to become relevant eventually, the angular resolution achievable by Bayesian spherical harmonic ASGWB searches in LISA is currently limited at high amplitudes by the choice of $\la$, and therefore driven by current computational limitations. As continued development of software like \blip and computational advancements will reduce the cost of higher $\la$ analyses, we expect the eventual angular resolution of Bayesian searches for high-amplitude ASGWBs in LISA to be better than what we report here.  Notably, this result exceeds the angular sensitivity expected for frequentist approaches to characterizing the LISA angular power spectrum \citep[e.g.,][]{kudoh_probing_2005,taruya_probing_2005,breivik_constraining_2020,contaldi_maximum_2020} for the frequency range considered. The upper limit achieved in this work of $\la=16$ exceeds even the $\la\lesssim15$ of \citet{contaldi_maximum_2020} for the equivalent case of a high-SNR ASGWB in well-understood noise. It is additionally worth noting that when we reconstitute the $a_{\ell m}$s from the $b_{\ell m}$s, we observe an insensitivity to odd $\ell_a$ modes, consistent with the calculations presented in e.g., \citet{kudoh_probing_2005,bartolo_probing_2022}; see Fig.~\ref{fig:alm_corners} and discussion in Appendix~\ref{appendix:alms}.

% Bayesian approaches to LISA ASGWB searches like that of \blips have several advantages that may lead to this relative improvement in resolving power. First is that the spherical harmonic formalism of \citet{banagiri_mapping_2021b}  not only mathematically ensures that all considered spatial distributions have non-negative power at every point, but also, by inferring the $b_{\ell m}$s of the distribution, avoids LISA's insensitivity to odd spherical harmonic modes. Another strength of the Bayesian approach is that it explicitly models the time-dependence of the LISA response to an ASGWB, thus leveraging the significant information provided by the motion of LISA's constellation. While \citet{mentasti_probing_2024a} show that for the frequentist map-making method of \citet{contaldi_maximum_2020} this does not strongly impact the multipole characterization uncertainty for LISA, in a Bayesian setting it is crucial to properly model the ASGWB's interaction with the time-dependent LISA response, and mismodelling of this time-dependence may even result in biased recoveries.

Finally, it is important to note that we do not consider additional complicating factors such as breathing modes of the LISA constellation or sources of nonstationary noise in the LISA detector, both of which will affect LISA's overall sensitivity to SGWBs \citep{hartwig_stochastic_2023,muratore_impact_2023}. Nor do we consider a realistic treatment of searches for underlying ASGWBs in the presence of the Galactic foreground. The impact of these factors on the efficacy and angular resolution of Bayesian LISA ASGWB searches should be investigated in future work. As such, this work does not extend to investigation of realistic astrophysical ASGWBs in LISA. While the \blip anisotropic analysis has been applied to local ASGWBs such as that of the Galaxy \citep{banagiri_mapping_2021b} and the Large Magellanic Cloud \citep{rieck_lisa_2024}, a crucial direction of future exploration will be the extragalactic SGWB from stellar-origin black hole binaries \citep[e.g.,][]{babak_stochastic_2023}. This SGWB will possess a kinematic dipole and small intrinsic anisotropies on the level of the large-scale structure of the Universe; these latter anisotropies may even hold the key to spectral separation between this astrophysical SGWB and an underlying signal of cosmological origin. It will therefore be important in the future to explore LISA's ability (or lack thereof) to characterize the anisotropies of low-amplitude, extragalactic ASGWBs in the presence of the Galactic foreground over its full nominal (4 year) and extended (10 year) mission durations.

\section*{Acknowledgements}
The authors would like to thank Erik Floden, Sharan Banagiri, Steven Rieck, Jessica Lawrence, and Joseph Romano  for many helpful conversations. This work is supported by the NASA grant 90NSSC19K0318, and utilized computing resources provided by the Minnesota Supercomputing Institute at the University of Minnesota.

% The Acknowledgements section is not numbered. Here you can thank helpful
% colleagues, acknowledge funding agencies, telescopes and facilities used etc.

%%%%%%%%%%%%%%%%%%%%%%%%%%%%%%%%%%%%%%%%%%%%%%%%%%
\section*{Data Availability}
\blip is open source and is available at \href{https://github.com/sharanbngr/blip}{https://github.com/sharanbngr/blip}. Simulation parameters and posterior samples for all analyses are available on Zenodo \citep{bloom_datasets_2024}. 
% \ac{We need to do this and update the text accordingly before submitting.}
 
% The inclusion of a Data Availability Statement is a requirement for articles published in MNRAS. Data Availability Statements provide a standardised format for readers to understand the availability of data underlying the research results described in the article. The statement may refer to original data generated in the course of the study or to third-party data analysed in the article. The statement should describe and provide means of access, where possible, by linking to the data or providing the required accession numbers for the relevant databases or DOIs.



\bibliography{LISA, banagiri_2021, LISA_Angular_Resolution} % if your bibtex file is called example.bib

\begin{thebibliography}{56}%
\makeatletter
\providecommand \@ifxundefined [1]{%
 \@ifx{#1\undefined}
}%
\providecommand \@ifnum [1]{%
 \ifnum #1\expandafter \@firstoftwo
 \else \expandafter \@secondoftwo
 \fi
}%
\providecommand \@ifx [1]{%
 \ifx #1\expandafter \@firstoftwo
 \else \expandafter \@secondoftwo
 \fi
}%
\providecommand \natexlab [1]{#1}%
\providecommand \enquote  [1]{``#1''}%
\providecommand \bibnamefont  [1]{#1}%
\providecommand \bibfnamefont [1]{#1}%
\providecommand \citenamefont [1]{#1}%
\providecommand \href@noop [0]{\@secondoftwo}%
\providecommand \href [0]{\begingroup \@sanitize@url \@href}%
\providecommand \@href[1]{\@@startlink{#1}\@@href}%
\providecommand \@@href[1]{\endgroup#1\@@endlink}%
\providecommand \@sanitize@url [0]{\catcode `\\12\catcode `\$12\catcode `\&12\catcode `\#12\catcode `\^12\catcode `\_12\catcode `\%12\relax}%
\providecommand \@@startlink[1]{}%
\providecommand \@@endlink[0]{}%
\providecommand \url  [0]{\begingroup\@sanitize@url \@url }%
\providecommand \@url [1]{\endgroup\@href {#1}{\urlprefix }}%
\providecommand \urlprefix  [0]{URL }%
\providecommand \Eprint [0]{\href }%
\providecommand \doibase [0]{https://doi.org/}%
\providecommand \selectlanguage [0]{\@gobble}%
\providecommand \bibinfo  [0]{\@secondoftwo}%
\providecommand \bibfield  [0]{\@secondoftwo}%
\providecommand \translation [1]{[#1]}%
\providecommand \BibitemOpen [0]{}%
\providecommand \bibitemStop [0]{}%
\providecommand \bibitemNoStop [0]{.\EOS\space}%
\providecommand \EOS [0]{\spacefactor3000\relax}%
\providecommand \BibitemShut  [1]{\csname bibitem#1\endcsname}%
\let\auto@bib@innerbib\@empty
%</preamble>
\bibitem [{\citenamefont {{Amaro-Seoane}}\ \emph {et~al.}(2017)\citenamefont {{Amaro-Seoane}}, \citenamefont {Audley}, \citenamefont {Babak}, \citenamefont {Baker}, \citenamefont {Barausse}, \citenamefont {Bender}, \citenamefont {Berti}, \citenamefont {Binetruy}, \citenamefont {Born}, \citenamefont {Bortoluzzi}, \citenamefont {Camp}, \citenamefont {Caprini}, \citenamefont {Cardoso}, \citenamefont {Colpi}, \citenamefont {Conklin}, \citenamefont {Cornish}, \citenamefont {Cutler}, \citenamefont {Danzmann}, \citenamefont {Dolesi}, \citenamefont {Ferraioli}, \citenamefont {Ferroni}, \citenamefont {Fitzsimons}, \citenamefont {Gair}, \citenamefont {Gesa~Bote}, \citenamefont {Giardini}, \citenamefont {Gibert}, \citenamefont {Grimani}, \citenamefont {Halloin}, \citenamefont {Heinzel}, \citenamefont {Hertog}, \citenamefont {Hewitson}, \citenamefont {{Holley-Bockelmann}}, \citenamefont {Hollington}, \citenamefont {Hueller}, \citenamefont {Inchauspe}, \citenamefont {Jetzer}, \citenamefont {Karnesis}, \citenamefont
  {Killow}, \citenamefont {Klein}, \citenamefont {Klipstein}, \citenamefont {Korsakova}, \citenamefont {Larson}, \citenamefont {Livas}, \citenamefont {Lloro}, \citenamefont {Man}, \citenamefont {Mance}, \citenamefont {Martino}, \citenamefont {Mateos}, \citenamefont {McKenzie}, \citenamefont {McWilliams}, \citenamefont {Miller}, \citenamefont {Mueller}, \citenamefont {Nardini}, \citenamefont {Nelemans}, \citenamefont {Nofrarias}, \citenamefont {Petiteau}, \citenamefont {Pivato}, \citenamefont {Plagnol}, \citenamefont {Porter}, \citenamefont {Reiche}, \citenamefont {Robertson}, \citenamefont {Robertson}, \citenamefont {Rossi}, \citenamefont {Russano}, \citenamefont {Schutz}, \citenamefont {Sesana}, \citenamefont {Shoemaker}, \citenamefont {Slutsky}, \citenamefont {Sopuerta}, \citenamefont {Sumner}, \citenamefont {Tamanini}, \citenamefont {Thorpe}, \citenamefont {Troebs}, \citenamefont {Vallisneri}, \citenamefont {Vecchio}, \citenamefont {Vetrugno}, \citenamefont {Vitale}, \citenamefont {Volonteri},
  \citenamefont {Wanner}, \citenamefont {Ward}, \citenamefont {Wass}, \citenamefont {Weber}, \citenamefont {Ziemer},\ and\ \citenamefont {Zweifel}}]{amaro-seoane_laser_2017}%
  \BibitemOpen
  \bibfield  {author} {\bibinfo {author} {\bibfnamefont {P.}~\bibnamefont {{Amaro-Seoane}}}, \bibinfo {author} {\bibfnamefont {H.}~\bibnamefont {Audley}}, \bibinfo {author} {\bibfnamefont {S.}~\bibnamefont {Babak}}, \emph {et~al.},\ }\href {https://doi.org/10.48550/arXiv.1702.00786} {\bibinfo {title} {Laser {{Interferometer Space Antenna}}}} (\bibinfo {year} {2017})\BibitemShut {NoStop}%
\bibitem [{\citenamefont {{Amaro-Seoane}}\ \emph {et~al.}(2023)\citenamefont {{Amaro-Seoane}}, \citenamefont {Andrews}, \citenamefont {Arca~Sedda}, \citenamefont {Askar}, \citenamefont {Baghi}, \citenamefont {Balasov}, \citenamefont {Bartos}, \citenamefont {Bavera}, \citenamefont {Bellovary}, \citenamefont {Berry}, \citenamefont {Berti}, \citenamefont {Bianchi}, \citenamefont {Blecha}, \citenamefont {Blondin}, \citenamefont {Bogdanovi{\'c}}, \citenamefont {Boissier}, \citenamefont {Bonetti}, \citenamefont {Bonoli}, \citenamefont {Bortolas}, \citenamefont {Breivik}, \citenamefont {Capelo}, \citenamefont {Caramete}, \citenamefont {Cattorini}, \citenamefont {Charisi}, \citenamefont {Chaty}, \citenamefont {Chen}, \citenamefont {Chru{\'s}li{\'n}ska}, \citenamefont {Chua}, \citenamefont {Church}, \citenamefont {Colpi}, \citenamefont {D'Orazio}, \citenamefont {Danielski}, \citenamefont {Davies}, \citenamefont {Dayal}, \citenamefont {De~Rosa}, \citenamefont {Derdzinski}, \citenamefont {Destounis}, \citenamefont
  {Dotti}, \citenamefont {Du{\c t}an}, \citenamefont {Dvorkin}, \citenamefont {Fabj}, \citenamefont {Foglizzo}, \citenamefont {Ford}, \citenamefont {Fouvry}, \citenamefont {Franchini}, \citenamefont {Fragos}, \citenamefont {Fryer}, \citenamefont {Gaspari}, \citenamefont {Gerosa}, \citenamefont {Graziani}, \citenamefont {Groot}, \citenamefont {Habouzit}, \citenamefont {Haggard}, \citenamefont {Haiman}, \citenamefont {Han}, \citenamefont {Istrate}, \citenamefont {Johansson}, \citenamefont {Khan}, \citenamefont {Kimpson}, \citenamefont {Kokkotas}, \citenamefont {Kong}, \citenamefont {Korol}, \citenamefont {Kremer}, \citenamefont {Kupfer}, \citenamefont {Lamberts}, \citenamefont {Larson}, \citenamefont {Lau}, \citenamefont {Liu}, \citenamefont {{Lloyd-Ronning}}, \citenamefont {Lodato}, \citenamefont {Lupi}, \citenamefont {Ma}, \citenamefont {Maccarone}, \citenamefont {Mandel}, \citenamefont {Mangiagli}, \citenamefont {Mapelli}, \citenamefont {Mathis}, \citenamefont {Mayer}, \citenamefont {McGee}, \citenamefont
  {McKernan}, \citenamefont {Miller}, \citenamefont {Mota}, \citenamefont {Mumpower}, \citenamefont {Nasim}, \citenamefont {Nelemans}, \citenamefont {Noble}, \citenamefont {Pacucci}, \citenamefont {Panessa}, \citenamefont {Paschalidis}, \citenamefont {Pfister}, \citenamefont {Porquet}, \citenamefont {Quenby}, \citenamefont {Ricarte}, \citenamefont {R{\"o}pke}, \citenamefont {Regan}, \citenamefont {Rosswog}, \citenamefont {Ruiter}, \citenamefont {Ruiz}, \citenamefont {Runnoe}, \citenamefont {Schneider}, \citenamefont {Schnittman}, \citenamefont {Secunda}, \citenamefont {Sesana}, \citenamefont {Seto}, \citenamefont {Shao}, \citenamefont {Shapiro}, \citenamefont {Sopuerta}, \citenamefont {Stone}, \citenamefont {Suvorov}, \citenamefont {Tamanini}, \citenamefont {Tamfal}, \citenamefont {Tauris}, \citenamefont {Temmink}, \citenamefont {Tomsick}, \citenamefont {Toonen}, \citenamefont {{Torres-Orjuela}}, \citenamefont {Toscani}, \citenamefont {Tsokaros}, \citenamefont {Unal}, \citenamefont {{V{\'a}zquez-Aceves}},
  \citenamefont {Valiante}, \citenamefont {{van Putten}}, \citenamefont {{van Roestel}}, \citenamefont {Vignali}, \citenamefont {Volonteri}, \citenamefont {Wu}, \citenamefont {Younsi}, \citenamefont {Yu}, \citenamefont {Zane}, \citenamefont {Zwick}, \citenamefont {Antonini}, \citenamefont {Baibhav}, \citenamefont {Barausse}, \citenamefont {Bonilla~Rivera}, \citenamefont {Branchesi}, \citenamefont {{Branduardi-Raymont}}, \citenamefont {Burdge}, \citenamefont {Chakraborty}, \citenamefont {Cuadra}, \citenamefont {Dage}, \citenamefont {Davis}, \citenamefont {{de Mink}}, \citenamefont {Decarli}, \citenamefont {Doneva}, \citenamefont {Escoffier}, \citenamefont {Gandhi}, \citenamefont {Haardt}, \citenamefont {Lousto}, \citenamefont {Nissanke}, \citenamefont {Nordhaus}, \citenamefont {O'Shaughnessy}, \citenamefont {Portegies~Zwart}, \citenamefont {Pound}, \citenamefont {Schussler}, \citenamefont {Sergijenko}, \citenamefont {Spallicci}, \citenamefont {Vernieri},\ and\ \citenamefont
  {{Vigna-G{\'o}mez}}}]{amaro-seoane_astrophysics_2023}%
  \BibitemOpen
  \bibfield  {author} {\bibinfo {author} {\bibfnamefont {P.}~\bibnamefont {{Amaro-Seoane}}}, \bibinfo {author} {\bibfnamefont {J.}~\bibnamefont {Andrews}}, \bibinfo {author} {\bibfnamefont {M.}~\bibnamefont {Arca~Sedda}}, \emph {et~al.},\ }\href {https://doi.org/10.1007/s41114-022-00041-y} {\bibfield  {journal} {\bibinfo  {journal} {Living Reviews in Relativity}\ }\textbf {\bibinfo {volume} {26}},\ \bibinfo {pages} {2} (\bibinfo {year} {2023})}\BibitemShut {NoStop}%
\bibitem [{\citenamefont {Agazie}\ \emph {et~al.}(2023{\natexlab{a}})\citenamefont {Agazie}, \citenamefont {Anumarlapudi}, \citenamefont {Archibald}, \citenamefont {Arzoumanian}, \citenamefont {Baker}, \citenamefont {B{\'e}csy}, \citenamefont {Blecha}, \citenamefont {Brazier}, \citenamefont {Brook}, \citenamefont {{Burke-Spolaor}}, \citenamefont {Burnette}, \citenamefont {Case}, \citenamefont {Charisi}, \citenamefont {Chatterjee}, \citenamefont {Chatziioannou}, \citenamefont {Cheeseboro}, \citenamefont {Chen}, \citenamefont {Cohen}, \citenamefont {Cordes}, \citenamefont {Cornish}, \citenamefont {Crawford}, \citenamefont {Cromartie}, \citenamefont {Crowter}, \citenamefont {Cutler}, \citenamefont {DeCesar}, \citenamefont {DeGan}, \citenamefont {Demorest}, \citenamefont {Deng}, \citenamefont {Dolch}, \citenamefont {Drachler}, \citenamefont {Ellis}, \citenamefont {Ferrara}, \citenamefont {Fiore}, \citenamefont {Fonseca}, \citenamefont {Freedman}, \citenamefont {{Garver-Daniels}}, \citenamefont {Gentile},
  \citenamefont {Gersbach}, \citenamefont {Glaser}, \citenamefont {Good}, \citenamefont {G{\"u}ltekin}, \citenamefont {Hazboun}, \citenamefont {Hourihane}, \citenamefont {Islo}, \citenamefont {Jennings}, \citenamefont {Johnson}, \citenamefont {Jones}, \citenamefont {Kaiser}, \citenamefont {Kaplan}, \citenamefont {Kelley}, \citenamefont {Kerr}, \citenamefont {Key}, \citenamefont {Klein}, \citenamefont {Laal}, \citenamefont {Lam}, \citenamefont {Lamb}, \citenamefont {Lazio}, \citenamefont {Lewandowska}, \citenamefont {Littenberg}, \citenamefont {Liu}, \citenamefont {Lommen}, \citenamefont {Lorimer}, \citenamefont {Luo}, \citenamefont {Lynch}, \citenamefont {Ma}, \citenamefont {Madison}, \citenamefont {Mattson}, \citenamefont {McEwen}, \citenamefont {McKee}, \citenamefont {McLaughlin}, \citenamefont {McMann}, \citenamefont {Meyers}, \citenamefont {Meyers}, \citenamefont {Mingarelli}, \citenamefont {Mitridate}, \citenamefont {Natarajan}, \citenamefont {Ng}, \citenamefont {Nice}, \citenamefont {Ocker},
  \citenamefont {Olum}, \citenamefont {Pennucci}, \citenamefont {Perera}, \citenamefont {Petrov}, \citenamefont {Pol}, \citenamefont {Radovan}, \citenamefont {Ransom}, \citenamefont {Ray}, \citenamefont {Romano}, \citenamefont {Sardesai}, \citenamefont {Schmiedekamp}, \citenamefont {Schmiedekamp}, \citenamefont {Schmitz}, \citenamefont {Schult}, \citenamefont {{Shapiro-Albert}}, \citenamefont {Siemens}, \citenamefont {Simon}, \citenamefont {Siwek}, \citenamefont {Stairs}, \citenamefont {Stinebring}, \citenamefont {Stovall}, \citenamefont {Sun}, \citenamefont {Susobhanan}, \citenamefont {Swiggum}, \citenamefont {Taylor}, \citenamefont {Taylor}, \citenamefont {Turner}, \citenamefont {Unal}, \citenamefont {Vallisneri}, \citenamefont {van Haasteren}, \citenamefont {Vigeland}, \citenamefont {Wahl}, \citenamefont {Wang}, \citenamefont {Witt}, \citenamefont {Young},\ and\ \citenamefont {Collaboration}}]{agazie_the_2023a}%
  \BibitemOpen
  \bibfield  {author} {\bibinfo {author} {\bibfnamefont {G.}~\bibnamefont {Agazie}}, \bibinfo {author} {\bibfnamefont {A.}~\bibnamefont {Anumarlapudi}}, \bibinfo {author} {\bibfnamefont {A.~M.}\ \bibnamefont {Archibald}}, \emph {et~al.},\ }\href {https://doi.org/10.3847/2041-8213/acdac6} {\bibfield  {journal} {\bibinfo  {journal} {The Astrophysical Journal Letters}\ }\textbf {\bibinfo {volume} {951}},\ \bibinfo {pages} {L8} (\bibinfo {year} {2023}{\natexlab{a}})}\BibitemShut {NoStop}%
\bibitem [{\citenamefont {Edlund}\ \emph {et~al.}(2005)\citenamefont {Edlund}, \citenamefont {Tinto}, \citenamefont {Krolak},\ and\ \citenamefont {Nelemans}}]{edlund_white_2005}%
  \BibitemOpen
  \bibfield  {author} {\bibinfo {author} {\bibfnamefont {J.~A.}\ \bibnamefont {Edlund}}, \bibinfo {author} {\bibfnamefont {M.}~\bibnamefont {Tinto}}, \bibinfo {author} {\bibfnamefont {A.}~\bibnamefont {Krolak}},\ and\ \bibinfo {author} {\bibfnamefont {G.}~\bibnamefont {Nelemans}},\ }\href {https://doi.org/10.1103/PhysRevD.71.122003} {\bibfield  {journal} {\bibinfo  {journal} {Physical Review D}\ }\textbf {\bibinfo {volume} {71}},\ \bibinfo {pages} {122003} (\bibinfo {year} {2005})},\ \Eprint {https://arxiv.org/abs/gr-qc/0504112} {arxiv:gr-qc/0504112} \BibitemShut {NoStop}%
\bibitem [{\citenamefont {Ruiter}\ \emph {et~al.}(2010)\citenamefont {Ruiter}, \citenamefont {Belczynski}, \citenamefont {Benacquista}, \citenamefont {Larson},\ and\ \citenamefont {Williams}}]{ruiter_lisa_2010}%
  \BibitemOpen
  \bibfield  {author} {\bibinfo {author} {\bibfnamefont {A.~J.}\ \bibnamefont {Ruiter}}, \bibinfo {author} {\bibfnamefont {K.}~\bibnamefont {Belczynski}}, \bibinfo {author} {\bibfnamefont {M.}~\bibnamefont {Benacquista}}, \emph {et~al.},\ }\href {https://doi.org/10.1088/0004-637X/717/2/1006} {\bibfield  {journal} {\bibinfo  {journal} {The Astrophysical Journal}\ }\textbf {\bibinfo {volume} {717}},\ \bibinfo {pages} {1006} (\bibinfo {year} {2010})},\ \Eprint {https://arxiv.org/abs/0705.3272} {arxiv:0705.3272} \BibitemShut {NoStop}%
\bibitem [{\citenamefont {Rieck}\ \emph {et~al.}(2024)\citenamefont {Rieck}, \citenamefont {Criswell}, \citenamefont {Korol}, \citenamefont {Keim}, \citenamefont {Bloom},\ and\ \citenamefont {Mandic}}]{rieck_lisa_2024}%
  \BibitemOpen
  \bibfield  {author} {\bibinfo {author} {\bibfnamefont {S.}~\bibnamefont {Rieck}}, \bibinfo {author} {\bibfnamefont {A.~W.}\ \bibnamefont {Criswell}}, \bibinfo {author} {\bibfnamefont {V.}~\bibnamefont {Korol}}, \emph {et~al.},\ }\href {https://doi.org/10.48550/arXiv.2308.12437} {\bibinfo {title} {{{A Stochastic Gravitational Wave Background in LISA from Unresolved White Dwarf Binaries in the Large Magellanic Cloud}}}} (\bibinfo {year} {2024}),\ \Eprint {https://arxiv.org/abs/2308.12437} {arxiv:2308.12437 [astro-ph, physics:gr-qc]} \BibitemShut {NoStop}%
\bibitem [{\citenamefont {Pozzoli}\ \emph {et~al.}(2023)\citenamefont {Pozzoli}, \citenamefont {Babak}, \citenamefont {Sesana}, \citenamefont {Bonetti},\ and\ \citenamefont {Karnesis}}]{pozzoli_computation_2023c}%
  \BibitemOpen
  \bibfield  {author} {\bibinfo {author} {\bibfnamefont {F.}~\bibnamefont {Pozzoli}}, \bibinfo {author} {\bibfnamefont {S.}~\bibnamefont {Babak}}, \bibinfo {author} {\bibfnamefont {A.}~\bibnamefont {Sesana}}, \emph {et~al.},\ }\href {https://doi.org/10.48550/arXiv.2302.07043} {\bibinfo {title} {Computation of stochastic background from extreme mass ratio inspiral populations for {{LISA}}}} (\bibinfo {year} {2023}),\ \Eprint {https://arxiv.org/abs/2302.07043} {arxiv:2302.07043 [astro-ph]} \BibitemShut {NoStop}%
\bibitem [{\citenamefont {Babak}\ \emph {et~al.}(2023)\citenamefont {Babak}, \citenamefont {Caprini}, \citenamefont {Figueroa}, \citenamefont {Karnesis}, \citenamefont {Marcoccia}, \citenamefont {Nardini}, \citenamefont {Pieroni}, \citenamefont {Ricciardone}, \citenamefont {Sesana},\ and\ \citenamefont {Torrado}}]{babak_stochastic_2023}%
  \BibitemOpen
  \bibfield  {author} {\bibinfo {author} {\bibfnamefont {S.}~\bibnamefont {Babak}}, \bibinfo {author} {\bibfnamefont {C.}~\bibnamefont {Caprini}}, \bibinfo {author} {\bibfnamefont {D.~G.}\ \bibnamefont {Figueroa}}, \emph {et~al.},\ }\href {https://doi.org/10.48550/arXiv.2304.06368} {\bibinfo {title} {Stochastic gravitational wave background from stellar origin binary black holes in {{LISA}}}} (\bibinfo {year} {2023}),\ \Eprint {https://arxiv.org/abs/2304.06368} {arxiv:2304.06368 [astro-ph, physics:gr-qc]} \BibitemShut {NoStop}%
\bibitem [{\citenamefont {Caprini}\ and\ \citenamefont {Figueroa}(2018)}]{caprini_cosmological_2018}%
  \BibitemOpen
  \bibfield  {author} {\bibinfo {author} {\bibfnamefont {C.}~\bibnamefont {Caprini}}\ and\ \bibinfo {author} {\bibfnamefont {D.~G.}\ \bibnamefont {Figueroa}},\ }\href {https://doi.org/10.1088/1361-6382/aac608} {\bibfield  {journal} {\bibinfo  {journal} {Classical and Quantum Gravity}\ }\textbf {\bibinfo {volume} {35}},\ \bibinfo {pages} {163001} (\bibinfo {year} {2018})}\BibitemShut {NoStop}%
\bibitem [{\citenamefont {Bethke}\ \emph {et~al.}(2013)\citenamefont {Bethke}, \citenamefont {Figueroa},\ and\ \citenamefont {Rajantie}}]{bethke_anisotropies_2013}%
  \BibitemOpen
  \bibfield  {author} {\bibinfo {author} {\bibfnamefont {L.}~\bibnamefont {Bethke}}, \bibinfo {author} {\bibfnamefont {D.~G.}\ \bibnamefont {Figueroa}},\ and\ \bibinfo {author} {\bibfnamefont {A.}~\bibnamefont {Rajantie}},\ }\href {https://doi.org/10.1103/PhysRevLett.111.011301} {\bibfield  {journal} {\bibinfo  {journal} {Physical Review Letters}\ }\textbf {\bibinfo {volume} {111}},\ \bibinfo {pages} {011301} (\bibinfo {year} {2013})}\BibitemShut {NoStop}%
\bibitem [{\citenamefont {Geller}\ \emph {et~al.}(2018)\citenamefont {Geller}, \citenamefont {Hook}, \citenamefont {Sundrum},\ and\ \citenamefont {Tsai}}]{geller_primordial_2018}%
  \BibitemOpen
  \bibfield  {author} {\bibinfo {author} {\bibfnamefont {M.}~\bibnamefont {Geller}}, \bibinfo {author} {\bibfnamefont {A.}~\bibnamefont {Hook}}, \bibinfo {author} {\bibfnamefont {R.}~\bibnamefont {Sundrum}},\ and\ \bibinfo {author} {\bibfnamefont {Y.}~\bibnamefont {Tsai}},\ }\href {https://doi.org/10.1103/PhysRevLett.121.201303} {\bibfield  {journal} {\bibinfo  {journal} {Physical Review Letters}\ }\textbf {\bibinfo {volume} {121}},\ \bibinfo {pages} {201303} (\bibinfo {year} {2018})}\BibitemShut {NoStop}%
\bibitem [{\citenamefont {Bartolo}\ \emph {et~al.}(2020{\natexlab{a}})\citenamefont {Bartolo}, \citenamefont {Bertacca}, \citenamefont {Luca}, \citenamefont {Franciolini}, \citenamefont {Matarrese}, \citenamefont {Peloso}, \citenamefont {Ricciardone}, \citenamefont {Riotto},\ and\ \citenamefont {Tasinato}}]{bartolo_gravitational_2020}%
  \BibitemOpen
  \bibfield  {author} {\bibinfo {author} {\bibfnamefont {N.}~\bibnamefont {Bartolo}}, \bibinfo {author} {\bibfnamefont {D.}~\bibnamefont {Bertacca}}, \bibinfo {author} {\bibfnamefont {V.~D.}\ \bibnamefont {Luca}}, \emph {et~al.},\ }\href {https://doi.org/10.1088/1475-7516/2020/02/028} {\bibfield  {journal} {\bibinfo  {journal} {Journal of Cosmology and Astroparticle Physics}\ }\textbf {\bibinfo {volume} {2020}}\bibinfo  {number} { (02)},\ \bibinfo {pages} {028}}\BibitemShut {NoStop}%
\bibitem [{\citenamefont {Contaldi}(2017)}]{contaldi_anisotropies_2017}%
  \BibitemOpen
\bibfield  {number} {  }\bibfield  {author} {\bibinfo {author} {\bibfnamefont {C.~R.}\ \bibnamefont {Contaldi}},\ }\href {https://doi.org/10.1016/j.physletb.2017.05.020} {\bibfield  {journal} {\bibinfo  {journal} {Physics Letters B}\ }\textbf {\bibinfo {volume} {771}},\ \bibinfo {pages} {9} (\bibinfo {year} {2017})}\BibitemShut {NoStop}%
\bibitem [{\citenamefont {Bartolo}\ \emph {et~al.}(2019)\citenamefont {Bartolo}, \citenamefont {Bertacca}, \citenamefont {Matarrese}, \citenamefont {Peloso}, \citenamefont {Ricciardone}, \citenamefont {Riotto},\ and\ \citenamefont {Tasinato}}]{bartolo_anisotropies_2019}%
  \BibitemOpen
  \bibfield  {author} {\bibinfo {author} {\bibfnamefont {N.}~\bibnamefont {Bartolo}}, \bibinfo {author} {\bibfnamefont {D.}~\bibnamefont {Bertacca}}, \bibinfo {author} {\bibfnamefont {S.}~\bibnamefont {Matarrese}}, \emph {et~al.},\ }\href {https://doi.org/10.1103/PhysRevD.100.121501} {\bibfield  {journal} {\bibinfo  {journal} {Physical Review D}\ }\textbf {\bibinfo {volume} {100}},\ \bibinfo {pages} {121501} (\bibinfo {year} {2019})}\BibitemShut {NoStop}%
\bibitem [{\citenamefont {Bartolo}\ \emph {et~al.}(2020{\natexlab{b}})\citenamefont {Bartolo}, \citenamefont {Bertacca}, \citenamefont {Matarrese}, \citenamefont {Peloso}, \citenamefont {Ricciardone}, \citenamefont {Riotto},\ and\ \citenamefont {Tasinato}}]{bartolo_characterizing_2020}%
  \BibitemOpen
  \bibfield  {author} {\bibinfo {author} {\bibfnamefont {N.}~\bibnamefont {Bartolo}}, \bibinfo {author} {\bibfnamefont {D.}~\bibnamefont {Bertacca}}, \bibinfo {author} {\bibfnamefont {S.}~\bibnamefont {Matarrese}}, \emph {et~al.},\ }\href {https://doi.org/10.1103/PhysRevD.102.023527} {\bibfield  {journal} {\bibinfo  {journal} {Physical Review D}\ }\textbf {\bibinfo {volume} {102}},\ \bibinfo {pages} {023527} (\bibinfo {year} {2020}{\natexlab{b}})},\ \Eprint {https://arxiv.org/abs/1912.09433} {arxiv:1912.09433 [astro-ph]} \BibitemShut {NoStop}%
\bibitem [{\citenamefont {Breivik}\ \emph {et~al.}(2020)\citenamefont {Breivik}, \citenamefont {Mingarelli},\ and\ \citenamefont {Larson}}]{breivik_constraining_2020}%
  \BibitemOpen
  \bibfield  {author} {\bibinfo {author} {\bibfnamefont {K.}~\bibnamefont {Breivik}}, \bibinfo {author} {\bibfnamefont {C.~M.~F.}\ \bibnamefont {Mingarelli}},\ and\ \bibinfo {author} {\bibfnamefont {S.~L.}\ \bibnamefont {Larson}},\ }\href {https://doi.org/10.3847/1538-4357/abab99} {\bibfield  {journal} {\bibinfo  {journal} {The Astrophysical Journal}\ }\textbf {\bibinfo {volume} {901}},\ \bibinfo {pages} {4} (\bibinfo {year} {2020})}\BibitemShut {NoStop}%
\bibitem [{\citenamefont {Cusin}\ \emph {et~al.}(2018{\natexlab{a}})\citenamefont {Cusin}, \citenamefont {Pitrou},\ and\ \citenamefont {Uzan}}]{cusin_the_2018}%
  \BibitemOpen
  \bibfield  {author} {\bibinfo {author} {\bibfnamefont {G.}~\bibnamefont {Cusin}}, \bibinfo {author} {\bibfnamefont {C.}~\bibnamefont {Pitrou}},\ and\ \bibinfo {author} {\bibfnamefont {J.-P.}\ \bibnamefont {Uzan}},\ }\href {https://doi.org/10.1103/PhysRevD.97.123527} {\bibfield  {journal} {\bibinfo  {journal} {Physical Review D}\ }\textbf {\bibinfo {volume} {97}},\ \bibinfo {pages} {123527} (\bibinfo {year} {2018}{\natexlab{a}})},\ \Eprint {https://arxiv.org/abs/1711.11345} {arxiv:1711.11345 [astro-ph, physics:gr-qc, physics:hep-th]} \BibitemShut {NoStop}%
\bibitem [{\citenamefont {Cusin}\ \emph {et~al.}(2018{\natexlab{b}})\citenamefont {Cusin}, \citenamefont {Dvorkin}, \citenamefont {Pitrou},\ and\ \citenamefont {Uzan}}]{cusin_first_2018}%
  \BibitemOpen
  \bibfield  {author} {\bibinfo {author} {\bibfnamefont {G.}~\bibnamefont {Cusin}}, \bibinfo {author} {\bibfnamefont {I.}~\bibnamefont {Dvorkin}}, \bibinfo {author} {\bibfnamefont {C.}~\bibnamefont {Pitrou}},\ and\ \bibinfo {author} {\bibfnamefont {J.-P.}\ \bibnamefont {Uzan}},\ }\href {https://doi.org/10.1103/PhysRevLett.120.231101} {\bibfield  {journal} {\bibinfo  {journal} {Physical Review Letters}\ }\textbf {\bibinfo {volume} {120}},\ \bibinfo {pages} {231101} (\bibinfo {year} {2018}{\natexlab{b}})},\ \Eprint {https://arxiv.org/abs/1803.03236} {arxiv:1803.03236 [astro-ph, physics:gr-qc, physics:hep-th]} \BibitemShut {NoStop}%
\bibitem [{\citenamefont {Bartolo}\ \emph {et~al.}(2022)\citenamefont {Bartolo}, \citenamefont {Bertacca}, \citenamefont {Caldwell}, \citenamefont {Contaldi}, \citenamefont {Cusin}, \citenamefont {De~Luca}, \citenamefont {Dimastrogiovanni}, \citenamefont {Fasiello}, \citenamefont {Figueroa}, \citenamefont {Franciolini}, \citenamefont {Jenkins}, \citenamefont {Peloso}, \citenamefont {Pieroni}, \citenamefont {Renzini}, \citenamefont {Ricciardone}, \citenamefont {Riotto}, \citenamefont {Sakellariadou}, \citenamefont {Sorbo}, \citenamefont {Tasinato}, \citenamefont {Torrado}, \citenamefont {Clesse},\ and\ \citenamefont {Kuroyanagi}}]{bartolo_probing_2022}%
  \BibitemOpen
  \bibfield  {author} {\bibinfo {author} {\bibfnamefont {N.}~\bibnamefont {Bartolo}}, \bibinfo {author} {\bibfnamefont {D.}~\bibnamefont {Bertacca}}, \bibinfo {author} {\bibfnamefont {R.}~\bibnamefont {Caldwell}}, \emph {et~al.},\ }\href {https://doi.org/10.1088/1475-7516/2022/11/009} {\bibfield  {journal} {\bibinfo  {journal} {Journal of Cosmology and Astroparticle Physics}\ }\textbf {\bibinfo {volume} {2022}}\bibfield  {number} {\bibinfo  {number} { (11)},\ \bibinfo {pages} {009}},\ }\Eprint {https://arxiv.org/abs/2201.08782} {arxiv:2201.08782 [astro-ph, physics:gr-qc]} \BibitemShut {NoStop}%
\bibitem [{\citenamefont {Floden}\ \emph {et~al.}(2022)\citenamefont {Floden}, \citenamefont {Mandic}, \citenamefont {Matas},\ and\ \citenamefont {Tsukada}}]{floden_angular_2022}%
  \BibitemOpen
  \bibfield  {author} {\bibinfo {author} {\bibfnamefont {E.}~\bibnamefont {Floden}}, \bibinfo {author} {\bibfnamefont {V.}~\bibnamefont {Mandic}}, \bibinfo {author} {\bibfnamefont {A.}~\bibnamefont {Matas}},\ and\ \bibinfo {author} {\bibfnamefont {L.}~\bibnamefont {Tsukada}},\ }\href {https://doi.org/10.1103/PhysRevD.106.023010} {\bibfield  {journal} {\bibinfo  {journal} {Physical Review D}\ }\textbf {\bibinfo {volume} {106}},\ \bibinfo {pages} {023010} (\bibinfo {year} {2022})},\ \Eprint {https://arxiv.org/abs/2203.17141} {arxiv:2203.17141 [astro-ph, physics:gr-qc]} \BibitemShut {NoStop}%
\bibitem [{\citenamefont {Abbott}\ \emph {et~al.}(2021)\citenamefont {Abbott}, \citenamefont {Abbott}, \citenamefont {Abraham}, \citenamefont {Acernese}, \citenamefont {Ackley}, \citenamefont {Adams}, \citenamefont {Adams}, \citenamefont {Adhikari}, \citenamefont {Adya}, \citenamefont {Affeldt}, \citenamefont {Agarwal}, \citenamefont {Agathos}, \citenamefont {Agatsuma}, \citenamefont {Aggarwal}, \citenamefont {Aguiar}, \citenamefont {Aiello}, \citenamefont {Ain}, \citenamefont {Ajith}, \citenamefont {Akutsu}, \citenamefont {Aleman}, \citenamefont {Allen}, \citenamefont {Allocca}, \citenamefont {Altin}, \citenamefont {Amato}, \citenamefont {Anand}, \citenamefont {Ananyeva}, \citenamefont {Anderson}, \citenamefont {Anderson}, \citenamefont {Ando}, \citenamefont {Angelova}, \citenamefont {Ansoldi}, \citenamefont {Antelis}, \citenamefont {Antier}, \citenamefont {Appert}, \citenamefont {Arai}, \citenamefont {Arai}, \citenamefont {Arai}, \citenamefont {Araki}, \citenamefont {Araya}, \citenamefont {Araya},
  \citenamefont {Areeda}, \citenamefont {Ar{\`e}ne}, \citenamefont {Aritomi}, \citenamefont {Arnaud}, \citenamefont {Aronson}, \citenamefont {Asada}, \citenamefont {Asali}, \citenamefont {Ashton}, \citenamefont {Aso}, \citenamefont {Aston}, \citenamefont {Astone}, \citenamefont {Aubin}, \citenamefont {Aufmuth}, \citenamefont {AultONeal}, \citenamefont {Austin}, \citenamefont {Babak}, \citenamefont {Badaracco}, \citenamefont {Bader}, \citenamefont {Bae}, \citenamefont {Bae}, \citenamefont {Baer}, \citenamefont {Bagnasco}, \citenamefont {Bai}, \citenamefont {Baiotti}, \citenamefont {Baird}, \citenamefont {Bajpai}, \citenamefont {Ball}, \citenamefont {Ballardin}, \citenamefont {Ballmer}, \citenamefont {Bals}, \citenamefont {Balsamo}, \citenamefont {Baltus}, \citenamefont {Banagiri}, \citenamefont {Bankar}, \citenamefont {Bankar}, \citenamefont {Barayoga}, \citenamefont {Barbieri}, \citenamefont {Barish}, \citenamefont {Barker}, \citenamefont {Barneo}, \citenamefont {Barone}, \citenamefont {Barr}, \citenamefont
  {Barsotti}, \citenamefont {Barsuglia}, \citenamefont {Barta}, \citenamefont {Bartlett}, \citenamefont {Barton}, \citenamefont {Bartos}, \citenamefont {Bassiri}, \citenamefont {Basti}, \citenamefont {Bawaj}, \citenamefont {Bayley}, \citenamefont {Baylor}, \citenamefont {Bazzan}, \citenamefont {B{\'e}csy}, \citenamefont {Bedakihale}, \citenamefont {Bejger}, \citenamefont {Belahcene}, \citenamefont {Benedetto}, \citenamefont {Beniwal}, \citenamefont {Benjamin}, \citenamefont {Bennett}, \citenamefont {Bentley}, \citenamefont {BenYaala}, \citenamefont {Bergamin}, \citenamefont {Berger}, \citenamefont {Bernuzzi}, \citenamefont {Bersanetti}, \citenamefont {Bertolini}, \citenamefont {Betzwieser}, \citenamefont {Bhandare}, \citenamefont {Bhandari}, \citenamefont {Bhattacharjee}, \citenamefont {Bhaumik}, \citenamefont {Bidler}, \citenamefont {Bilenko}, \citenamefont {Billingsley}, \citenamefont {Birney}, \citenamefont {Birnholtz}, \citenamefont {Biscans}, \citenamefont {Bischi}, \citenamefont {Biscoveanu},
  \citenamefont {Bisht}, \citenamefont {Biswas}, \citenamefont {Bitossi}, \citenamefont {Bizouard}, \citenamefont {Blackburn}, \citenamefont {Blackman}, \citenamefont {Blair}, \citenamefont {Blair}, \citenamefont {Blair}, \citenamefont {Bobba}, \citenamefont {Bode}, \citenamefont {Boer}, \citenamefont {Bogaert}, \citenamefont {Boldrini}, \citenamefont {Bondu}, \citenamefont {Bonilla}, \citenamefont {Bonnand}, \citenamefont {Booker}, \citenamefont {Boom}, \citenamefont {Bork}, \citenamefont {Boschi}, \citenamefont {Bose}, \citenamefont {Bose}, \citenamefont {Bossilkov}, \citenamefont {Boudart}, \citenamefont {Bouffanais}, \citenamefont {Bozzi}, \citenamefont {Bradaschia}, \citenamefont {Brady}, \citenamefont {Bramley}, \citenamefont {Branch}, \citenamefont {Branchesi}, \citenamefont {Brau}, \citenamefont {Breschi}, \citenamefont {Briant}, \citenamefont {Briggs}, \citenamefont {Brillet}, \citenamefont {Brinkmann}, \citenamefont {Brockill}, \citenamefont {Brooks}, \citenamefont {Brooks}, \citenamefont {Brown},
  \citenamefont {Brunett}, \citenamefont {Bruno}, \citenamefont {Bruntz}, \citenamefont {Bryant}, \citenamefont {Buikema}, \citenamefont {Bulik}, \citenamefont {Bulten}, \citenamefont {Buonanno}, \citenamefont {Buscicchio}, \citenamefont {Buskulic}, \citenamefont {Byer}, \citenamefont {Cadonati}, \citenamefont {Caesar}, \citenamefont {Cagnoli}, \citenamefont {Cahillane}, \citenamefont {Cain}, \citenamefont {Bustillo}, \citenamefont {Callaghan}, \citenamefont {Callister}, \citenamefont {Calloni}, \citenamefont {Camp}, \citenamefont {Canepa}, \citenamefont {Cannavacciuolo}, \citenamefont {Cannon}, \citenamefont {Cao}, \citenamefont {Cao}, \citenamefont {Cao}, \citenamefont {Capocasa}, \citenamefont {Capote}, \citenamefont {Carapella}, \citenamefont {Carbognani}, \citenamefont {Carlin}, \citenamefont {Carney}, \citenamefont {Carpinelli}, \citenamefont {Carullo}, \citenamefont {Carver}, \citenamefont {Diaz}, \citenamefont {Casentini}, \citenamefont {Castaldi}, \citenamefont {Caudill}, \citenamefont
  {Cavagli{\`a}}, \citenamefont {Cavalier}, \citenamefont {Cavalieri}, \citenamefont {Cella}, \citenamefont {{Cerd{\'a}-Dur{\'a}n}}, \citenamefont {Cesarini}, \citenamefont {Chaibi}, \citenamefont {Chakravarti}, \citenamefont {Champion}, \citenamefont {Chan}, \citenamefont {Chan}, \citenamefont {Chan}, \citenamefont {Chan}, \citenamefont {Chandra}, \citenamefont {Chanial}, \citenamefont {Chao}, \citenamefont {Charlton}, \citenamefont {Chase}, \citenamefont {{Chassande-Mottin}}, \citenamefont {Chatterjee}, \citenamefont {Chaturvedi}, \citenamefont {Chen}, \citenamefont {Chen}, \citenamefont {Chen}, \citenamefont {Chen}, \citenamefont {Chen}, \citenamefont {Chen}, \citenamefont {Chen}, \citenamefont {Chen}, \citenamefont {Chen}, \citenamefont {Cheng}, \citenamefont {Cheong}, \citenamefont {Cheung}, \citenamefont {Chia}, \citenamefont {Chiadini}, \citenamefont {Chiang}, \citenamefont {Chierici}, \citenamefont {Chincarini}, \citenamefont {Chiofalo}, \citenamefont {Chiummo}, \citenamefont {Cho}, \citenamefont
  {Cho}, \citenamefont {Choate}, \citenamefont {Choudhary}, \citenamefont {Choudhary}, \citenamefont {Christensen}, \citenamefont {Chu}, \citenamefont {Chu}, \citenamefont {Chu}, \citenamefont {Chua}, \citenamefont {Chung}, \citenamefont {Ciani}, \citenamefont {Ciecielag}, \citenamefont {Cie{\'s}lar}, \citenamefont {Cifaldi}, \citenamefont {Ciobanu}, \citenamefont {Ciolfi}, \citenamefont {Cipriano}, \citenamefont {Cirone}, \citenamefont {Clara}, \citenamefont {Clark}, \citenamefont {Clark}, \citenamefont {Clarke}, \citenamefont {Clearwater}, \citenamefont {Clesse}, \citenamefont {Cleva}, \citenamefont {Coccia}, \citenamefont {Cohadon}, \citenamefont {Cohen}, \citenamefont {Cohen}, \citenamefont {Colleoni}, \citenamefont {Collette}, \citenamefont {Colpi}, \citenamefont {Compton}, \citenamefont {Constancio}, \citenamefont {Conti}, \citenamefont {Cooper}, \citenamefont {Corban}, \citenamefont {Corbitt}, \citenamefont {{Cordero-Carri{\'o}n}}, \citenamefont {Corezzi}, \citenamefont {Corley}, \citenamefont
  {Cornish}, \citenamefont {Corre}, \citenamefont {Corsi}, \citenamefont {Cortese}, \citenamefont {Costa}, \citenamefont {Cotesta}, \citenamefont {Coughlin}, \citenamefont {Coughlin}, \citenamefont {Coulon}, \citenamefont {Countryman}, \citenamefont {Cousins}, \citenamefont {Couvares}, \citenamefont {Covas}, \citenamefont {Coward}, \citenamefont {Cowart}, \citenamefont {Coyne}, \citenamefont {Coyne}, \citenamefont {Creighton}, \citenamefont {Creighton}, \citenamefont {Criswell}, \citenamefont {Croquette}, \citenamefont {Crowder}, \citenamefont {Cudell}, \citenamefont {Cullen}, \citenamefont {Cumming}, \citenamefont {Cummings}, \citenamefont {Cuoco}, \citenamefont {Cury{\l}o}, \citenamefont {Canton}, \citenamefont {D{\'a}lya}, \citenamefont {Dana}, \citenamefont {DaneshgaranBajastani}, \citenamefont {D'Angelo}, \citenamefont {Danilishin}, \citenamefont {D'Antonio}, \citenamefont {Danzmann}, \citenamefont {{Darsow-Fromm}}, \citenamefont {Dasgupta}, \citenamefont {Datrier}, \citenamefont {Dattilo}, \citenamefont
  {Dave}, \citenamefont {Davier}, \citenamefont {Davies}, \citenamefont {Davis}, \citenamefont {Daw}, \citenamefont {Dean}, \citenamefont {DeBra}, \citenamefont {Deenadayalan}, \citenamefont {Degallaix}, \citenamefont {De~Laurentis}, \citenamefont {Del{\'e}glise}, \citenamefont {Del~Favero}, \citenamefont {De~Lillo}, \citenamefont {De~Lillo}, \citenamefont {Del~Pozzo}, \citenamefont {DeMarchi}, \citenamefont {De~Matteis}, \citenamefont {D'Emilio}, \citenamefont {Demos}, \citenamefont {Dent}, \citenamefont {Depasse}, \citenamefont {De~Pietri}, \citenamefont {De~Rosa}, \citenamefont {De~Rossi}, \citenamefont {DeSalvo}, \citenamefont {De~Simone}, \citenamefont {Dhurandhar}, \citenamefont {D{\'i}az}, \citenamefont {{Diaz-Ortiz}}, \citenamefont {Didio}, \citenamefont {Dietrich}, \citenamefont {Di~Fiore}, \citenamefont {Di~Fronzo}, \citenamefont {Di~Giorgio}, \citenamefont {Di~Giovanni}, \citenamefont {Di~Girolamo}, \citenamefont {Di~Lieto}, \citenamefont {Ding}, \citenamefont {Di~Pace}, \citenamefont {Di~Palma},
  \citenamefont {Di~Renzo}, \citenamefont {Divakarla}, \citenamefont {Dmitriev}, \citenamefont {Doctor}, \citenamefont {D'Onofrio}, \citenamefont {Donovan}, \citenamefont {Dooley}, \citenamefont {Doravari}, \citenamefont {Dorrington}, \citenamefont {Drago}, \citenamefont {Driggers}, \citenamefont {Drori}, \citenamefont {Du}, \citenamefont {Ducoin}, \citenamefont {Dupej}, \citenamefont {Durante}, \citenamefont {D'Urso}, \citenamefont {Duverne}, \citenamefont {Dwyer}, \citenamefont {Easter}, \citenamefont {Ebersold}, \citenamefont {Eddolls}, \citenamefont {Edelman}, \citenamefont {Edo}, \citenamefont {Edy}, \citenamefont {Effler}, \citenamefont {Eguchi}, \citenamefont {Eichholz}, \citenamefont {Eikenberry}, \citenamefont {Eisenmann}, \citenamefont {Eisenstein}, \citenamefont {Ejlli}, \citenamefont {Enomoto}, \citenamefont {Errico}, \citenamefont {Essick}, \citenamefont {Estell{\'e}s}, \citenamefont {Estevez}, \citenamefont {Etienne}, \citenamefont {Etzel}, \citenamefont {Evans}, \citenamefont {Evans},
  \citenamefont {Ewing}, \citenamefont {Fafone}, \citenamefont {Fair}, \citenamefont {Fairhurst}, \citenamefont {Fan}, \citenamefont {Farah}, \citenamefont {Farinon}, \citenamefont {Farr}, \citenamefont {Farr}, \citenamefont {Farrow}, \citenamefont {{Fauchon-Jones}}, \citenamefont {Favata}, \citenamefont {Fays}, \citenamefont {Fazio}, \citenamefont {Feicht}, \citenamefont {Fejer}, \citenamefont {Feng}, \citenamefont {Fenyvesi}, \citenamefont {Ferguson}, \citenamefont {{Fernandez-Galiana}}, \citenamefont {Ferrante}, \citenamefont {Ferreira}, \citenamefont {Fidecaro}, \citenamefont {Figura}, \citenamefont {Fiori}, \citenamefont {Fishbach}, \citenamefont {Fisher}, \citenamefont {Fittipaldi}, \citenamefont {Fiumara}, \citenamefont {Flaminio}, \citenamefont {Floden}, \citenamefont {Flynn}, \citenamefont {Fong}, \citenamefont {Font}, \citenamefont {Fornal}, \citenamefont {Forsyth}, \citenamefont {Franke}, \citenamefont {Frasca}, \citenamefont {Frasconi}, \citenamefont {Frederick}, \citenamefont {Frei},
  \citenamefont {Freise}, \citenamefont {Frey}, \citenamefont {Fritschel}, \citenamefont {Frolov}, \citenamefont {Fronz{\'e}}, \citenamefont {Fujii}, \citenamefont {Fujikawa}, \citenamefont {Fukunaga}, \citenamefont {Fukushima}, \citenamefont {Fulda}, \citenamefont {Fyffe}, \citenamefont {Gabbard}, \citenamefont {Gadre}, \citenamefont {Gaebel}, \citenamefont {Gair}, \citenamefont {Gais}, \citenamefont {Galaudage}, \citenamefont {Gamba}, \citenamefont {Ganapathy}, \citenamefont {Ganguly}, \citenamefont {Gao}, \citenamefont {Gaonkar}, \citenamefont {Garaventa}, \citenamefont {{Garc{\'i}a-N{\'u}{\~n}ez}}, \citenamefont {{Garc{\'i}a-Quir{\'o}s}}, \citenamefont {Garufi}, \citenamefont {Gateley}, \citenamefont {Gaudio}, \citenamefont {Gayathri}, \citenamefont {Ge}, \citenamefont {Gemme}, \citenamefont {Gennai}, \citenamefont {George}, \citenamefont {Gergely}, \citenamefont {Gewecke}, \citenamefont {Ghonge}, \citenamefont {Ghosh}, \citenamefont {Ghosh}, \citenamefont {Ghosh}, \citenamefont {Ghosh}, \citenamefont
  {Ghosh}, \citenamefont {Giacomazzo}, \citenamefont {Giacoppo}, \citenamefont {Giaime}, \citenamefont {Giardina}, \citenamefont {Gibson}, \citenamefont {Gier}, \citenamefont {Giesler}, \citenamefont {Giri}, \citenamefont {Gissi}, \citenamefont {Glanzer}, \citenamefont {Gleckl}, \citenamefont {Godwin}, \citenamefont {Goetz}, \citenamefont {Goetz}, \citenamefont {Gohlke}, \citenamefont {Goncharov}, \citenamefont {Gonz{\'a}lez}, \citenamefont {Gopakumar}, \citenamefont {Gosselin}, \citenamefont {Gouaty}, \citenamefont {Grace}, \citenamefont {Grado}, \citenamefont {Granata}, \citenamefont {Granata}, \citenamefont {Grant}, \citenamefont {Gras}, \citenamefont {Grassia}, \citenamefont {Gray}, \citenamefont {Gray}, \citenamefont {Greco}, \citenamefont {Green}, \citenamefont {Green}, \citenamefont {Gretarsson}, \citenamefont {Gretarsson}, \citenamefont {Griffith}, \citenamefont {Griffiths}, \citenamefont {Griggs}, \citenamefont {Grignani}, \citenamefont {Grimaldi}, \citenamefont {Grimes}, \citenamefont {Grimm},
  \citenamefont {Grote}, \citenamefont {Grunewald}, \citenamefont {Gruning}, \citenamefont {Guerrero}, \citenamefont {Guidi}, \citenamefont {Guimaraes}, \citenamefont {Guix{\'e}}, \citenamefont {Gulati}, \citenamefont {Guo}, \citenamefont {Guo}, \citenamefont {Gupta}, \citenamefont {Gupta}, \citenamefont {Gupta}, \citenamefont {Gustafson}, \citenamefont {Gustafson}, \citenamefont {Guzman}, \citenamefont {Ha}, \citenamefont {Haegel}, \citenamefont {Hagiwara}, \citenamefont {Haino}, \citenamefont {Halim}, \citenamefont {Hall}, \citenamefont {Hamilton}, \citenamefont {Hammond}, \citenamefont {Han}, \citenamefont {Haney}, \citenamefont {Hanks}, \citenamefont {Hanna}, \citenamefont {Hannam}, \citenamefont {Hannuksela}, \citenamefont {Hansen}, \citenamefont {Hansen}, \citenamefont {Hanson}, \citenamefont {Harder}, \citenamefont {Hardwick}, \citenamefont {Haris}, \citenamefont {Harms}, \citenamefont {Harry}, \citenamefont {Harry}, \citenamefont {Hartwig}, \citenamefont {Hasegawa}, \citenamefont {Haskell},
  \citenamefont {Hasskew}, \citenamefont {Haster}, \citenamefont {Hattori}, \citenamefont {Haughian}, \citenamefont {Hayakawa}, \citenamefont {Hayama}, \citenamefont {Hayes}, \citenamefont {Healy}, \citenamefont {Heidmann}, \citenamefont {Heintze}, \citenamefont {Heinze}, \citenamefont {Heinzel}, \citenamefont {Heitmann}, \citenamefont {Hellman}, \citenamefont {Hello}, \citenamefont {{Helmling-Cornell}}, \citenamefont {Hemming}, \citenamefont {Hendry}, \citenamefont {Heng}, \citenamefont {Hennes}, \citenamefont {Hennig}, \citenamefont {Hennig}, \citenamefont {Vivanco}, \citenamefont {Heurs}, \citenamefont {Hild}, \citenamefont {Hill}, \citenamefont {Himemoto}, \citenamefont {Hines}, \citenamefont {Hiranuma}, \citenamefont {Hirata}, \citenamefont {Hirose}, \citenamefont {Hochheim}, \citenamefont {Hofman}, \citenamefont {Hohmann}, \citenamefont {Holgado}, \citenamefont {Holland}, \citenamefont {Hollows}, \citenamefont {Holmes}, \citenamefont {Holt}, \citenamefont {Holz}, \citenamefont {Hong}, \citenamefont
  {Hopkins}, \citenamefont {Hough}, \citenamefont {Howell}, \citenamefont {Hoy}, \citenamefont {Hoyland}, \citenamefont {Hreibi}, \citenamefont {Hsieh}, \citenamefont {Hsu}, \citenamefont {Huang}, \citenamefont {Huang}, \citenamefont {Huang}, \citenamefont {Huang}, \citenamefont {Huang}, \citenamefont {Huang}, \citenamefont {H{\"u}bner}, \citenamefont {Huddart}, \citenamefont {Huerta}, \citenamefont {Hughey}, \citenamefont {Hui}, \citenamefont {Hui}, \citenamefont {Husa}, \citenamefont {Huttner}, \citenamefont {Huxford}, \citenamefont {{Huynh-Dinh}}, \citenamefont {Ide}, \citenamefont {Idzkowski}, \citenamefont {Iess}, \citenamefont {Ikenoue}, \citenamefont {Imam}, \citenamefont {Inayoshi}, \citenamefont {Inchauspe}, \citenamefont {Ingram}, \citenamefont {Inoue}, \citenamefont {Intini}, \citenamefont {Ioka}, \citenamefont {Isi}, \citenamefont {Isleif}, \citenamefont {Ito}, \citenamefont {Itoh}, \citenamefont {Iyer}, \citenamefont {Izumi}, \citenamefont {JaberianHamedan}, \citenamefont {Jacqmin}, \citenamefont
  {Jadhav}, \citenamefont {Jadhav}, \citenamefont {James}, \citenamefont {Jan}, \citenamefont {Jani}, \citenamefont {Janssens}, \citenamefont {Janthalur}, \citenamefont {Jaranowski}, \citenamefont {Jariwala}, \citenamefont {Jaume}, \citenamefont {Jenkins}, \citenamefont {Jeon}, \citenamefont {Jeunon}, \citenamefont {Jia}, \citenamefont {Jiang}, \citenamefont {Jin}, \citenamefont {Johns}, \citenamefont {Jones}, \citenamefont {Jones}, \citenamefont {Jones}, \citenamefont {Jones}, \citenamefont {Jones}, \citenamefont {Jonker}, \citenamefont {Ju}, \citenamefont {Jung}, \citenamefont {Jung}, \citenamefont {Junker}, \citenamefont {Kaihotsu}, \citenamefont {Kajita}, \citenamefont {Kakizaki}, \citenamefont {Kalaghatgi}, \citenamefont {Kalogera}, \citenamefont {Kamai}, \citenamefont {Kamiizumi}, \citenamefont {Kanda}, \citenamefont {Kandhasamy}, \citenamefont {Kang}, \citenamefont {Kanner}, \citenamefont {Kao}, \citenamefont {Kapadia}, \citenamefont {Kapasi}, \citenamefont {Karat}, \citenamefont {Karathanasis},
  \citenamefont {Karki}, \citenamefont {Kashyap}, \citenamefont {Kasprzack}, \citenamefont {Kastaun}, \citenamefont {Katsanevas}, \citenamefont {Katsavounidis}, \citenamefont {Katzman}, \citenamefont {Kaur}, \citenamefont {Kawabe}, \citenamefont {Kawaguchi}, \citenamefont {Kawai}, \citenamefont {Kawasaki}, \citenamefont {K{\'e}f{\'e}lian}, \citenamefont {Keitel}, \citenamefont {Key}, \citenamefont {Khadka}, \citenamefont {Khalili}, \citenamefont {Khan}, \citenamefont {Khan}, \citenamefont {Khazanov}, \citenamefont {Khetan}, \citenamefont {Khursheed}, \citenamefont {Kijbunchoo}, \citenamefont {Kim}, \citenamefont {Kim}, \citenamefont {Kim}, \citenamefont {Kim}, \citenamefont {Kim}, \citenamefont {Kim}, \citenamefont {Kimball}, \citenamefont {Kimura}, \citenamefont {King}, \citenamefont {{Kinley-Hanlon}}, \citenamefont {Kirchhoff}, \citenamefont {Kissel}, \citenamefont {Kita}, \citenamefont {Kitazawa}, \citenamefont {Kleybolte}, \citenamefont {Klimenko}, \citenamefont {Knee}, \citenamefont {Knowles},
  \citenamefont {Knyazev}, \citenamefont {Koch}, \citenamefont {Koekoek}, \citenamefont {Kojima}, \citenamefont {Kokeyama}, \citenamefont {Koley}, \citenamefont {Kolitsidou}, \citenamefont {Kolstein}, \citenamefont {Komori}, \citenamefont {Kondrashov}, \citenamefont {Kong}, \citenamefont {Kontos}, \citenamefont {Koper}, \citenamefont {Korobko}, \citenamefont {Kotake}, \citenamefont {Kovalam}, \citenamefont {Kozak}, \citenamefont {Kozakai}, \citenamefont {Kozu}, \citenamefont {Kringel}, \citenamefont {Krishnendu}, \citenamefont {Kr{\'o}lak}, \citenamefont {Kuehn}, \citenamefont {Kuei}, \citenamefont {Kumar}, \citenamefont {Kumar}, \citenamefont {Kumar}, \citenamefont {Kumar}, \citenamefont {Kume}, \citenamefont {Kuns}, \citenamefont {Kuo}, \citenamefont {Kuo}, \citenamefont {Kuromiya}, \citenamefont {Kuroyanagi}, \citenamefont {Kusayanagi}, \citenamefont {Kwak}, \citenamefont {Kwang}, \citenamefont {Laghi}, \citenamefont {Lalande}, \citenamefont {Lam}, \citenamefont {Lamberts}, \citenamefont {Landry},
  \citenamefont {Lane}, \citenamefont {Lang}, \citenamefont {Lange}, \citenamefont {Lantz}, \citenamefont {La~Rosa}, \citenamefont {{Lartaux-Vollard}}, \citenamefont {Lasky}, \citenamefont {Laxen}, \citenamefont {Lazzarini}, \citenamefont {Lazzaro}, \citenamefont {Leaci}, \citenamefont {Leavey}, \citenamefont {Lecoeuche}, \citenamefont {Lee}, \citenamefont {Lee}, \citenamefont {Lee}, \citenamefont {Lee}, \citenamefont {Lee}, \citenamefont {Lee}, \citenamefont {Lehmann}, \citenamefont {Lema{\^i}tre}, \citenamefont {Leon}, \citenamefont {Leonardi}, \citenamefont {Leroy}, \citenamefont {Letendre}, \citenamefont {Levin}, \citenamefont {Leviton}, \citenamefont {Li}, \citenamefont {Li}, \citenamefont {Li}, \citenamefont {Li}, \citenamefont {Li}, \citenamefont {Li}, \citenamefont {Lin}, \citenamefont {Lin}, \citenamefont {Lin}, \citenamefont {Lin}, \citenamefont {Lin}, \citenamefont {Linde}, \citenamefont {Linker}, \citenamefont {Linley}, \citenamefont {Littenberg}, \citenamefont {Liu}, \citenamefont {Liu},
  \citenamefont {Liu}, \citenamefont {Liu}, \citenamefont {{Llorens-Monteagudo}}, \citenamefont {Lo}, \citenamefont {Lockwood}, \citenamefont {Lollie}, \citenamefont {London}, \citenamefont {Longo}, \citenamefont {Lopez}, \citenamefont {Lorenzini}, \citenamefont {Loriette}, \citenamefont {Lormand}, \citenamefont {Losurdo}, \citenamefont {Lough}, \citenamefont {Lousto}, \citenamefont {Lovelace}, \citenamefont {L{\"u}ck}, \citenamefont {Lumaca}, \citenamefont {Lundgren}, \citenamefont {Luo}, \citenamefont {Macas}, \citenamefont {MacInnis}, \citenamefont {Macleod}, \citenamefont {MacMillan}, \citenamefont {Macquet}, \citenamefont {Hernandez}, \citenamefont {{Maga{\~n}a-Sandoval}}, \citenamefont {Magazz{\`u}}, \citenamefont {Magee}, \citenamefont {Maggiore}, \citenamefont {Majorana}, \citenamefont {Makarem}, \citenamefont {Maksimovic}, \citenamefont {Maliakal}, \citenamefont {Malik}, \citenamefont {Man}, \citenamefont {Mandic}, \citenamefont {Mangano}, \citenamefont {Mango}, \citenamefont {Mansell}, \citenamefont
  {Manske}, \citenamefont {Mantovani}, \citenamefont {Mapelli}, \citenamefont {Marchesoni}, \citenamefont {Marchio}, \citenamefont {Marion}, \citenamefont {Mark}, \citenamefont {M{\'a}rka}, \citenamefont {M{\'a}rka}, \citenamefont {Markakis}, \citenamefont {Markosyan}, \citenamefont {Markowitz}, \citenamefont {Maros}, \citenamefont {Marquina}, \citenamefont {Marsat}, \citenamefont {Martelli}, \citenamefont {Martin}, \citenamefont {Martin}, \citenamefont {Martinez}, \citenamefont {Martinez}, \citenamefont {Martinovic}, \citenamefont {Martynov}, \citenamefont {Marx}, \citenamefont {Masalehdan}, \citenamefont {Mason}, \citenamefont {Massera}, \citenamefont {Masserot}, \citenamefont {Massinger}, \citenamefont {{Masso-Reid}}, \citenamefont {Mastrogiovanni}, \citenamefont {Matas}, \citenamefont {{Mateu-Lucena}}, \citenamefont {Matichard}, \citenamefont {Matiushechkina}, \citenamefont {Mavalvala}, \citenamefont {McCann}, \citenamefont {McCarthy}, \citenamefont {McClelland}, \citenamefont {McClincy}, \citenamefont
  {McCormick}, \citenamefont {McCuller}, \citenamefont {McGhee}, \citenamefont {McGuire}, \citenamefont {McIsaac}, \citenamefont {McIver}, \citenamefont {McManus}, \citenamefont {McRae}, \citenamefont {McWilliams}, \citenamefont {Meacher}, \citenamefont {Mehmet}, \citenamefont {Mehta}, \citenamefont {Melatos}, \citenamefont {Melchor}, \citenamefont {Mendell}, \citenamefont {{Menendez-Vazquez}}, \citenamefont {Menoni}, \citenamefont {Mercer}, \citenamefont {Mereni}, \citenamefont {Merfeld}, \citenamefont {Merilh}, \citenamefont {Merritt}, \citenamefont {Merzougui}, \citenamefont {Meshkov}, \citenamefont {Messenger}, \citenamefont {Messick}, \citenamefont {Meyers}, \citenamefont {Meylahn}, \citenamefont {Mhaske}, \citenamefont {Miani}, \citenamefont {Miao}, \citenamefont {Michaloliakos}, \citenamefont {Michel}, \citenamefont {Michimura}, \citenamefont {Middleton}, \citenamefont {Milano}, \citenamefont {Miller}, \citenamefont {Millhouse}, \citenamefont {Mills}, \citenamefont {Milotti}, \citenamefont
  {{Milovich-Goff}}, \citenamefont {Minazzoli}, \citenamefont {Minenkov}, \citenamefont {Mio}, \citenamefont {Mir}, \citenamefont {Mishkin}, \citenamefont {Mishra}, \citenamefont {Mishra}, \citenamefont {Mistry}, \citenamefont {Mitra}, \citenamefont {Mitrofanov}, \citenamefont {Mitselmakher}, \citenamefont {Mittleman}, \citenamefont {Miyakawa}, \citenamefont {Miyamoto}, \citenamefont {Miyazaki}, \citenamefont {Miyo}, \citenamefont {Miyoki}, \citenamefont {Mo}, \citenamefont {Mogushi}, \citenamefont {Mohapatra}, \citenamefont {Mohite}, \citenamefont {Molina}, \citenamefont {{Molina-Ruiz}}, \citenamefont {Mondin}, \citenamefont {Montani}, \citenamefont {Moore}, \citenamefont {Moraru}, \citenamefont {Morawski}, \citenamefont {More}, \citenamefont {Moreno}, \citenamefont {Moreno}, \citenamefont {Mori}, \citenamefont {Morisaki}, \citenamefont {Moriwaki}, \citenamefont {Mours}, \citenamefont {{Mow-Lowry}}, \citenamefont {Mozzon}, \citenamefont {Muciaccia}, \citenamefont {Mukherjee}, \citenamefont {Mukherjee},
  \citenamefont {Mukherjee}, \citenamefont {Mukherjee}, \citenamefont {Mukund}, \citenamefont {Mullavey}, \citenamefont {Munch}, \citenamefont {Mu{\~n}iz}, \citenamefont {Murray}, \citenamefont {Musenich}, \citenamefont {Nadji}, \citenamefont {Nagano}, \citenamefont {Nagano}, \citenamefont {Nakamura}, \citenamefont {Nakano}, \citenamefont {Nakano}, \citenamefont {Nakashima}, \citenamefont {Nakayama}, \citenamefont {Nardecchia}, \citenamefont {Narikawa}, \citenamefont {Naticchioni}, \citenamefont {Nayak}, \citenamefont {Nayak}, \citenamefont {Negishi}, \citenamefont {Neil}, \citenamefont {Neilson}, \citenamefont {Nelemans}, \citenamefont {Nelson}, \citenamefont {Nery}, \citenamefont {Neunzert}, \citenamefont {Ng}, \citenamefont {Ng}, \citenamefont {Nguyen}, \citenamefont {Nguyen}, \citenamefont {Nguyen}, \citenamefont {Quynh}, \citenamefont {Ni}, \citenamefont {Nichols}, \citenamefont {Nishizawa}, \citenamefont {Nissanke}, \citenamefont {Nocera}, \citenamefont {Noh}, \citenamefont {Norman}, \citenamefont
  {North}, \citenamefont {Nozaki}, \citenamefont {Nuttall}, \citenamefont {Oberling}, \citenamefont {O'Brien}, \citenamefont {Obuchi}, \citenamefont {O'Dell}, \citenamefont {Ogaki}, \citenamefont {Oganesyan}, \citenamefont {Oh}, \citenamefont {Oh}, \citenamefont {Oh}, \citenamefont {Ohashi}, \citenamefont {Ohishi}, \citenamefont {Ohkawa}, \citenamefont {Ohme}, \citenamefont {Ohta}, \citenamefont {Okada}, \citenamefont {Okutani}, \citenamefont {Okutomi}, \citenamefont {Olivetto}, \citenamefont {Oohara}, \citenamefont {Ooi}, \citenamefont {Oram}, \citenamefont {O'Reilly}, \citenamefont {Ormiston}, \citenamefont {Ormsby}, \citenamefont {Ortega}, \citenamefont {O'Shaughnessy}, \citenamefont {O'Shea}, \citenamefont {Oshino}, \citenamefont {Ossokine}, \citenamefont {Osthelder}, \citenamefont {Otabe}, \citenamefont {Ottaway}, \citenamefont {Overmier}, \citenamefont {Pace}, \citenamefont {Pagano}, \citenamefont {Page}, \citenamefont {Pagliaroli}, \citenamefont {Pai}, \citenamefont {Pai}, \citenamefont {Palamos},
  \citenamefont {Palashov}, \citenamefont {Palomba}, \citenamefont {Pan}, \citenamefont {Panda}, \citenamefont {Pang}, \citenamefont {Pang}, \citenamefont {Pankow}, \citenamefont {Pannarale}, \citenamefont {Pant}, \citenamefont {Paoletti}, \citenamefont {Paoli}, \citenamefont {Paolone}, \citenamefont {Parisi}, \citenamefont {Park}, \citenamefont {Parker}, \citenamefont {Pascucci}, \citenamefont {Pasqualetti}, \citenamefont {Passaquieti}, \citenamefont {Passuello}, \citenamefont {Patel}, \citenamefont {Patricelli}, \citenamefont {Payne}, \citenamefont {Pechsiri}, \citenamefont {Pedraza}, \citenamefont {Pegoraro}, \citenamefont {Pele}, \citenamefont {Arellano}, \citenamefont {Penn}, \citenamefont {Perego}, \citenamefont {Pereira}, \citenamefont {Pereira}, \citenamefont {Perez}, \citenamefont {P{\'e}rigois}, \citenamefont {Perreca}, \citenamefont {Perri{\`e}s}, \citenamefont {Petermann}, \citenamefont {Petterson}, \citenamefont {Pfeiffer}, \citenamefont {Pham}, \citenamefont {Phukon}, \citenamefont {Piccinni},
  \citenamefont {Pichot}, \citenamefont {Piendibene}, \citenamefont {Piergiovanni}, \citenamefont {Pierini}, \citenamefont {Pierro}, \citenamefont {Pillant}, \citenamefont {Pilo}, \citenamefont {Pinard}, \citenamefont {Pinto}, \citenamefont {Piotrzkowski}, \citenamefont {Piotrzkowski}, \citenamefont {Pirello}, \citenamefont {Pitkin}, \citenamefont {Placidi}, \citenamefont {Plastino}, \citenamefont {Pluchar}, \citenamefont {Poggiani}, \citenamefont {Polini}, \citenamefont {Pong}, \citenamefont {Ponrathnam}, \citenamefont {Popolizio}, \citenamefont {Porter}, \citenamefont {Powell}, \citenamefont {Pracchia}, \citenamefont {Pradier}, \citenamefont {Prajapati}, \citenamefont {Prasai}, \citenamefont {Prasanna}, \citenamefont {Pratten}, \citenamefont {Prestegard}, \citenamefont {Principe}, \citenamefont {Prodi}, \citenamefont {Prokhorov}, \citenamefont {Prosposito}, \citenamefont {Prudenzi}, \citenamefont {Puecher}, \citenamefont {Punturo}, \citenamefont {Puosi}, \citenamefont {Puppo}, \citenamefont {P{\"u}rrer},
  \citenamefont {Qi}, \citenamefont {Quetschke}, \citenamefont {Quinonez}, \citenamefont {{Quitzow-James}}, \citenamefont {Raab}, \citenamefont {Raaijmakers}, \citenamefont {Radkins}, \citenamefont {Radulesco}, \citenamefont {Raffai}, \citenamefont {Rail}, \citenamefont {Raja}, \citenamefont {Rajan}, \citenamefont {Ramirez}, \citenamefont {Ramirez}, \citenamefont {{Ramos-Buades}}, \citenamefont {Rana}, \citenamefont {Rapagnani}, \citenamefont {Rapol}, \citenamefont {Ratto}, \citenamefont {Raymond}, \citenamefont {Raza}, \citenamefont {Razzano}, \citenamefont {Read}, \citenamefont {Rees}, \citenamefont {Regimbau}, \citenamefont {Rei}, \citenamefont {Reid}, \citenamefont {Reitze}, \citenamefont {Relton}, \citenamefont {Renzini}, \citenamefont {Rettegno}, \citenamefont {Ricci}, \citenamefont {Richardson}, \citenamefont {Richardson}, \citenamefont {Richardson}, \citenamefont {Ricker}, \citenamefont {Riemenschneider}, \citenamefont {Riles}, \citenamefont {Rizzo}, \citenamefont {Robertson}, \citenamefont {Robie},
  \citenamefont {Robinet}, \citenamefont {Rocchi}, \citenamefont {Rocha}, \citenamefont {Rodriguez}, \citenamefont {{Rodriguez-Soto}}, \citenamefont {Rolland}, \citenamefont {Rollins}, \citenamefont {Roma}, \citenamefont {Romanelli}, \citenamefont {Romano}, \citenamefont {Romano}, \citenamefont {Romel}, \citenamefont {Romero}, \citenamefont {{Romero-Shaw}}, \citenamefont {Romie}, \citenamefont {Rose}, \citenamefont {Rosi{\'n}ska}, \citenamefont {Rosofsky}, \citenamefont {Ross}, \citenamefont {Rowan}, \citenamefont {Rowlinson}, \citenamefont {Roy}, \citenamefont {Roy}, \citenamefont {Rozza}, \citenamefont {Ruggi}, \citenamefont {Ryan}, \citenamefont {Sachdev}, \citenamefont {Sadecki}, \citenamefont {Sadiq}, \citenamefont {Sago}, \citenamefont {Saito}, \citenamefont {Saito}, \citenamefont {Sakai}, \citenamefont {Sakai}, \citenamefont {Sakellariadou}, \citenamefont {Sakuno}, \citenamefont {Salafia}, \citenamefont {Salconi}, \citenamefont {Saleem}, \citenamefont {Salemi}, \citenamefont {Samajdar}, \citenamefont
  {Sanchez}, \citenamefont {Sanchez}, \citenamefont {Sanchez}, \citenamefont {{Sanchis-Gual}}, \citenamefont {Sanders}, \citenamefont {Sanuy}, \citenamefont {Saravanan}, \citenamefont {Sarin}, \citenamefont {Sassolas}, \citenamefont {Satari}, \citenamefont {Sato}, \citenamefont {Sato}, \citenamefont {Sauter}, \citenamefont {Savage}, \citenamefont {Savant}, \citenamefont {Sawada}, \citenamefont {Sawant}, \citenamefont {Sawant}, \citenamefont {Sayah}, \citenamefont {Schaetzl}, \citenamefont {Scheel}, \citenamefont {Scheuer}, \citenamefont {{Schindler-Tyka}}, \citenamefont {Schmidt}, \citenamefont {Schnabel}, \citenamefont {Schneewind}, \citenamefont {Schofield}, \citenamefont {Sch{\"o}nbeck}, \citenamefont {Schulte}, \citenamefont {Schutz}, \citenamefont {Schwartz}, \citenamefont {Scott}, \citenamefont {Scott}, \citenamefont {{Seglar-Arroyo}}, \citenamefont {Seidel}, \citenamefont {Sekiguchi}, \citenamefont {Sekiguchi}, \citenamefont {Sellers}, \citenamefont {Sergeev}, \citenamefont {Sengupta}, \citenamefont
  {Sennett}, \citenamefont {Sentenac}, \citenamefont {Seo}, \citenamefont {Sequino}, \citenamefont {Setyawati}, \citenamefont {Shaffer}, \citenamefont {Shahriar}, \citenamefont {Shams}, \citenamefont {Shao}, \citenamefont {Sharifi}, \citenamefont {Sharma}, \citenamefont {Sharma}, \citenamefont {Shawhan}, \citenamefont {Shcheblanov}, \citenamefont {Shen}, \citenamefont {Shibagaki}, \citenamefont {Shikauchi}, \citenamefont {Shimizu}, \citenamefont {Shimoda}, \citenamefont {Shimode}, \citenamefont {Shink}, \citenamefont {Shinkai}, \citenamefont {Shishido}, \citenamefont {Shoda}, \citenamefont {Shoemaker}, \citenamefont {Shoemaker}, \citenamefont {Shukla}, \citenamefont {ShyamSundar}, \citenamefont {Sieniawska}, \citenamefont {Sigg}, \citenamefont {Singer}, \citenamefont {Singh}, \citenamefont {Singh}, \citenamefont {Singha}, \citenamefont {Sintes}, \citenamefont {Sipala}, \citenamefont {Skliris}, \citenamefont {Slagmolen}, \citenamefont {{Slaven-Blair}}, \citenamefont {Smetana}, \citenamefont {Smith},
  \citenamefont {Smith}, \citenamefont {Somala}, \citenamefont {Somiya}, \citenamefont {Son}, \citenamefont {Soni}, \citenamefont {Soni}, \citenamefont {Sorazu}, \citenamefont {Sordini}, \citenamefont {Sorrentino}, \citenamefont {Sorrentino}, \citenamefont {Sotani}, \citenamefont {Soulard}, \citenamefont {Souradeep}, \citenamefont {Sowell}, \citenamefont {Spagnuolo}, \citenamefont {Spencer}, \citenamefont {Spera}, \citenamefont {Srivastava}, \citenamefont {Srivastava}, \citenamefont {Staats}, \citenamefont {Stachie}, \citenamefont {Steer}, \citenamefont {Steinlechner}, \citenamefont {Steinlechner}, \citenamefont {Stops}, \citenamefont {Stover}, \citenamefont {Strain}, \citenamefont {Strang}, \citenamefont {Stratta}, \citenamefont {Strunk}, \citenamefont {Sturani}, \citenamefont {Stuver}, \citenamefont {S{\"u}dbeck}, \citenamefont {Sudhagar}, \citenamefont {Sudhir}, \citenamefont {Sugimoto}, \citenamefont {Suh}, \citenamefont {Summerscales}, \citenamefont {Sun}, \citenamefont {Sun}, \citenamefont {Sunil},
  \citenamefont {Sur}, \citenamefont {Suresh}, \citenamefont {Sutton}, \citenamefont {Suzuki}, \citenamefont {Suzuki}, \citenamefont {Swinkels}, \citenamefont {Szczepa{\'n}czyk}, \citenamefont {Szewczyk}, \citenamefont {Tacca}, \citenamefont {Tagoshi}, \citenamefont {Tait}, \citenamefont {Takahashi}, \citenamefont {Takahashi}, \citenamefont {Takamori}, \citenamefont {Takano}, \citenamefont {Takeda}, \citenamefont {Takeda}, \citenamefont {Talbot}, \citenamefont {Tanaka}, \citenamefont {Tanaka}, \citenamefont {Tanaka}, \citenamefont {Tanaka}, \citenamefont {Tanaka}, \citenamefont {Tanasijczuk}, \citenamefont {Tanioka}, \citenamefont {Tanner}, \citenamefont {Tao}, \citenamefont {Tapia}, \citenamefont {Martin}, \citenamefont {Martin}, \citenamefont {Tasson}, \citenamefont {Telada}, \citenamefont {Tenorio}, \citenamefont {Terkowski}, \citenamefont {Test}, \citenamefont {Thirugnanasambandam}, \citenamefont {Thomas}, \citenamefont {Thomas}, \citenamefont {Thompson}, \citenamefont {Thondapu}, \citenamefont {Thorne},
  \citenamefont {Thrane}, \citenamefont {Tiwari}, \citenamefont {Tiwari}, \citenamefont {Tiwari}, \citenamefont {Toland}, \citenamefont {Tolley}, \citenamefont {Tomaru}, \citenamefont {Tomigami}, \citenamefont {Tomura}, \citenamefont {Tonelli}, \citenamefont {{Torres-Forn{\'e}}}, \citenamefont {Torrie}, \citenamefont {{e Melo}}, \citenamefont {T{\"o}yr{\"a}}, \citenamefont {Trapananti}, \citenamefont {Travasso}, \citenamefont {Traylor}, \citenamefont {Tringali}, \citenamefont {Tripathee}, \citenamefont {Troiano}, \citenamefont {Trovato}, \citenamefont {Trozzo}, \citenamefont {Trudeau}, \citenamefont {Tsai}, \citenamefont {Tsai}, \citenamefont {Tsang}, \citenamefont {Tsang}, \citenamefont {Tsao}, \citenamefont {Tse}, \citenamefont {Tso}, \citenamefont {Tsubono}, \citenamefont {Tsuchida}, \citenamefont {Tsukada}, \citenamefont {Tsuna}, \citenamefont {Tsutsui}, \citenamefont {Tsuzuki}, \citenamefont {Turconi}, \citenamefont {Tuyenbayev}, \citenamefont {Ubhi}, \citenamefont {Uchikata}, \citenamefont {Uchiyama},
  \citenamefont {Udall}, \citenamefont {Ueda}, \citenamefont {Uehara}, \citenamefont {Ueno}, \citenamefont {Ueshima}, \citenamefont {Ugolini}, \citenamefont {Unnikrishnan}, \citenamefont {Uraguchi}, \citenamefont {Urban}, \citenamefont {Ushiba}, \citenamefont {Usman}, \citenamefont {Utina}, \citenamefont {Vahlbruch}, \citenamefont {Vajente}, \citenamefont {Vajpeyi}, \citenamefont {Valdes}, \citenamefont {Valentini}, \citenamefont {Valsan}, \citenamefont {{van Bakel}}, \citenamefont {{van Beuzekom}}, \citenamefont {{van den Brand}}, \citenamefont {Van Den~Broeck}, \citenamefont {{Vander-Hyde}}, \citenamefont {{van der Schaaf}}, \citenamefont {{van Heijningen}}, \citenamefont {Vanosky}, \citenamefont {{van Putten}}, \citenamefont {{van Remortel}}, \citenamefont {Vardaro}, \citenamefont {Vargas}, \citenamefont {Varma}, \citenamefont {Vas{\'u}th}, \citenamefont {Vecchio}, \citenamefont {Vedovato}, \citenamefont {Veitch}, \citenamefont {Veitch}, \citenamefont {Venkateswara}, \citenamefont {Venneberg},
  \citenamefont {Venugopalan}, \citenamefont {Verkindt}, \citenamefont {Verma}, \citenamefont {Veske}, \citenamefont {Vetrano}, \citenamefont {Vicer{\'e}}, \citenamefont {Viets}, \citenamefont {{Villa-Ortega}}, \citenamefont {Vinet}, \citenamefont {Vitale}, \citenamefont {Vo}, \citenamefont {Vocca}, \citenamefont {{von Reis}}, \citenamefont {{von Wrangel}}, \citenamefont {Vorvick}, \citenamefont {Vyatchanin}, \citenamefont {Wade}, \citenamefont {Wade}, \citenamefont {Wagner}, \citenamefont {Walet}, \citenamefont {Walker}, \citenamefont {Wallace}, \citenamefont {Wallace}, \citenamefont {Walsh}, \citenamefont {Wang}, \citenamefont {Wang}, \citenamefont {Wang}, \citenamefont {Ward}, \citenamefont {Warner}, \citenamefont {Was}, \citenamefont {Washimi}, \citenamefont {Washington}, \citenamefont {Watchi}, \citenamefont {Weaver}, \citenamefont {Wei}, \citenamefont {Weinert}, \citenamefont {Weinstein}, \citenamefont {Weiss}, \citenamefont {Weller}, \citenamefont {Wellmann}, \citenamefont {Wen}, \citenamefont
  {We{\ss}els}, \citenamefont {Westhouse}, \citenamefont {Wette}, \citenamefont {Whelan}, \citenamefont {White}, \citenamefont {Whiting}, \citenamefont {Whittle}, \citenamefont {Wilken}, \citenamefont {Williams}, \citenamefont {Williams}, \citenamefont {Williamson}, \citenamefont {Willis}, \citenamefont {Willke}, \citenamefont {Wilson}, \citenamefont {Winkler}, \citenamefont {Wipf}, \citenamefont {Wlodarczyk}, \citenamefont {Woan}, \citenamefont {Woehler}, \citenamefont {Wofford}, \citenamefont {Wong}, \citenamefont {Wu}, \citenamefont {Wu}, \citenamefont {Wu}, \citenamefont {Wu}, \citenamefont {Wysocki}, \citenamefont {Xiao}, \citenamefont {Xu}, \citenamefont {Yamada}, \citenamefont {Yamamoto}, \citenamefont {Yamamoto}, \citenamefont {Yamamoto}, \citenamefont {Yamamoto}, \citenamefont {Yamashita}, \citenamefont {Yamazaki}, \citenamefont {Yang}, \citenamefont {Yang}, \citenamefont {Yang}, \citenamefont {Yang}, \citenamefont {Yang}, \citenamefont {Yap}, \citenamefont {Yeeles}, \citenamefont {Yelikar},
  \citenamefont {Ying}, \citenamefont {Yokogawa}, \citenamefont {Yokoyama}, \citenamefont {Yokozawa}, \citenamefont {Yoon}, \citenamefont {Yoshioka}, \citenamefont {Yu}, \citenamefont {Yu}, \citenamefont {Yuzurihara}, \citenamefont {Zadro{\.z}ny}, \citenamefont {Zanolin}, \citenamefont {Zeidler}, \citenamefont {Zelenova}, \citenamefont {Zendri}, \citenamefont {Zevin}, \citenamefont {Zhan}, \citenamefont {Zhang}, \citenamefont {Zhang}, \citenamefont {Zhang}, \citenamefont {Zhang}, \citenamefont {Zhang}, \citenamefont {Zhao}, \citenamefont {Zhao}, \citenamefont {Zhao}, \citenamefont {Zhao}, \citenamefont {Zhou}, \citenamefont {Zhu}, \citenamefont {Zhu}, \citenamefont {Zimmerman}, \citenamefont {Zucker},\ and\ \citenamefont {Zweizig}}]{abbott_search_2021}%
  \BibitemOpen
  \bibfield  {author} {\bibinfo {author} {\bibfnamefont {R.}~\bibnamefont {Abbott}}, \bibinfo {author} {\bibfnamefont {T.~D.}\ \bibnamefont {Abbott}}, \bibinfo {author} {\bibfnamefont {S.}~\bibnamefont {Abraham}}, \emph {et~al.},\ }\href {https://doi.org/10.1103/PhysRevD.104.022005} {\bibfield  {journal} {\bibinfo  {journal} {Physical Review D}\ }\textbf {\bibinfo {volume} {104}},\ \bibinfo {pages} {022005} (\bibinfo {year} {2021})}\BibitemShut {NoStop}%
\bibitem [{\citenamefont {{Ali-Ha{\"i}moud}}\ \emph {et~al.}(2020)\citenamefont {{Ali-Ha{\"i}moud}}, \citenamefont {Smith},\ and\ \citenamefont {Mingarelli}}]{ali-haimoud_fisher_2020}%
  \BibitemOpen
  \bibfield  {author} {\bibinfo {author} {\bibfnamefont {Y.}~\bibnamefont {{Ali-Ha{\"i}moud}}}, \bibinfo {author} {\bibfnamefont {T.~L.}\ \bibnamefont {Smith}},\ and\ \bibinfo {author} {\bibfnamefont {C.~M.~F.}\ \bibnamefont {Mingarelli}},\ }\href {https://doi.org/10.1103/PhysRevD.102.122005} {\bibfield  {journal} {\bibinfo  {journal} {Physical Review D}\ }\textbf {\bibinfo {volume} {102}},\ \bibinfo {pages} {122005} (\bibinfo {year} {2020})},\ \Eprint {https://arxiv.org/abs/2006.14570} {arxiv:2006.14570 [astro-ph, physics:gr-qc, physics:physics]} \BibitemShut {NoStop}%
\bibitem [{\citenamefont {Mingarelli}\ \emph {et~al.}(2013)\citenamefont {Mingarelli}, \citenamefont {Sidery}, \citenamefont {Mandel},\ and\ \citenamefont {Vecchio}}]{mingarelli_characterizing_2013}%
  \BibitemOpen
  \bibfield  {author} {\bibinfo {author} {\bibfnamefont {C.~M.~F.}\ \bibnamefont {Mingarelli}}, \bibinfo {author} {\bibfnamefont {T.}~\bibnamefont {Sidery}}, \bibinfo {author} {\bibfnamefont {I.}~\bibnamefont {Mandel}},\ and\ \bibinfo {author} {\bibfnamefont {A.}~\bibnamefont {Vecchio}},\ }\href {https://doi.org/10.1103/PhysRevD.88.062005} {\bibfield  {journal} {\bibinfo  {journal} {Physical Review D}\ }\textbf {\bibinfo {volume} {88}},\ \bibinfo {pages} {062005} (\bibinfo {year} {2013})}\BibitemShut {NoStop}%
\bibitem [{\citenamefont {Taylor}\ and\ \citenamefont {Gair}(2013)}]{taylor_searching_2013}%
  \BibitemOpen
  \bibfield  {author} {\bibinfo {author} {\bibfnamefont {S.~R.}\ \bibnamefont {Taylor}}\ and\ \bibinfo {author} {\bibfnamefont {J.~R.}\ \bibnamefont {Gair}},\ }\href {https://doi.org/10.1103/PhysRevD.88.084001} {\bibfield  {journal} {\bibinfo  {journal} {Physical Review D}\ }\textbf {\bibinfo {volume} {88}},\ \bibinfo {pages} {084001} (\bibinfo {year} {2013})}\BibitemShut {NoStop}%
\bibitem [{\citenamefont {Taylor}\ \emph {et~al.}(2020)\citenamefont {Taylor}, \citenamefont {{van Haasteren}},\ and\ \citenamefont {Sesana}}]{taylor_from_2020}%
  \BibitemOpen
  \bibfield  {author} {\bibinfo {author} {\bibfnamefont {S.~R.}\ \bibnamefont {Taylor}}, \bibinfo {author} {\bibfnamefont {R.}~\bibnamefont {{van Haasteren}}},\ and\ \bibinfo {author} {\bibfnamefont {A.}~\bibnamefont {Sesana}},\ }\href {https://doi.org/10.1103/PhysRevD.102.084039} {\bibfield  {journal} {\bibinfo  {journal} {Physical Review D}\ }\textbf {\bibinfo {volume} {102}},\ \bibinfo {pages} {084039} (\bibinfo {year} {2020})}\BibitemShut {NoStop}%
\bibitem [{\citenamefont {Cornish}\ and\ \citenamefont {{van Haasteren}}(2014)}]{cornish_mapping_2014}%
  \BibitemOpen
  \bibfield  {author} {\bibinfo {author} {\bibfnamefont {N.~J.}\ \bibnamefont {Cornish}}\ and\ \bibinfo {author} {\bibfnamefont {R.}~\bibnamefont {{van Haasteren}}},\ }\href {https://doi.org/10.48550/arXiv.1406.4511} {\bibinfo {title} {Mapping the nano-{{Hertz}} gravitational wave sky}} (\bibinfo {year} {2014}),\ \Eprint {https://arxiv.org/abs/1406.4511} {arxiv:1406.4511 [astro-ph, physics:gr-qc]} \BibitemShut {NoStop}%
\bibitem [{\citenamefont {Agazie}\ \emph {et~al.}(2023{\natexlab{b}})\citenamefont {Agazie}, \citenamefont {Anumarlapudi}, \citenamefont {Archibald}, \citenamefont {Arzoumanian}, \citenamefont {Baker}, \citenamefont {B{\'e}csy}, \citenamefont {Blecha}, \citenamefont {Brazier}, \citenamefont {Brook}, \citenamefont {{Burke-Spolaor}}, \citenamefont {{Casey-Clyde}}, \citenamefont {Charisi}, \citenamefont {Chatterjee}, \citenamefont {Cohen}, \citenamefont {Cordes}, \citenamefont {Cornish}, \citenamefont {Crawford}, \citenamefont {Cromartie}, \citenamefont {Crowter}, \citenamefont {DeCesar}, \citenamefont {Demorest}, \citenamefont {Dolch}, \citenamefont {Drachler}, \citenamefont {Ferrara}, \citenamefont {Fiore}, \citenamefont {Fonseca}, \citenamefont {Freedman}, \citenamefont {Gardiner}, \citenamefont {{Garver-Daniels}}, \citenamefont {Gentile}, \citenamefont {Glaser}, \citenamefont {Good}, \citenamefont {G{\"u}ltekin}, \citenamefont {Hazboun}, \citenamefont {Jennings}, \citenamefont {Johnson}, \citenamefont
  {Jones}, \citenamefont {Kaiser}, \citenamefont {Kaplan}, \citenamefont {Kelley}, \citenamefont {Kerr}, \citenamefont {Key}, \citenamefont {Laal}, \citenamefont {Lam}, \citenamefont {Lamb}, \citenamefont {Lazio}, \citenamefont {Lewandowska}, \citenamefont {Liu}, \citenamefont {Lorimer}, \citenamefont {Luo}, \citenamefont {Lynch}, \citenamefont {Ma}, \citenamefont {Madison}, \citenamefont {McEwen}, \citenamefont {McKee}, \citenamefont {McLaughlin}, \citenamefont {McMann}, \citenamefont {Meyers}, \citenamefont {Mingarelli}, \citenamefont {Mitridate}, \citenamefont {Ng}, \citenamefont {Nice}, \citenamefont {Ocker}, \citenamefont {Olum}, \citenamefont {Pennucci}, \citenamefont {Perera}, \citenamefont {Pol}, \citenamefont {Radovan}, \citenamefont {Ransom}, \citenamefont {Ray}, \citenamefont {Romano}, \citenamefont {Sardesai}, \citenamefont {Schmiedekamp}, \citenamefont {Schmiedekamp}, \citenamefont {Schmitz}, \citenamefont {Schult}, \citenamefont {{Shapiro-Albert}}, \citenamefont {Siemens}, \citenamefont {Simon},
  \citenamefont {Siwek}, \citenamefont {Stairs}, \citenamefont {Stinebring}, \citenamefont {Stovall}, \citenamefont {Susobhanan}, \citenamefont {Swiggum}, \citenamefont {Taylor}, \citenamefont {Turner}, \citenamefont {Unal}, \citenamefont {Vallisneri}, \citenamefont {Vigeland}, \citenamefont {Wahl}, \citenamefont {Witt},\ and\ \citenamefont {Young}}]{agazie_the_2023d}%
  \BibitemOpen
  \bibfield  {author} {\bibinfo {author} {\bibfnamefont {G.}~\bibnamefont {Agazie}}, \bibinfo {author} {\bibfnamefont {A.}~\bibnamefont {Anumarlapudi}}, \bibinfo {author} {\bibfnamefont {A.~M.}\ \bibnamefont {Archibald}}, \emph {et~al.},\ }\href {https://doi.org/10.3847/2041-8213/acdac6,10.3847/2041-8213/acda9a,10.3847/2041-8213/acda88} {\bibinfo {title} {The {{NANOGrav}} 15-year {{Data Set}}: {{Search}} for {{Anisotropy}} in the {{Gravitational-Wave Background}}}} (\bibinfo {year} {2023}{\natexlab{b}}),\ \Eprint {https://arxiv.org/abs/2306.16221} {arxiv:2306.16221 [astro-ph, physics:gr-qc]} \BibitemShut {NoStop}%
\bibitem [{\citenamefont {Cornish}(2001{\natexlab{a}})}]{Cornish:2001hg}%
  \BibitemOpen
  \bibfield  {author} {\bibinfo {author} {\bibfnamefont {N.~J.}\ \bibnamefont {Cornish}},\ }\href {https://doi.org/10.1088/0264-9381/18/20/307} {\bibfield  {journal} {\bibinfo  {journal} {Class. Quant. Grav.}\ }\textbf {\bibinfo {volume} {18}},\ \bibinfo {pages} {4277} (\bibinfo {year} {2001}{\natexlab{a}})},\ \Eprint {https://arxiv.org/abs/astro-ph/0105374} {arXiv:astro-ph/0105374} \BibitemShut {NoStop}%
\bibitem [{\citenamefont {Ungarelli}\ and\ \citenamefont {Vecchio}(2001)}]{Ungarelli:2001xu}%
  \BibitemOpen
  \bibfield  {author} {\bibinfo {author} {\bibfnamefont {C.}~\bibnamefont {Ungarelli}}\ and\ \bibinfo {author} {\bibfnamefont {A.}~\bibnamefont {Vecchio}},\ }\href {https://doi.org/10.1103/PhysRevD.64.121501} {\bibfield  {journal} {\bibinfo  {journal} {Phys. Rev. D}\ }\textbf {\bibinfo {volume} {64}},\ \bibinfo {pages} {121501} (\bibinfo {year} {2001})},\ \Eprint {https://arxiv.org/abs/astro-ph/0106538} {arXiv:astro-ph/0106538} \BibitemShut {NoStop}%
\bibitem [{\citenamefont {Kudoh}\ and\ \citenamefont {Taruya}(2005{\natexlab{a}})}]{Kudoh:2004he}%
  \BibitemOpen
  \bibfield  {author} {\bibinfo {author} {\bibfnamefont {H.}~\bibnamefont {Kudoh}}\ and\ \bibinfo {author} {\bibfnamefont {A.}~\bibnamefont {Taruya}},\ }\href {https://doi.org/10.1103/PhysRevD.71.024025} {\bibfield  {journal} {\bibinfo  {journal} {Phys. Rev. D}\ }\textbf {\bibinfo {volume} {71}},\ \bibinfo {pages} {024025} (\bibinfo {year} {2005}{\natexlab{a}})},\ \Eprint {https://arxiv.org/abs/gr-qc/0411017} {arXiv:gr-qc/0411017} \BibitemShut {NoStop}%
\bibitem [{\citenamefont {Taruya}\ and\ \citenamefont {Kudoh}(2005{\natexlab{a}})}]{Taruya:2005yf}%
  \BibitemOpen
  \bibfield  {author} {\bibinfo {author} {\bibfnamefont {A.}~\bibnamefont {Taruya}}\ and\ \bibinfo {author} {\bibfnamefont {H.}~\bibnamefont {Kudoh}},\ }\href {https://doi.org/10.1103/PhysRevD.72.104015} {\bibfield  {journal} {\bibinfo  {journal} {Phys. Rev. D}\ }\textbf {\bibinfo {volume} {72}},\ \bibinfo {pages} {104015} (\bibinfo {year} {2005}{\natexlab{a}})},\ \Eprint {https://arxiv.org/abs/gr-qc/0507114} {arXiv:gr-qc/0507114} \BibitemShut {NoStop}%
\bibitem [{\citenamefont {Taruya}(2006)}]{Taruya:2006kqa}%
  \BibitemOpen
  \bibfield  {author} {\bibinfo {author} {\bibfnamefont {A.}~\bibnamefont {Taruya}},\ }\href {https://doi.org/10.1103/PhysRevD.74.104022} {\bibfield  {journal} {\bibinfo  {journal} {Phys. Rev. D}\ }\textbf {\bibinfo {volume} {74}},\ \bibinfo {pages} {104022} (\bibinfo {year} {2006})},\ \Eprint {https://arxiv.org/abs/gr-qc/0607080} {arXiv:gr-qc/0607080} \BibitemShut {NoStop}%
\bibitem [{\citenamefont {Renzini}\ and\ \citenamefont {Contaldi}(2018)}]{Renzini:2018vkx}%
  \BibitemOpen
  \bibfield  {author} {\bibinfo {author} {\bibfnamefont {A.}~\bibnamefont {Renzini}}\ and\ \bibinfo {author} {\bibfnamefont {C.}~\bibnamefont {Contaldi}},\ }\href {https://doi.org/10.1093/mnras/sty2546} {\bibfield  {journal} {\bibinfo  {journal} {Mon. Not. Roy. Astron. Soc.}\ }\textbf {\bibinfo {volume} {481}},\ \bibinfo {pages} {4650} (\bibinfo {year} {2018})},\ \Eprint {https://arxiv.org/abs/1806.11360} {arXiv:1806.11360 [astro-ph.IM]} \BibitemShut {NoStop}%
\bibitem [{\citenamefont {Buscicchio}\ \emph {et~al.}(2024)\citenamefont {Buscicchio}, \citenamefont {Klein}, \citenamefont {Korol}, \citenamefont {Renzo}, \citenamefont {Moore}, \citenamefont {Gerosa},\ and\ \citenamefont {Carzaniga}}]{buscicchio_a_2024a}%
  \BibitemOpen
  \bibfield  {author} {\bibinfo {author} {\bibfnamefont {R.}~\bibnamefont {Buscicchio}}, \bibinfo {author} {\bibfnamefont {A.}~\bibnamefont {Klein}}, \bibinfo {author} {\bibfnamefont {V.}~\bibnamefont {Korol}}, \emph {et~al.},\ }\href {https://doi.org/10.48550/arXiv.2410.08263} {\bibinfo {title} {A test for {{LISA}} foreground {{Gaussianity}} and stationarity. {{I}}. {{Galactic}} white-dwarf binaries}} (\bibinfo {year} {2024}),\ \Eprint {https://arxiv.org/abs/2410.08263} {arXiv:2410.08263 [astro-ph]} \BibitemShut {NoStop}%
\bibitem [{\citenamefont {Piarulli}\ \emph {et~al.}(2024)\citenamefont {Piarulli}, \citenamefont {Buscicchio}, \citenamefont {Pozzoli}, \citenamefont {Burke}, \citenamefont {Bonetti},\ and\ \citenamefont {Sesana}}]{piarulli_a_2024b}%
  \BibitemOpen
  \bibfield  {author} {\bibinfo {author} {\bibfnamefont {M.}~\bibnamefont {Piarulli}}, \bibinfo {author} {\bibfnamefont {R.}~\bibnamefont {Buscicchio}}, \bibinfo {author} {\bibfnamefont {F.}~\bibnamefont {Pozzoli}}, \emph {et~al.},\ }\href {https://doi.org/10.48550/arXiv.2410.08862} {\bibinfo {title} {A test for {{LISA}} foreground {{Gaussianity}} and stationarity. {{II}}. {{Extreme}} mass-ratio inspirals}} (\bibinfo {year} {2024}),\ \Eprint {https://arxiv.org/abs/2410.08862} {arXiv:2410.08862 [astro-ph]} \BibitemShut {NoStop}%
\bibitem [{\citenamefont {Pozzoli}\ \emph {et~al.}(2024)\citenamefont {Pozzoli}, \citenamefont {Buscicchio}, \citenamefont {Klein}, \citenamefont {Korol}, \citenamefont {Sesana},\ and\ \citenamefont {Haardt}}]{pozzoli_cyclostationary_2024}%
  \BibitemOpen
  \bibfield  {author} {\bibinfo {author} {\bibfnamefont {F.}~\bibnamefont {Pozzoli}}, \bibinfo {author} {\bibfnamefont {R.}~\bibnamefont {Buscicchio}}, \bibinfo {author} {\bibfnamefont {A.}~\bibnamefont {Klein}}, \emph {et~al.},\ }\href {https://doi.org/10.48550/arXiv.2410.08274} {\bibinfo {title} {Cyclostationary signals in {{LISA}}: A practical application to {{Milky Way}} satellites}} (\bibinfo {year} {2024}),\ \Eprint {https://arxiv.org/abs/2410.08274} {arXiv:2410.08274} \BibitemShut {NoStop}%
\bibitem [{\citenamefont {Criswell}\ \emph {et~al.}(2024)\citenamefont {Criswell}, \citenamefont {Rieck},\ and\ \citenamefont {Mandic}}]{criswell_templated_2024}%
  \BibitemOpen
  \bibfield  {author} {\bibinfo {author} {\bibfnamefont {A.~W.}\ \bibnamefont {Criswell}}, \bibinfo {author} {\bibfnamefont {S.}~\bibnamefont {Rieck}},\ and\ \bibinfo {author} {\bibfnamefont {V.}~\bibnamefont {Mandic}},\ }\href {https://doi.org/10.48550/arXiv.2410.23260} {\bibinfo {title} {Templated {{Anisotropic Analyses}} of the {{LISA Galactic Foreground}}}} (\bibinfo {year} {2024}),\ \Eprint {https://arxiv.org/abs/2410.23260} {arXiv:2410.23260 [astro-ph]} \BibitemShut {NoStop}%
\bibitem [{\citenamefont {Banagiri}\ \emph {et~al.}(2021)\citenamefont {Banagiri}, \citenamefont {Criswell}, \citenamefont {Kuan}, \citenamefont {Mandic}, \citenamefont {Romano},\ and\ \citenamefont {Taylor}}]{banagiri_mapping_2021b}%
  \BibitemOpen
  \bibfield  {author} {\bibinfo {author} {\bibfnamefont {S.}~\bibnamefont {Banagiri}}, \bibinfo {author} {\bibfnamefont {A.}~\bibnamefont {Criswell}}, \bibinfo {author} {\bibfnamefont {T.}~\bibnamefont {Kuan}}, \emph {et~al.},\ }\href {https://doi.org/10.1093/mnras/stab2479} {\bibfield  {journal} {\bibinfo  {journal} {Monthly Notices of the Royal Astronomical Society}\ }\textbf {\bibinfo {volume} {507}},\ \bibinfo {pages} {5451} (\bibinfo {year} {2021})}\BibitemShut {NoStop}%
\bibitem [{\citenamefont {Peterseim}\ \emph {et~al.}(1997)\citenamefont {Peterseim}, \citenamefont {Jennrich}, \citenamefont {Danzmann},\ and\ \citenamefont {Schutz}}]{peterseim_angular_1997}%
  \BibitemOpen
  \bibfield  {author} {\bibinfo {author} {\bibfnamefont {M.}~\bibnamefont {Peterseim}}, \bibinfo {author} {\bibfnamefont {O.}~\bibnamefont {Jennrich}}, \bibinfo {author} {\bibfnamefont {K.}~\bibnamefont {Danzmann}},\ and\ \bibinfo {author} {\bibfnamefont {B.~F.}\ \bibnamefont {Schutz}},\ }\href {https://doi.org/10.1088/0264-9381/14/6/019} {\bibfield  {journal} {\bibinfo  {journal} {Classical and Quantum Gravity}\ }\textbf {\bibinfo {volume} {14}},\ \bibinfo {pages} {1507} (\bibinfo {year} {1997})}\BibitemShut {NoStop}%
\bibitem [{\citenamefont {Cutler}(1998)}]{cutler_angular_1998}%
  \BibitemOpen
  \bibfield  {author} {\bibinfo {author} {\bibfnamefont {C.}~\bibnamefont {Cutler}},\ }\href {https://doi.org/10.1103/PhysRevD.57.7089} {\bibfield  {journal} {\bibinfo  {journal} {Physical Review D}\ }\textbf {\bibinfo {volume} {57}},\ \bibinfo {pages} {7089} (\bibinfo {year} {1998})}\BibitemShut {NoStop}%
\bibitem [{\citenamefont {Moore}\ and\ \citenamefont {Hellings}(2002)}]{moore_angular_2002}%
  \BibitemOpen
  \bibfield  {author} {\bibinfo {author} {\bibfnamefont {T.~A.}\ \bibnamefont {Moore}}\ and\ \bibinfo {author} {\bibfnamefont {R.~W.}\ \bibnamefont {Hellings}},\ }\href {https://doi.org/10.1103/PhysRevD.65.062001} {\bibfield  {journal} {\bibinfo  {journal} {Physical Review D}\ }\textbf {\bibinfo {volume} {65}},\ \bibinfo {pages} {062001} (\bibinfo {year} {2002})}\BibitemShut {NoStop}%
\bibitem [{\citenamefont {Kudoh}\ and\ \citenamefont {Taruya}(2005{\natexlab{b}})}]{kudoh_probing_2005}%
  \BibitemOpen
  \bibfield  {author} {\bibinfo {author} {\bibfnamefont {H.}~\bibnamefont {Kudoh}}\ and\ \bibinfo {author} {\bibfnamefont {A.}~\bibnamefont {Taruya}},\ }\href {https://doi.org/10.1103/PhysRevD.71.024025} {\bibfield  {journal} {\bibinfo  {journal} {Physical Review D}\ }\textbf {\bibinfo {volume} {71}},\ \bibinfo {pages} {024025} (\bibinfo {year} {2005}{\natexlab{b}})}\BibitemShut {NoStop}%
\bibitem [{\citenamefont {Taruya}\ and\ \citenamefont {Kudoh}(2005{\natexlab{b}})}]{taruya_probing_2005}%
  \BibitemOpen
  \bibfield  {author} {\bibinfo {author} {\bibfnamefont {A.}~\bibnamefont {Taruya}}\ and\ \bibinfo {author} {\bibfnamefont {H.}~\bibnamefont {Kudoh}},\ }\href {https://doi.org/10.1103/PhysRevD.72.104015} {\bibfield  {journal} {\bibinfo  {journal} {Physical Review D}\ }\textbf {\bibinfo {volume} {72}},\ \bibinfo {pages} {104015} (\bibinfo {year} {2005}{\natexlab{b}})}\BibitemShut {NoStop}%
\bibitem [{\citenamefont {Contaldi}\ \emph {et~al.}(2020)\citenamefont {Contaldi}, \citenamefont {Pieroni}, \citenamefont {Renzini}, \citenamefont {Cusin}, \citenamefont {Karnesis}, \citenamefont {Peloso}, \citenamefont {Ricciardone},\ and\ \citenamefont {Tasinato}}]{contaldi_maximum_2020}%
  \BibitemOpen
  \bibfield  {author} {\bibinfo {author} {\bibfnamefont {C.~R.}\ \bibnamefont {Contaldi}}, \bibinfo {author} {\bibfnamefont {M.}~\bibnamefont {Pieroni}}, \bibinfo {author} {\bibfnamefont {A.~I.}\ \bibnamefont {Renzini}}, \emph {et~al.},\ }\href {https://doi.org/10.1103/PhysRevD.102.043502} {\bibfield  {journal} {\bibinfo  {journal} {Physical Review D}\ }\textbf {\bibinfo {volume} {102}},\ \bibinfo {pages} {043502} (\bibinfo {year} {2020})},\ \Eprint {https://arxiv.org/abs/2006.03313} {arxiv:2006.03313 [astro-ph]} \BibitemShut {NoStop}%
\bibitem [{\citenamefont {Mentasti}\ \emph {et~al.}(2024)\citenamefont {Mentasti}, \citenamefont {Contaldi},\ and\ \citenamefont {Peloso}}]{mentasti_probing_2024a}%
  \BibitemOpen
  \bibfield  {author} {\bibinfo {author} {\bibfnamefont {G.}~\bibnamefont {Mentasti}}, \bibinfo {author} {\bibfnamefont {C.~R.}\ \bibnamefont {Contaldi}},\ and\ \bibinfo {author} {\bibfnamefont {M.}~\bibnamefont {Peloso}},\ }\href {https://doi.org/10.48550/arXiv.2312.10792} {\bibinfo {title} {Probing the galactic and extragalactic gravitational wave backgrounds with space-based interferometers}} (\bibinfo {year} {2024}),\ \Eprint {https://arxiv.org/abs/2312.10792} {arXiv:2312.10792 [astro-ph, physics:gr-qc]} \BibitemShut {NoStop}%
\bibitem [{\citenamefont {Littenberg}\ and\ \citenamefont {Cornish}(2023)}]{littenberg_prototype_2023}%
  \BibitemOpen
  \bibfield  {author} {\bibinfo {author} {\bibfnamefont {T.~B.}\ \bibnamefont {Littenberg}}\ and\ \bibinfo {author} {\bibfnamefont {N.~J.}\ \bibnamefont {Cornish}},\ }\href {https://doi.org/10.48550/arXiv.2301.03673} {\bibinfo {title} {Prototype {{Global Analysis}} of {{LISA Data}} with {{Multiple Source Types}}}} (\bibinfo {year} {2023}),\ \Eprint {https://arxiv.org/abs/2301.03673} {arxiv:2301.03673 [astro-ph, physics:gr-qc]} \BibitemShut {NoStop}%
\bibitem [{\citenamefont {Katz}\ \emph {et~al.}(2024)\citenamefont {Katz}, \citenamefont {Karnesis}, \citenamefont {Korsakova}, \citenamefont {Gair},\ and\ \citenamefont {Stergioulas}}]{katz_an_2024}%
  \BibitemOpen
  \bibfield  {author} {\bibinfo {author} {\bibfnamefont {M.~L.}\ \bibnamefont {Katz}}, \bibinfo {author} {\bibfnamefont {N.}~\bibnamefont {Karnesis}}, \bibinfo {author} {\bibfnamefont {N.}~\bibnamefont {Korsakova}}, \emph {et~al.},\ }\href {https://doi.org/10.48550/arXiv.2405.04690} {\bibinfo {title} {An efficient {{GPU-accelerated}} multi-source global fit pipeline for {{LISA}} data analysis}} (\bibinfo {year} {2024}),\ \Eprint {https://arxiv.org/abs/2405.04690} {arXiv:2405.04690 [astro-ph, physics:gr-qc]} \BibitemShut {NoStop}%
\bibitem [{\citenamefont {Cornish}\ and\ \citenamefont {Larson}(2001)}]{cornish_space_2001}%
  \BibitemOpen
  \bibfield  {author} {\bibinfo {author} {\bibfnamefont {N.~J.}\ \bibnamefont {Cornish}}\ and\ \bibinfo {author} {\bibfnamefont {S.~L.}\ \bibnamefont {Larson}},\ }\href {https://doi.org/10.1088/0264-9381/18/17/308} {\bibfield  {journal} {\bibinfo  {journal} {Classical and Quantum Gravity}\ }\textbf {\bibinfo {volume} {18}},\ \bibinfo {pages} {3473} (\bibinfo {year} {2001})},\ \Eprint {https://arxiv.org/abs/gr-qc/0103075} {arxiv:gr-qc/0103075} \BibitemShut {NoStop}%
\bibitem [{\citenamefont {Cornish}(2001{\natexlab{b}})}]{cornish_detecting_2001}%
  \BibitemOpen
  \bibfield  {author} {\bibinfo {author} {\bibfnamefont {N.~J.}\ \bibnamefont {Cornish}},\ }\href {https://doi.org/10.1103/PhysRevD.65.022004} {\bibfield  {journal} {\bibinfo  {journal} {Physical Review D}\ }\textbf {\bibinfo {volume} {65}},\ \bibinfo {pages} {022004} (\bibinfo {year} {2001}{\natexlab{b}})}\BibitemShut {NoStop}%
\bibitem [{\citenamefont {Speagle}(2020)}]{speagle_dynesty:_2020}%
  \BibitemOpen
  \bibfield  {author} {\bibinfo {author} {\bibfnamefont {J.~S.}\ \bibnamefont {Speagle}},\ }\href {https://doi.org/10.1093/mnras/staa278} {\bibfield  {journal} {\bibinfo  {journal} {Monthly Notices of the Royal Astronomical Society}\ }\textbf {\bibinfo {volume} {493}},\ \bibinfo {pages} {3132} (\bibinfo {year} {2020})}\BibitemShut {NoStop}%
\bibitem [{\citenamefont {Gorski}\ \emph {et~al.}(2005)\citenamefont {Gorski}, \citenamefont {Hivon}, \citenamefont {Banday}, \citenamefont {Wandelt}, \citenamefont {Hansen}, \citenamefont {Reinecke},\ and\ \citenamefont {Bartelman}}]{gorski_healpix_2005}%
  \BibitemOpen
  \bibfield  {author} {\bibinfo {author} {\bibfnamefont {K.~M.}\ \bibnamefont {Gorski}}, \bibinfo {author} {\bibfnamefont {E.}~\bibnamefont {Hivon}}, \bibinfo {author} {\bibfnamefont {A.~J.}\ \bibnamefont {Banday}}, \emph {et~al.},\ }\href {https://doi.org/10.1086/427976} {\bibfield  {journal} {\bibinfo  {journal} {The Astrophysical Journal}\ }\textbf {\bibinfo {volume} {622}},\ \bibinfo {pages} {759} (\bibinfo {year} {2005})},\ \bibinfo {note} {arXiv:astro-ph/0409513}\BibitemShut {NoStop}%
\bibitem [{\citenamefont {Tinto}\ and\ \citenamefont {Dhurandhar}(2020)}]{tinto_time-delay_2020}%
  \BibitemOpen
  \bibfield  {author} {\bibinfo {author} {\bibfnamefont {M.}~\bibnamefont {Tinto}}\ and\ \bibinfo {author} {\bibfnamefont {S.~V.}\ \bibnamefont {Dhurandhar}},\ }\href {https://doi.org/10.1007/s41114-020-00029-6} {\bibfield  {journal} {\bibinfo  {journal} {Living Reviews in Relativity}\ }\textbf {\bibinfo {volume} {24}},\ \bibinfo {pages} {1} (\bibinfo {year} {2020})}\BibitemShut {NoStop}%
\bibitem [{\citenamefont {Adams}\ and\ \citenamefont {Cornish}(2010)}]{adams_discriminating_2010}%
  \BibitemOpen
  \bibfield  {author} {\bibinfo {author} {\bibfnamefont {M.~R.}\ \bibnamefont {Adams}}\ and\ \bibinfo {author} {\bibfnamefont {N.~J.}\ \bibnamefont {Cornish}},\ }\href {https://doi.org/10.1103/PhysRevD.82.022002} {\bibfield  {journal} {\bibinfo  {journal} {Physical Review D}\ }\textbf {\bibinfo {volume} {82}},\ \bibinfo {pages} {022002} (\bibinfo {year} {2010})},\ \Eprint {https://arxiv.org/abs/1002.1291} {arxiv:1002.1291} \BibitemShut {NoStop}%
\bibitem [{\citenamefont {Hartwig}\ \emph {et~al.}(2023)\citenamefont {Hartwig}, \citenamefont {Lilley}, \citenamefont {Muratore},\ and\ \citenamefont {Pieroni}}]{hartwig_stochastic_2023}%
  \BibitemOpen
  \bibfield  {author} {\bibinfo {author} {\bibfnamefont {O.}~\bibnamefont {Hartwig}}, \bibinfo {author} {\bibfnamefont {M.}~\bibnamefont {Lilley}}, \bibinfo {author} {\bibfnamefont {M.}~\bibnamefont {Muratore}},\ and\ \bibinfo {author} {\bibfnamefont {M.}~\bibnamefont {Pieroni}},\ }\href {https://doi.org/10.1103/PhysRevD.107.123531} {\bibfield  {journal} {\bibinfo  {journal} {Physical Review D}\ }\textbf {\bibinfo {volume} {107}},\ \bibinfo {pages} {123531} (\bibinfo {year} {2023})}\BibitemShut {NoStop}%
\bibitem [{\citenamefont {Muratore}\ \emph {et~al.}(2023)\citenamefont {Muratore}, \citenamefont {Gair},\ and\ \citenamefont {Speri}}]{muratore_impact_2023}%
  \BibitemOpen
  \bibfield  {author} {\bibinfo {author} {\bibfnamefont {M.}~\bibnamefont {Muratore}}, \bibinfo {author} {\bibfnamefont {J.}~\bibnamefont {Gair}},\ and\ \bibinfo {author} {\bibfnamefont {L.}~\bibnamefont {Speri}},\ }\href {https://doi.org/10.48550/arXiv.2308.01056} {\bibinfo {title} {Impact of the noise knowledge uncertainty for the science exploitation of cosmological and astrophysical stochastic gravitational wave background with {{LISA}}}} (\bibinfo {year} {2023}),\ \Eprint {https://arxiv.org/abs/2308.01056} {arxiv:2308.01056 [astro-ph, physics:gr-qc]} \BibitemShut {NoStop}%
\bibitem [{\citenamefont {Bloom}\ \emph {et~al.}(2024)\citenamefont {Bloom}, \citenamefont {Criswell},\ and\ \citenamefont {Mandic}}]{bloom_datasets_2024}%
  \BibitemOpen
  \bibfield  {author} {\bibinfo {author} {\bibfnamefont {M.}~\bibnamefont {Bloom}}, \bibinfo {author} {\bibfnamefont {A.}~\bibnamefont {Criswell}},\ and\ \bibinfo {author} {\bibfnamefont {V.}~\bibnamefont {Mandic}},\ }\href {https://doi.org/10.5281/zenodo.14537985} {\bibinfo {title} {Datasets for "{{Angular Resolution}} of a {{Bayesian Search}} for {{Anisotropic Stochastic Gravitational Wave Backgrounds}} with {{LISA}}"}} (\bibinfo {year} {2024})\BibitemShut {NoStop}%
\end{thebibliography}%
\bibliographystyle{apsrev4-2.bst}

%\begin{thebibliography}{99}
%\end{thebibliography}

%%%%%%%%%%%%%%%%%%%%%%%%%%%%%%%%%%%%%%%%%%%%%%%%%%

%%%%%%%%%%%%%%%%% APPENDICES %%%%%%%%%%%%%%%%%%%%%

\appendix

\section{Insensitivity to Odd-$\ell_a$ Modes}\label{appendix:alms}
\begin{figure*}
    \centering
    \includegraphics[width=\linewidth]{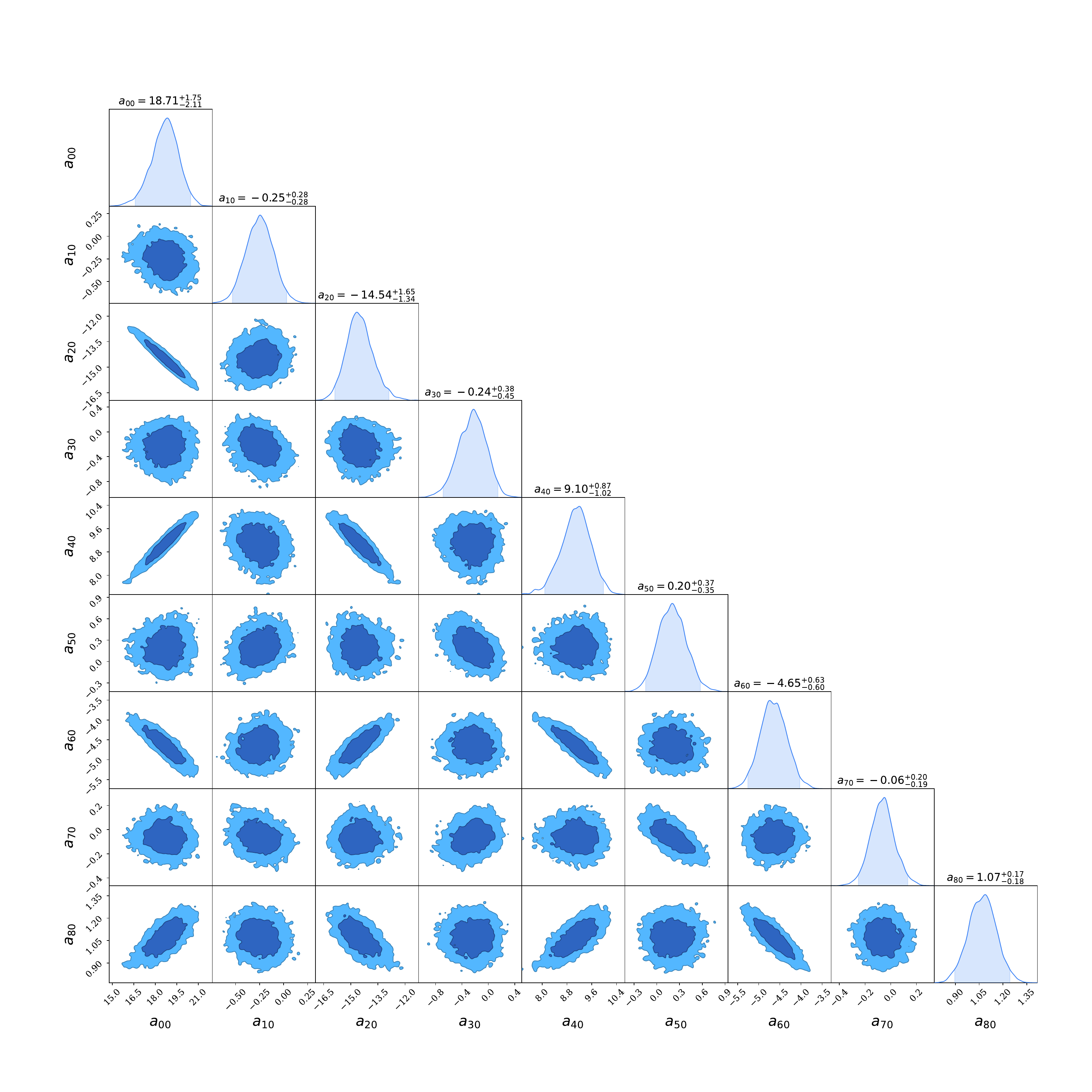}
    \caption{Transformed $a_{\ell m}$ posterior samples for simulation 116 ($\la=8$). Only $m=0$ terms are shown so as to preserve the legibility of the corner plot. Quoted bounds are the mean and 95\% C.I. of the posterior samples. The shading in the one-dimensional posterior distribution denotes the 95\% C.I.; the dark and light shaded regions of the two-dimensional distributions denote $1$- and $2\sigma$ bounds, respectively. Note that while the even-$\ell_a$ modes are well-recovered, the odd-$\ell_a$ mode posterior distributions are all consistent with zero. }
    \label{fig:alm_corners}
\end{figure*}

We consider in brief the insensitivity of LISA to odd-$\ell_a$ spherical harmonic modes discussed in \citep{kudoh_probing_2005,bartolo_probing_2022}. Transforming our $b_{\ell m}$ posterior samples back to the $a_{\ell m}$ basis, we find that the odd $a_{\ell m}$s are recovered to be consistent with zero (whereas the even modes are, generally, not). Fig.~\ref{fig:alm_corners} shows the transformed $a_{\ell m}$ posterior distribution for two-point simulation no. 116 (analyzed with $\la=8$; see Table~\ref{table:tp_sims}), including only $m=0$ terms for legibility.

\section{Simulation Details}\label{appendix:simdetails}
Table~\ref{table:sp_sims} (\ref{table:tp_sims}) includes the values of all simulation/analysis settings for the single-point (two-point) simulations discussed in this work. These parameters are described in \S\ref{sec:sims}.

\clearpage

\setlength{\tabcolsep}{8pt}
\setlength{\extrarowheight}{2pt}
\begin{center}
\begin{longtable*}{|l|c|c|c|c|c|c|}
\caption{Single-point simulations.} \label{table:sp_sims} \\

\hline \multicolumn{1}{|c|}{\textbf{Simulation}} & \multicolumn{1}{c|}{\textbf{$\la$}} & \multicolumn{1}{c|}{\textbf{$\Omega_{\text{ref}}$}} & \multicolumn{1}{c|}{\textbf{$ T_{\mathrm{obs}}$} (months)} &
\multicolumn{1}{c|}{\textbf{$\ell^{a}_{\text{max, inj}}$}} &
\multicolumn{1}{c|}{\textbf{$\theta$}} &
\multicolumn{1}{c|}{\textbf{$\phi$}}
\\ \hline 
\endfirsthead

\multicolumn{7}{c}%
{{\bfseries \tablename\ \thetable{} -- continued from previous page}} \\
\hline \multicolumn{1}{|c|}{\textbf{Simulation}} & 
\multicolumn{1}{c|}{\textbf{$\la$}} & \multicolumn{1}{c|}{\textbf{$\Omega_{\text{ref}}$}} & \multicolumn{1}{c|}{\textbf{$ T_{\mathrm{obs}}$}} &
\multicolumn{1}{c|}{\textbf{$\ell^{a}_{\text{max, inj}}$}} &
\multicolumn{1}{c|}{\textbf{$\theta$}} &
\multicolumn{1}{c|}{\textbf{$\phi$}} \\ \hline 
\endhead

\hline \multicolumn{7}{|r|}{{Continued on next page}} \\ \hline
\endfoot

\hline \hline
\endlastfoot

\hline
1 & 4 & $8.0\times10^{-9}$ & 3  & 16 & $\pi/2$ & $\pi/2$ \\
2 & 6 & $8.0\times10^{-9}$ & 3  & 16 & $\pi/2$ & $\pi/2$ \\
3 & 8 & $8.0\times10^{-9}$ & 3  & 16 & $\pi/2$ & $\pi/2$ \\
4 & 10 & $8.0\times10^{-9}$ & 3  & 16 & $\pi/2$ & $\pi/2$ \\
5 & 12 & $8.0\times10^{-9}$ & 3  & 16 & $\pi/2$ & $\pi/2$ \\
6 & 14 & $8.0\times10^{-9}$ & 3  & 16 & $\pi/2$ & $\pi/2$ \\
7 & 16 & $8.0\times10^{-9}$ & 3  & 16 & $\pi/2$ & $\pi/2$ \\
8 & 4 & $4.0\times10^{-8}$ & 3  & 16 & $\pi/2$ & $\pi/2$ \\
9 & 6 & $4.0\times10^{-8}$ & 3  & 16 & $\pi/2$ & $\pi/2$ \\
10 & 8 & $4.0\times10^{-8}$ & 3  & 16 & $\pi/2$ & $\pi/2$ \\
11 & 10 & $4.0\times10^{-8}$ & 3  & 16 & $\pi/2$ & $\pi/2$ \\
12 & 12 & $4.0\times10^{-8}$ & 3  & 16 & $\pi/2$ & $\pi/2$ \\
13 & 14 & $4.0\times10^{-8}$ & 3  & 16 & $\pi/2$ & $\pi/2$ \\
14 & 16 & $4.0\times10^{-8}$ & 3  & 16 & $\pi/2$ & $\pi/2$ \\
15 & 4 & $8.0\times10^{-8}$ & 3  & 16 & $\pi/2$ & $\pi/2$ \\
16 & 6 & $8.0\times10^{-8}$ & 3  & 16 & $\pi/2$ & $\pi/2$ \\
17 & 8 & $8.0\times10^{-8}$ & 3  & 16 & $\pi/2$ & $\pi/2$ \\
18 & 10 & $8.0\times10^{-8}$ & 3  & 16 & $\pi/2$ & $\pi/2$ \\
19 & 12 & $8.0\times10^{-8}$ & 3  & 16 & $\pi/2$ & $\pi/2$ \\
20 & 14 & $8.0\times10^{-8}$ & 3  & 16 & $\pi/2$ & $\pi/2$ \\
21 & 16 & $8.0\times10^{-8}$ & 3  & 16 & $\pi/2$ & $\pi/2$ \\
22 & 4 & $4.0\times10^{-7}$ & 3  & 16 & $\pi/2$ & $\pi/2$ \\
23 & 6 & $4.0\times10^{-7}$ & 3  &16 & $\pi/2$ & $\pi/2$ \\
24 & 8 & $4.0\times10^{-7}$ & 3  & 16 & $\pi/2$ & $\pi/2$ \\
25 & 10 & $4.0\times10^{-7}$ & 3  & 16 & $\pi/2$ & $\pi/2$ \\
26 & 12 & $4.0\times10^{-7}$ & 3  & 16 & $\pi/2$ & $\pi/2$ \\
27 & 14 & $4.0\times10^{-7}$ & 3  & 16 & $\pi/2$ & $\pi/2$ \\
28 & 16 & $4.0\times10^{-7}$ & 3  & 16 & $\pi/2$ & $\pi/2$ \\
\hline
29 & 4 & $4.8\times10^{-9}$ & 3  & 4 & $\pi/2$ & $\pi/2$ \\
30 & 4 & $5.6\times10^{-9}$ & 3  & 4 & $\pi/2$ & $\pi/2$ \\
31 & 4 & $6.4\times10^{-9}$ & 3  & 4 & $\pi/2$ & $\pi/2$ \\
32 & 4 & $7.2\times10^{-9}$ & 3  & 4 & $\pi/2$ & $\pi/2$ \\
33 & 4 & $8.0\times10^{-9}$ & 3  & 4 & $\pi/2$ & $\pi/2$ \\
34 & 4 & $1.6\times10^{-8}$ & 3  & 4 & $\pi/2$ & $\pi/2$ \\
35 & 4 & $2.4\times10^{-8}$ & 3  & 4 & $\pi/2$ & $\pi/2$ \\
36 & 4 & $3.2\times10^{-8}$ & 3  & 4 & $\pi/2$ & $\pi/2$ \\
37 & 4 & $4.0\times10^{-8}$ & 3  & 4 & $\pi/2$ & $\pi/2$ \\
38 & 4 & $4.8\times10^{-8}$ & 3  & 4 & $\pi/2$ & $\pi/2$ \\
39 & 4 & $5.6\times10^{-8}$ & 3  & 4 & $\pi/2$ & $\pi/2$ \\
40 & 4 & $6.4\times10^{-8}$ & 3  & 4 & $\pi/2$ & $\pi/2$ \\
41 & 4 & $7.2\times10^{-8}$ & 3  & 4 & $\pi/2$ & $\pi/2$ \\
42 & 4 & $8.0\times10^{-8}$ & 3  & 4 & $\pi/2$ & $\pi/2$ \\
43 & 4 & $1.6\times10^{-7}$ & 3  & 4 & $\pi/2$ & $\pi/2$ \\
44 & 4 & $2.4\times10^{-7}$ & 3  & 4 & $\pi/2$ & $\pi/2$ \\
45 & 4 & $3.2\times10^{-7}$ & 3  & 4 & $\pi/2$ & $\pi/2$ \\
46 & 4 & $4.0\times10^{-7}$ & 3  & 4 & $\pi/2$ & $\pi/2$ \\
47 & 4 & $2.4\times10^{-9}$ & 6  & 4 & $\pi/2$ & $\pi/2$ \\
48 & 4 & $3.2\times10^{-9}$ & 6  & 4 & $\pi/2$ & $\pi/2$ \\
49 & 4 & $4.0\times10^{-9}$ & 6  & 4 & $\pi/2$ & $\pi/2$ \\
50 & 4 & $4.8\times10^{-9}$ & 6  & 4 & $\pi/2$ & $\pi/2$ \\
51 & 4 & $5.6\times10^{-9}$ & 6  & 4 & $\pi/2$ & $\pi/2$ \\
52 & 4 & $6.4\times10^{-9}$ & 6  & 4 & $\pi/2$ & $\pi/2$ \\
53 & 4 & $7.2\times10^{-9}$ & 6  & 4 & $\pi/2$ & $\pi/2$ \\
54 & 4 & $8.0\times10^{-9}$ & 6  & 4 & $\pi/2$ & $\pi/2$ \\
55 & 4 & $1.6\times10^{-8}$ & 6  & 4 & $\pi/2$ & $\pi/2$ \\
56 & 4 & $2.4\times10^{-8}$ & 6  & 4 & $\pi/2$ & $\pi/2$ \\
57 & 4 & $3.2\times10^{-8}$ & 6  & 4 & $\pi/2$ & $\pi/2$ \\
58 & 4 & $4.0\times10^{-8}$ & 6  & 4 & $\pi/2$ & $\pi/2$ \\
59 & 4 & $4.8\times10^{-8}$ & 6  & 4 & $\pi/2$ & $\pi/2$ \\
60 & 4 & $5.6\times10^{-8}$ & 6  & 4 & $\pi/2$ & $\pi/2$ \\
61 & 4 & $6.4\times10^{-8}$ & 6  & 4 & $\pi/2$ & $\pi/2$ \\
62 & 4 & $7.2\times10^{-8}$ & 6  & 4 & $\pi/2$ & $\pi/2$ \\
63 & 4 & $8.0\times10^{-8}$ & 6  & 4 & $\pi/2$ & $\pi/2$ \\
64 & 4 & $1.6\times10^{-7}$ & 6  & 4 & $\pi/2$ & $\pi/2$ \\
65 & 4 & $2.4\times10^{-7}$ & 6  & 4 & $\pi/2$ & $\pi/2$ \\
66 & 4 & $3.2\times10^{-7}$ & 6  & 4 & $\pi/2$ & $\pi/2$ \\
67 & 4 & $4.0\times10^{-8}$ & 6  & 4 & $\pi/2$ & $\pi/2$ \\
68 & 4 & $1.6\times10^{-9}$ & 12  & 4 & $\pi/2$ & $\pi/2$ \\
69 & 4 & $2.4\times10^{-9}$ & 12  & 4 & $\pi/2$ & $\pi/2$ \\
70 & 4 & $3.2\times10^{-9}$ & 12  & 4 & $\pi/2$ & $\pi/2$ \\
71 & 4 & $4.0\times10^{-9}$ & 12  & 4 & $\pi/2$ & $\pi/2$ \\
72 & 4 & $4.8\times10^{-9}$ & 12  & 4 & $\pi/2$ & $\pi/2$ \\
73 & 4 & $5.6\times10^{-9}$ & 12  & 4 & $\pi/2$ & $\pi/2$ \\
74 & 4 & $6.4\times10^{-9}$ & 12  & 4 & $\pi/2$ & $\pi/2$ \\
75 & 4 & $7.2\times10^{-9}$ & 12  & 4 & $\pi/2$ & $\pi/2$ \\
76 & 4 & $8.0\times10^{-9}$ & 12 & 4 & $\pi/2$ & $\pi/2$ \\
77 & 4 & $1.6\times10^{-8}$ & 12 & 4 & $\pi/2$ & $\pi/2$ \\
78 & 4 & $2.4\times10^{-8}$ & 12 & 4 & $\pi/2$ & $\pi/2$ \\
79 & 4 & $3.2\times10^{-8}$ & 12 & 4 & $\pi/2$ & $\pi/2$ \\
80 & 4 & $4.0\times10^{-8}$ & 12 & 4 & $\pi/2$ & $\pi/2$ \\
81 & 4 & $4.8\times10^{-8}$ & 12 & 4 & $\pi/2$ & $\pi/2$ \\
82 & 4 & $5.6\times10^{-8}$ & 12 & 4 & $\pi/2$ & $\pi/2$ \\
83 & 4 & $6.4\times10^{-8}$ & 12 & 4 & $\pi/2$ & $\pi/2$ \\
84 & 4 & $7.2\times10^{-8}$ & 12 & 4 & $\pi/2$ & $\pi/2$ \\
85 & 4 & $8.0\times10^{-8}$ & 12 & 4 & $\pi/2$ & $\pi/2$ \\
86 & 4 & $1.6\times10^{-7}$ & 12 & 4 & $\pi/2$ & $\pi/2$ \\
87 & 4 & $2.4\times10^{-7}$ & 12 & 4 & $\pi/2$ & $\pi/2$ \\
88 & 4 & $3.2\times10^{-7}$ & 12 & 4 & $\pi/2$ & $\pi/2$ \\
89 & 4 & $4.0\times10^{-7}$ & 12 & 4 & $\pi/2$ & $\pi/2$ \\
\hline
\caption{List of single point source simulations used in this paper. Simulations 1-28 are reported in Fig. \ref{fig:sp_svslmax}. Simulations 29-89 are reported in Fig. \ref{fig:sp_svsomega}. }
\end{longtable*}
\end{center}

\clearpage

\setlength{\tabcolsep}{8pt}
\setlength{\extrarowheight}{2pt}
\begin{center}
\begin{longtable*}{|l|c|c|c|c|c|c|c|c|c|}
\caption{Two-point simulations.} \label{table:tp_sims} \\
\hline 
\multicolumn{1}{|c|}{\textbf{Simulation}} & 
\multicolumn{1}{c|}{\textbf{$\la$}} & 
\multicolumn{1}{c|}{\textbf{$\Omega_{\text{ref}}$}} & 
\multicolumn{1}{c|}{\textbf{$ T_{\mathrm{obs}}$} (months)} &
\multicolumn{1}{c|}{\textbf{$\ell^{a}_{\text{max, inj}}$}} &
\multicolumn{1}{c|}{\textbf{$\theta_1$}} &
\multicolumn{1}{c|}{\textbf{$\phi_1$}} &
\multicolumn{1}{c|}{\textbf{$\theta_2$}} &
\multicolumn{1}{c|}{\textbf{$\phi_2$}} &
\multicolumn{1}{c|}{\textbf{$\Delta \phi$}}
\\ \hline 
\endfirsthead
\multicolumn{10}{c}%
{{\bfseries \tablename\ \thetable{} -- continued from previous page}} \\
\hline \multicolumn{1}{|c|}{\textbf{Simulation}} & 
\multicolumn{1}{c|}{\textbf{$\la$}} & \multicolumn{1}{c|}{\textbf{$\Omega_{\text{ref}}$}} & \multicolumn{1}{c|}{\textbf{$ T_{\mathrm{obs}}$}} &
\multicolumn{1}{c|}{\textbf{$\ell^{a}_{\text{max, inj}}$}} &
\multicolumn{1}{c|}{\textbf{$\theta$}} &
\multicolumn{1}{c|}{\textbf{$\phi$}} \\ \hline 
\endhead
\hline \multicolumn{7}{|r|}{{Continued on next page}} \\ \hline
\endfoot
\hline \hline
\endlastfoot
\hline
90 & 4 & $8.0\times10^{-9}$ & 6  & 12 & $\pi/2$ & $\pi/5$ & $\pi/2 $ & $-\pi/5$ & $2\pi/5$\\
91 & 6 & $8.0\times10^{-9}$ & 6  & 12 & $\pi/2$ & $\pi/5$ & $\pi/2 $ & $-\pi/5$ & $2\pi/5$\\
92 & 8 & $8.0\times10^{-9}$ & 6  & 12 & $\pi/2$ & $\pi/5$ & $\pi/2 $ & $-\pi/5$ & $2\pi/5$\\
93 & 10 & $8.0\times10^{-9}$ & 6  & 12 & $\pi/2$ & $\pi/5$ & $\pi/2 $ & $-\pi/5$ & $2\pi/5$\\
94 & 4 & $8.0\times10^{-9}$ & 6  & 12 & $\pi/2$ & $2\pi/5$ & $\pi/2 $ & $-2\pi/5$ & $4\pi/5$\\
95 & 6 & $8.0\times10^{-9}$ & 6  & 12 & $\pi/2$ & $2\pi/5$ & $\pi/2 $ & $-2\pi/5$ & $4\pi/5$\\
96 & 8 & $8.0\times10^{-9}$ & 6  & 12 & $\pi/2$ & $2\pi/5$ & $\pi/2 $ & $-2\pi/5$ & $4\pi/5$\\
97 & 10 & $8.0\times10^{-9}$ & 6  & 12 & $\pi/2$ & $2\pi/5$ & $\pi/2 $ & $-2\pi/5$ & $4\pi/5$\\
98$^{\dagger}$ & 10 & $8.0\times10^{-9}$ & 6  & 12 & $\pi/2$ & $0$ & $\pi/2 $ & $-4\pi/5$ & $4\pi/5$\\
99 & 4 & $8.0\times10^{-9}$ & 6  & 12 & $\pi/2$ & $3\pi/5$ & $\pi/2 $ & $-3\pi/5$ & $6\pi/5$\\
100 & 6 & $8.0\times10^{-9}$ & 6  & 12 & $\pi/2$ & $3\pi/5$ & $\pi/2 $ & $-3\pi/5$ & $6\pi/5$\\
101 & 8 & $8.0\times10^{-9}$ & 6  & 12 & $\pi/2$ & $3\pi/5$ & $\pi/2 $ & $-3\pi/5$ & $6\pi/5$\\
102 & 10 & $8.0\times10^{-9}$ & 6  & 12 & $\pi/2$ & $3\pi/5$ & $\pi/2 $ & $-3\pi/5$ & $6\pi/5$\\
103 & 4 & $8.0\times10^{-9}$ & 6  & 12 & $\pi/2$ & $4\pi/5$ & $\pi/2 $ & $-4\pi/5$ & $8\pi/5$\\
104 & 6 & $8.0\times10^{-9}$ & 6  & 12 & $\pi/2$ & $4\pi/5$ & $\pi/2 $ & $-4\pi/5$ & $8\pi/5$\\
105 & 8 & $8.0\times10^{-9}$ & 6  & 12 & $\pi/2$ & $4\pi/5$ & $\pi/2 $ & $-4\pi/5$ & $8\pi/5$\\
106 & 10 & $8.0\times10^{-9}$ & 6  & 12 & $\pi/2$ & $4\pi/5$ & $\pi/2 $ & $-4\pi/5$ & $8\pi/5$\\
107 & 4 & $8.0\times10^{-9}$ & 6  & 12 & $\pi/2$ & $\pi/2$ & $\pi/2 $ & $-\pi/2$ & $\pi$\\
108 & 6 & $8.0\times10^{-9}$ & 6  & 12 & $\pi/2$ & $\pi/2$ & $\pi/2 $ & $-\pi/2$ & $\pi$\\
109 & 8 & $8.0\times10^{-9}$ & 6  & 12 & $\pi/2$ & $\pi/2$ & $\pi/2 $ & $-\pi/2$ & $\pi$\\
110 & 10 & $8.0\times10^{-9}$ & 6  & 12 & $\pi/2$ & $\pi/2$ & $\pi/2 $ & $-\pi/2$ & $\pi$\\
\hline
111 & 8 & $4.0\times10^{-8}$ & 6  & 12 & $\pi/2$ & $\pi/5$ & $\pi/2 $ & $-\pi/5$ & $2\pi/5$\\
112 & 8 & $4.0\times10^{-8}$ & 6  & 12 & $\pi/2$ & $2\pi/5$ & $\pi/2 $ & $-2\pi/5$ & $4\pi/5$\\
113 & 8 & $4.0\times10^{-8}$ & 6  & 12 & $\pi/2$ & $3\pi/5$ & $\pi/2 $ & $-3\pi/5$ & $6\pi/5$\\
114 & 8 & $4.0\times10^{-8}$ & 6  & 12 & $\pi/2$ & $4\pi/5$ & $\pi/2 $ & $-4\pi/5$ & $8\pi/5$\\
115 & 8 & $4.0\times10^{-8}$ & 6  & 12 & $\pi/2$ & $\pi/2$ & $\pi/2 $ & $-\pi/2$ & $\pi$\\
116 & 8 & $8.0\times10^{-8}$ & 6  & 12 & $\pi/2$ & $\pi/5$ & $\pi/2 $ & $-\pi/5$ & $2\pi/5$\\
117 & 8 & $8.0\times10^{-8}$ & 6  & 12 & $\pi/2$ & $2\pi/5$ & $\pi/2 $ & $-2\pi/5$ & $4\pi/5$\\
118 & 8 & $8.0\times10^{-8}$ & 6  & 12 & $\pi/2$ & $3\pi/5$ & $\pi/2 $ & $-3\pi/5$ & $6\pi/5$\\
119 & 8 & $8.0\times10^{-8}$ & 6  & 12 & $\pi/2$ & $4\pi/5$ & $\pi/2 $ & $-4\pi/5$ & $8\pi/5$\\
120 & 8 & $8.0\times10^{-8}$ & 6  & 12 & $\pi/2$ & $\pi/2$ & $\pi/2 $ & $-\pi/2$ & $\pi$\\

\hline
\caption{List of two point source simulations used in this paper. Simulations 90-110 are reported in Fig. \ref{fig:tp_metric_vs_sepangl}. Simulations 92, 96, 101, 105, 109, and 111-120 are reported in Fig. \ref{fig:tp_metric_vs_sepang}. }
\end{longtable*}
\end{center}

\label{lastpage}

% Produces the bibliography via BibTeX.

\end{document}